\documentclass{article}

\usepackage[svgnames, x11names]{xcolor}

\usepackage[normalem]{ulem} 

\usepackage{listings}

\lstdefinelanguage{XML}
{
	basicstyle=\ttfamily\footnotesize,
	morestring=[b]",
	moredelim=[s][\bfseries\color{Maroon}]{<}{\ },
	moredelim=[s][\bfseries\color{Maroon}]{</}{>},
	moredelim=[l][\bfseries\color{Maroon}]{/>},
	moredelim=[l][\bfseries\color{Maroon}]{>},
	morecomment=[s]{<?}{?>},
	morecomment=[s]{<!--}{-->},
	commentstyle=\color{gray},
	stringstyle=\color{blue},
	identifierstyle=\color{red}
}
\usepackage[pdftex]{graphicx}
\graphicspath{{./figures/}}
\DeclareGraphicsExtensions{.pdf}

\usepackage[cmex10]{amsmath}
\usepackage{amssymb}
\usepackage{mathtools}

\usepackage[]{subfig}

\usepackage{algorithmicx}
\usepackage{algpseudocode}
\usepackage[ruled]{algorithm}
\definecolor{light-gray}{gray}{0.75}
\algrenewcommand{\algorithmiccomment}[1]{\hskip3em{{\footnotesize \textcolor{light-gray}{$\blacktriangleright$}}} #1}

\usepackage{multirow}

\usepackage{hyperref}

\usepackage{xspace}
\usepackage[nocompress]{cite}

\usepackage{enumitem}

\usepackage[english]{babel}
\usepackage{blindtext}

\begin{document}
	%
	\title{Model-driven Scheduling for Distributed Stream Processing Systems}
	%
	%
	%
	%
	
	\author{Anshu Shukla and Yogesh Simmhan \\
		\normalsize{\emph{Department of Computational and Data Sciences}}\\
		\normalsize{\emph{Indian Institute of Science (IISc), Bangalore 560012, India}}\\
		\normalsize{\emph{Email: shukla@grads.cds.iisc.ac.in, simmhan@cds.iisc.ac.in}}}
	
	\date{}
	\maketitle
	
	\begin{abstract}
	Distributed Stream Processing frameworks are being commonly used with the  evolution of Internet of Things(IoT). These frameworks are designed to adapt to the dynamic input  message rate by scaling in/out.Apache Storm, originally developed by Twitter is  a widely used stream processing engine while others includes Flink~\cite{flink:tcde:2015} Spark streaming~\cite{spark:usenix:2012}. For running the streaming applications successfully there is need to know the optimal resource requirement, as over-estimation of resources adds extra cost.So we need some strategy to come up with the optimal resource requirement for a given streaming application. In this article, we propose a model-driven approach for scheduling streaming applications that effectively utilizes \emph{a priori} knowledge of the applications to provide predictable scheduling behavior. Specifically, we use application performance models to offer reliable estimates of the resource allocation required. Further, this intuition also drives resource mapping, and helps narrow the estimated and actual dataflow performance and resource utilization. Together, this model-driven scheduling approach gives a predictable application performance and resource utilization behavior for executing a given DSPS application at a target input stream rate on distributed resources.
					
	\end{abstract}

\section{Introduction}

Big Data platforms have evolved over the last decade to address the unique challenges posed by the ability to collect data at vast scales, and the need to process them rapidly. These platforms have also leveraged the availability of distributed computing resources, such as commodity clusters and Clouds, to allow the application to scale out as data sizes grow. In particular, platforms like Apache Hadoop and Spark have allowed massive data volumes to be processed with high throughput, and NoSQL databases like Hive and HBase support low latency queries over semi-structured data at large scales. 

However, much of the research and innovation in Big Data platforms has skewed toward the \emph{volume} rather than the \emph{velocity} dimension~\cite{laney:gartner:2001}. On the other hand, the growing prevalence of Internet of Things (IoT) is contributing to the deployment of physical and virtual sensors to monitor and respond to infrastructure, nature and human activity is leading to a rapid influx of streaming data~\cite{buyya:fgcs:2013}. These emerging applications complement the existing needs of micro-blogs like Twitter that already contend with the need to rapidly process tweet streams for detecting trends or malicious activity~\cite{kulkarni:twitter:2015}. Such streaming applications require low-latency processing and analysis of data streams to take decisions that control the physical or digital eco-system they observe.

A \emph{Distributed Stream Processing System (DSPS)} is a Big Data platform designed for online processing of such data streams~\cite{balazinska:cidr:2003}. While early stream processing systems date back to applications on wireless sensor networks~\cite{chandrasekaran:cidr:2003}, contemporary DSPS's such as Apache Storm from Twitter, Flink and Spark Streaming are designed to execute complex dataflows over tuple streams using commodity clusters~\cite{toshniwal:sigmod:2014,flink:tcde:2015,spark:usenix:2012}. These dataflows are typically designed as Directed Acyclic Graphs (DAGs), where user tasks are vertices and streams are edges.  They can leverage data parallelism across tuples in the stream using multiple threads of execution per task, in addition to pipelined and task-parallel execution of the DAG, and have been shown to process $1000's$ of tuples per second~\cite{toshniwal:sigmod:2014,shukla:tpctc:2016}.

A DSPS executes streaming dataflow applications on distributed resources such as commodity clusters and Cloud Virtual Machines (VMs). In order to meet the required performance for these applications, the DSPS needs to schedule these dataflows efficiently over the resources. Scheduling for a DSPS has two parts: \emph{resource allocation} and \emph{resource mapping}. The former determines the appropriate degrees of parallelism per task (e.g., threads of execution) and quanta of computing resources (e.g., number and type of VMs) for the given dataflow. Here, care has to be taken to avoid both over-allocation, that can have monetary costs when using Cloud VMs, or under-allocation, that can impact performance. Resource mapping decides the specific assignment of the threads to the VMs to ensure that the expected performance behavior and resource utilization is met. 

Despite their growing use, resource scheduling for DSPSs tends to be done in an \emph{ad hoc} manner, favoring empirical and reactive approaches, rather than a model-driven and analytical approach. Such empirical approaches may arrive at an approximate resource allocation for the DSPS based on a linear extrapolation of the resource needs and performance of dataflow tasks, and hand-tune these to meet the Quality of Service (QoS)~\cite{schneider:ipdps:2009,xu:icdcs:2014}. Mapping of tasks to resources may be round-robin or consider data locality and resource capacities~\cite{khandekar:middleware:2009,rychly:cicic:2014}. More sophisticated research techniques support dynamic scheduling by monitoring the queue waiting times or tuple latencies to incrementally increase/decrease the degrees of parallelism for individual tasks or the number of VMs they run on~\cite{kumbhare:ccgrid:2014,fu:icdcs:2015}. While these dynamic techniques have the advantage of working for arbitrary dataflows and stream rates, such schedules can lead to local optima for individual tasks without regard to global efficiency of the dataflow, introduce latency and cost overheads due to constant changes to the mapping, or offer weaker guarantees for the QoS.

In this article, we propose a model-driven approach for scheduling streaming applications that effectively utilizes \emph{a priori} knowledge of the applications to provide predictable scheduling behavior. Specifically, we leverage our observation that dataflow tasks have diverse performance behavior as the degree of parallelism increases, and use performance models to offer reliable estimates of the resource allocation required. Further, this intuition also drives resource mapping to mitigate the impact of multi-tenancy of different tasks on the same resource, and helps narrow the estimated and actual dataflow performance and resource utilization. Together, this model-driven scheduling approach gives a predictable application performance and resource utilization behavior for executing a given DSPS application at a target input stream rate on distributed resources.
Often, importance is given to lower latency of resource usage rather than predictable behavior. But in stream processing that can be latency sensitive, it may be more important to offer tighter bounds rather than lower bounds.

We limit this article to static scheduling of the dataflow on distributed resources, before the application starts running. This is complementary to dynamic scheduling algorithms that can react to changes in the stream rates~\cite{xu:ic2e:2016}, application composition~\cite{kumbhare:ccgrid:2014} and make use of Cloud elasticity~\cite{esc:cloud:2011}. However, our work can be extended and applied to a dynamic context as well. Rather than incrementally increase or decrease resource allocation and the mapping until the QoS stabilizes, a dynamic algorithm can make use of our model to converge to a stable configuration more rapidly. Our work is of particular use for enterprises and service providers who have a large class of infrastructure applications that are run frequently~\cite{jha:cep:2016,simmhan:wbdb:2015}, or who reuse a library of common tasks when composing their applications~\cite{fb:sigmod:2016,filgueira:escience:2015,biem:sigmod:2010}, as is common in the scientific workflow community~\cite{myexperiment}. This amortizes the cost of building task-level performance models. Our approach also resembles scheduling in HPC centers that typically have a captive set of scientific applications that can benefit from such a model-driven approach ~\cite{quasar:sigplan:2014}~\cite{weidendorfer:hipeac:2016}.




Specifically, we make the following key contributions in this article:
\begin{enumerate}
	\item We highlight the gap between ideal and actual performance of dataflow tasks using performance models, that causes many existing DSPS scheduling algorithms to fail and motivates our model-based approach for reliable scheduling.
	\item We propose an allocation and a mapping algorithm that leverage these performance models to schedule DSPS dataflows for a fixed input rate, minimizing the distributed resources used and offering predictable performance behavior.
	\item We offer detailed experimental results and analysis evaluating our scheduling algorithm using the Apache Storm DSPS, and compare it against the state-of-the-art scheduling approaches, for micro and application dataflows.
\end{enumerate}

The rest of the article is organized as follows: \S~\ref{sec:bg} introduces the problem \emph{motivation} and \S~\ref{sec:problem} formalizes the scheduling \emph{problem}; \S~\ref{sec:approach} offers a high-level intuition of the analytical \emph{approach} taken to solving the problem; \S~\ref{sec:bm} offers evidence on the diversity of task's behavior using \emph{performance models}, leveraged in the solution; \S~\ref{sec:allocation} proposes a novel \emph{Model Based Allocation (MBA) algorithm} using these models, and also describes a Linear Scaling Allocation (LSA) used as a contemporary baseline; \S~\ref{sec:mapping} presents our \emph{Slot-Aware Mapping (SAM) algorithm} that leverages thread-locality in a resource slot, and lists two existing algorithms from literature and practice used as comparison; we offer detailed experimental \emph{results and analysis} of these allocation and mapping algorithms in \S~\ref{sec:results}; contrast our work against \emph{related literature} in \S~\ref{sec:related}; and lastly, present our \emph{conclusions} in \S~\ref{sec:conclusions}.

\section{Background and Motivation}
\label{sec:bg}
We offer an overview of the generic composition and execution model favored by contemporary DSPSs such as Apache Storm, Apache Spark Streaming, Apache Flink and IBM InfoSphere Streams in this section. We further use this to motivate the specific research challenges and technical problems we address in this article; a formal definition follows in the subsequent section, \S~\ref{sec:problem}. While we use features and limitations of the popular Apache Storm as a representative DSPS here, similar features and short-comings of other DSPSs are discussed in the related work, \S~\ref{sec:related}.

Streaming applications are composed as a dataflow in a DSPS, represented as a \textit{directed acyclic graph (DAG)}, where \emph{tasks} form vertices and \textit{tuple streams} are the edges connecting the output of one task to the input of its downstream task. Contents of the \emph{tuples} (also called \emph{events} or \emph{messages}) are generally opaque to the DSPS, except for special fields like IDs and keys that may be present for recovery and routing. \emph{Source tasks} in the DAG contain user logic responsible for acquiring and generating the initial input stream to the DAG, say by subscribing to a message broker or pulling events over the network from sensors. For other tasks, their logic is executed once for each input tuple arriving at that task, and may produce zero or more output tuples for each invocation. These output tuples are passed to downstream tasks in the DAG, and so on till the \emph{Sink tasks} are reached. These sinks do not emit any output stream but may store the tuples or notify external services. Apache Storm uses the terms \emph{topology} and \emph{component} for a DAG and a task, and more specifically \emph{spout} and \emph{bolt} for source tasks and non-source tasks, respectively.

Multiple outgoing edges connecting one task to downstream tasks may indicate different \emph{routing semantics} for output tuples on that edge, based on the application definition -- tuples may be \emph{duplicated} to all downstream tasks, passed in a \emph{round-robin} manner to the tasks, or \emph{mapped} based on an output key in the tuple. Likewise, multiple input streams incident on a task typically have their tuples \emph{interleaved} into a single logical stream, though semantics such as \emph{joins} across tuples from different input streams are possible as well. 
Further, the \emph{selectivity} of an outgoing edge of a task is the ratio between the average number of tuples generated on that output stream for each input tuple to the task.  

Streaming applications are designed to process tuples with low latency. The \emph{end-to-end latency} for processing an input tuple from the source to the sink task(s) is typically a measure of the \emph{Quality of Service (QoS)} expected. This QoS depends on both the \emph{input stream rate} at the source task(s) and the resource allocation to the tasks in the DSPS. A key requirement is that the execution performance of the streaming application remains \emph{stable}, i.e., the end-to-end latency is maintained within a narrow range over time and the queue size at each task does not grow. Otherwise, an unstable application can lead to an exponential growth in the latency and the queue size, causing hosts to run out of memory.

The execution model of a DSPS can naturally leverage \emph{pipelining} and \emph{task parallelism} due to the composition of linear and concurrent tasks in the DAG, respectively. These benefits are bound by the length and the number of tasks in the DAG. In addition, they also actively make use of \emph{data-parallel} execution for a single task by assigning multiple threads of execution that can each operate on an independent tuple in the input stream. This data parallelism is typically limited to stateless tasks, where threads of execution for a task do not share a global variable or state, such as a sum and a count for aggregation; stateful tasks require more involved distributed coordination~\cite{wu:icdes:2012}. Stateless tasks are common in streaming dataflows, allowing the DSPS to make use of this \emph{important dimension of parallelism that is not limited by the dataflow size but rather the stream rate and resource availability.}

In operational DSPSs such as Apache Storm, Yahoo S4~\cite{neumeyer:icdmw:2010}, Twitter Heron~\cite{kulkarni:twitter:2015} and IBM InfoSphere Streams~\cite{biem:sigmod:2010}, the platform expects the application developer to provide the \emph{number of threads} or degrees of data parallelism that should be exploited for a single task. As we show in \S~\ref{sec:bm}, general rules of thumb are inadequate for deciding this and both over- and under-allocation of threads can have a negative effect on performance. This value may also change with the input rate of the stream. Thread allocation is one of the challenges we tackle in this paper.

In addition, the user is responsible for deciding the \emph{number of compute resources} to be allocated to the DAG. Typically, as with many Big Data platforms, each host or Virtual Machine (VM) in the DSPS cluster exposes multiple \emph{resource slots}, and those many \emph{worker processes} can be run on the host. Typically, there are as many workers as the number of CPU cores in that host. Each worker process can internally run multiple threads for one or more tasks in the DAG. The user specifies the number of hosts or slots in the DSPS cluster to be used for the given DAG when it is submitted for execution.  For e.g., in Apache Storm, the threads for the dataflow can use all available resource slots in the DSPS cluster by default. 
In practice, this again ends up being a trial and error empirical process for the user or the system to decide the resource allocation for the DAG, and will change for each DAG or the input rate that it needs to support. Ascertaining the Cloud VM resource allocation for the given DAG and input rate is another problem that we address in this article, and this is equally applicable to commodity hosts in a cluster as well.

The DSPS, on receiving a DAG for scheduling, is responsible for deploying the dataflow on the cluster and coordinating its execution. As part of this deployment, it needs to decide the mapping from the threads of the tasks to the slots in the workers. There has been prior work on making this mapping efficient for stream processing platforms~\cite{eidenbenz:infocom:2016}~\cite{khandekar:middleware:2009}. For e.g., the native scheduler of Apache Storm uses a round-robin technique for assigning threads from different tasks to all available slots for the DAG. Its intuition is to avoid contention for the same resource by threads for the same task, and also balance the workload among the available workers. More recently, Storm has incorporated the R-Storm scheduler~\cite{peng:middleware:2015} that is both resource and network topology aware, and this offers better efficiency by considering the resource needs for an individual task thread. In InfoSphere Streams~\cite{biem:sigmod:2010} and our own prior work~\cite{kumbhare:sc:2013}, the mapping decision is dynamic and relies on the current load and previous resource utilization for the tasks.

This placement decision is important, as an optimal resource allocation for a given DAG and input rate may still under-perform if the task thread to worker mapping is not effective. This inefficiency can be due to additional costs for resource contention between threads of a task or different tasks on a VM, excess context switching between threads in a core, movement of tuples between distributed tasks over the network, among other reasons. This inefficiency is manifest in the form of additional latency for the messages to be processed, or a lower input rate that is supported in a stable manner for the DAG with the given resources and mapping. In this article, we enhance the mapping strategy for the DSPS by using a model-driven approach that goes beyond a resource-aware approach, such as used in R-Storm.


In summary, the aim of this paper is to \emph{use a model-driven approach to perform predictable and efficient resource scheduling for a given DAG and input event rate}. The specific goals are to determine:
\begin{itemize}
	\item the thread allocation per task,
	\item the VM allocation for the entire DAG, and 
	\item the resource mapping from a task thread to a resource slot.
\end{itemize}  

The outcome of this schedule will be to improve the efficiency and reduce the contention for VM resources, reduce the number of VM resources, and hence \emph{monetary costs}, for executing the DAG, and ensure a stable (and preferably low) latency for execution. The \emph{predictable performance} of the proposed schedule is also important as it reduces uncertainty and trial and error. Further, when using this scheduling approach for handling dynamism in the workload or resources, say when the input rate or the DAG structure changes, this predictability allows us to update the schedule and pay the overhead cost for the rebalancing once, rather than constantly tweak the schedule purely based on monitoring of the execution. 

While our article does not directly address dynamism, such as changing input rates, non-deterministic VM performance or modifications to the DAG, the approach we propose offers a valuable methodology for supporting it. Likewise, we limit our work here to scheduling a single streaming application on an exclusive cluster, that is common in on-demand Cloud environments, rather than multi-tenancy of applications in the same cluster. Our algorithm in fact helps determine the smallest number of VMs and their sizes required to meet the application's needs.

\section{Problem Definition}
\label{sec:problem}
\begin{figure}[t]
	\centering
	\parbox{1.0\textwidth}{
		\footnotesize
		\[\begin{array}{cl}
		\hline
		\textsc{Notation} & \textsc{Description}\\
		\hline
		\hline 
		\mathcal{G} = \langle \mathbb {T},\mathbb {E} \rangle & \text{DAG to be scheduled}\\
		\mathbb{T}=\{t_1,...,t_n\} & \text{Set of $n$ task vertices in the DAG}\\
		\mathbb {E} =\{e_{ij} | e_{ij}= \langle t_i,t_j \rangle\} & \text{Set of stream edges in the DAG}\\
		\Omega & \text{Input rate (tuples/sec) to be supported for DAG}\\
		\hline 
		v_j \in \mathbb{V} & \text{VMs available} \\
		p_j & \text{Number of resource slots available in VM $v_j$}\\
		\hline 
		q_i & \text{Number of threads allocated to task $t_i$}\\
		r_i^k \in R & \text{Set of task threads allocated to tasks in DAG}\\
		\rho & \text{Sum of the resource slots allocated to the DAG}\\
		\hline 
		v_j \in V  & \text{VMs allocated to the DAG} \\
		s_j^l \in S & \text{Resource slots in VMs allocated to DAG}\\
		\mathcal{M} : R \rightarrow S & \text{Mapping function from a thread to its slot}\\
		\hline 
		\mathcal{P}_i : \tau \rightarrow \langle \omega, c, m \rangle & \text{Performance model for task $t_i$. Maps from $\tau$ threads to the peak input rate} \\
		& \text{supported $\omega$, CPU\% $c$ and Memory\% $m$}\\
		\hline 
		\mathcal{C}_i(q),~~\mathcal{M}_i(q) & \text{Incremental CPU\%, memory\% used by task $t_i$ with $q$ threads on a single resource slot} \\
		\bar{c_i} = \mathcal{C}_i(1),~~\bar{m_i} = \mathcal{M}_i(1) & \text{CPU\% and memory\% required by 1 thread of the task } t_i \text{ on a single slot}\\
		\mathcal{I}_i(q) & \text{Peak input rate supported by the task $t_i$ with $q$ threads on a single slot} \\
		\mathcal{T}_i(\omega) & \text{Smallest thread count $q$ needed to satisfy the input rate $\omega$ for task $t_i$ on a single slot} \\
		\bar{\omega_i}  & \text{Peak rate sustained by 1 thread of task $t_i$ running in 1 slot} \\
		\widehat{\omega_i} & \text{Peak rate sustained across any number of threads of task $t_i$ running in 1 slot} \\
		\hline 
		C_j,~~M_j & \text{Available CPU\%, memory\% on a VM}\\
		M_j^l & \text{Available memory\% on single slot}\\
		\tau_i & \text{Number of threads allocated to task $t_i$} \\
		\widehat{v} & \text{Reference VM, VM on which last task thread was mapped  }\\
		\hline 
		C_j^l, M_j^l & \text{Available CPU\%, memory\% on single slot}\\
		\widehat{\tau_i} & \text{Number of threads needed to support peak rate $\widehat{\omega_{i}}$ for task $t_{i}$ on 1 slot}\\
		\hline
		\end{array}
		\]
	}
	\caption{Summary of notations used in article}
	\label{fig:terms}
\end{figure}



Let the input \emph{DAG} be given as $\mathcal{G} = \langle \mathbb {T},\mathbb {E} \rangle$, where  $\mathbb{T}=\{t_1,...,t_n\}$ is the set of $n$ \textit{tasks} that form the vertices of the DAG, and $\mathbb {E} =\{e_{ij} | e_{ij}= \langle t_i,t_j \rangle,~~t_i,t_j \in \mathbb{T}\}$ is the set of \textit{tuple stream} edges connecting the output of a task $t_i$ to the input of its downstream task $t_j$. Let $\sigma_{ij}$ be the \emph{selectivity} for the edge $e_{ij}$. We assume interleave semantics on the input streams and duplicate semantics on the output streams to and from a task, respectively.  


Further, we are given a set of VMs, $v_j \in \mathbb{V}$, with each VM $v_j$ having multiple identical resource slots, $s_j^1 .. s_j^p$. Each slot is homogeneous in resource capacity and corresponds to a single CPU core of a rated speed and a specific quanta of memory allocated to that core. Let $p_j$ be the number of processing slots present in VM $v_j$. The VMs themselves may be heterogeneous in terms of the number of slots that they have, even as we assume for simplicity that the slots themselves are homogeneous. This is consistent with the ``container'' or ``slot'' model followed by Big Data platforms like Apache Hadoop~\cite{vavilapalli:cc:2013} and Storm, though it can be extended to allow heterogeneous slots as well.

\emph{Task threads}, $r_i^1 .. r_i^q$, are responsible for executing the logic for a task $t_i$ on a tuple arriving on the input stream for the task. Each task has one or more threads allocated to it, and each thread is mapped to a resource slot. Different threads can execute different tuples on the input stream, in a data-parallel manner. The order of execution does not matter. Each resource slot can run multiple threads from one or more tasks concurrently.

We are given an input stream rate of $\Omega$~tuples/sec (or events/sec) that should be supported for the DAG $\mathcal{G}$. Our goal is to schedule the tasks of the DAG on \emph{VMs with the least number of cumulative resource slots, to support a stable execution of the DAG} for the given input rate. This problem can be divided into two phases:

\begin{enumerate}
	\item \emph{Resource Allocation:} Find the minimum number $q_i$ of task threads required per task $t_i$, and the cumulative resource slots $\rho$ required to meet the input rate to the DAG. Minimizing the slots translates to a minimization of costs for these on-demand VM resources from Cloud providers since the pricing for VMs are typically proportional to the number of slots they offer.
	
	\item \emph{Resource Mapping:} Given the set of task threads $r_i^k \in R$ for all tasks in the DAG, and the number of resource slots $\rho$ allocated to the DAG, determine the set of VMs $V$ such that they have an adequate number of slots, $\rho \le \sum_{v_j \in V}{p_j}$. Further, for resource slots $s_j^l \in S$ present in the VMs $v_j \in V$, find an optimal mapping function $\mathcal{M} : R \rightarrow S$ for the allocated task threads on to available slots, to match the resources needed to support the required input rate $\Omega$ in a stable manner. 
\end{enumerate}

There are several qualitative and quantitative measures we use to evaluate the solution to this problem.
\begin{enumerate}
	\item The number of resource slots and VMs estimated by the allocation algorithm should be minimized.
	\item The actual number of resource slots and VMs required by the mapping algorithm to successfully place the threads should be minimized. This is the actual cost paid for acquiring and using the VMs. Closeness of this value to the estimate from above indicates a reliable estimate.
	\item The actual stable input rate that is supported for the DAG by this allocation and mapping at runtime should be greater than or equal to the required input rate $\Omega$. A value close to $\Omega$ indicates better predictability.
	\item The expected resource usage as estimated by the scheduling algorithm and the actual resource usage for each slot should be proximate. The closer these two values, the better the predictability of the dataflow's performance and the scheduling algorithm's robustness under dynamic conditions. 
\end{enumerate}



\section{Solution Approach}
\label{sec:approach}
We propose a model-based approach to solving the two sub-problems of resource allocation and resource mapping, in order to arrive at an efficient and predictable schedule for the DAG to meet the required input rate. The intuition for this is as follows.

The stable input rate that can be supported by a task depends on the the number of concurrent threads for that task that can execute over tuples on the input stream in a data-parallel manner. The number of threads for a task in turn determines the 
resources required by the task. Traditional scheduling approaches for DSPSs assume that both these relationships -- between thread count and rate supported, and thread count and resources required -- are a linear function. That is, if we double the number of threads for a task, we can achieve twice the input rate and require twice the computing resources.

However, as we show later in \S~\ref{sec:bm} and Fig.~\ref{fig:bm}, this is not true. Instead, we observe that depending on the task, both these relationships may range anywhere from flat line to a bell curve to a linear function with a positive or a negative slope. The reason for this is understandable. As the number of threads increase in a single VM or resource slot, there is more contention for those specific resources by threads of a task that can mitigate their performance. This can also affect the actual resource used by the threads. For simplicity, we limit our analysis to CPU and memory resources, though it can be extended to disk IOPS and network bandwidth as well. As a result of this real-word behavior of tasks, scheduling that assumes a linear behavior can under-perform.

In our approach, we embrace this non-uniformity in the task performance and incorporate the empirical model of the performance into our schedule planning. First, we use micro-benchmarks on a single resource slot to develop a \emph{performance model} function $\mathcal{P}_i$ for each task $t_i$ which, given a certain number of threads $\tau$ for the task on a single resource slot, provides the peak input rate $\omega$ supported, and the CPU and memory utilization, $c$ and $m$, at that rate. This is discussed in \S~\ref{sec:bm}.

Second, we use these performance models to determine the number of threads $q_i$ for each task $t_i$ in the DAG that is required to support a given input rate, and the cumulative number of resource slots $\rho$ for all threads in the DAG. This \emph{Model Based Allocation (MBA)} described in \S~\ref{sec:allocation} offers an accurate estimate of the resource needs and task performance, for individual tasks and for the entire DAG. We also discuss a commonly used baseline approach, \emph{Linear Scaling Allocation (LSA)}, in that section. As mentioned before, it assumes that the performance of a single thread on a resource slot can be linearly extrapolated to multiple threads on that slot, as long as the resource capacity of the slot is not exhausted. It under-performs, as we show later.

This resource allocation can subsequently be used by different resource mapping algorithms that exist, such as the round-robin algorithm used by default in Apache Storm~\cite{toshniwal:sigmod:2014}, which we refer to as the \emph{Default Storm Mapping (DSM)}, or a resource-aware mapping proposed in R-Storm~\cite{peng:middleware:2015} and included in the latest Apache Storm distribution, which we call \emph{R-Storm Mapping (RSM)}. However, these mapping algorithms do not provide adequate co-location of threads onto the same slot to exploit the intuition of the model based allocation. We propose a novel \emph{Slot Aware Mapping (SAM)} algorithm that attempts to map threads from the same task as a group to individual slots, as a form of \emph{gang scheduling}~\cite{ousterhout:icdcs:1982}. Here, our goal is to maximize the peak event rate that can be exploited from that slot, minimize the interference between threads from different tasks, and ensure predictable performance. These allocation strategies are explored in \S~\ref{sec:mapping}.

\subsection{Illustration}
\begin{figure}[t]
	\centering
	\includegraphics[width=0.80\textwidth]{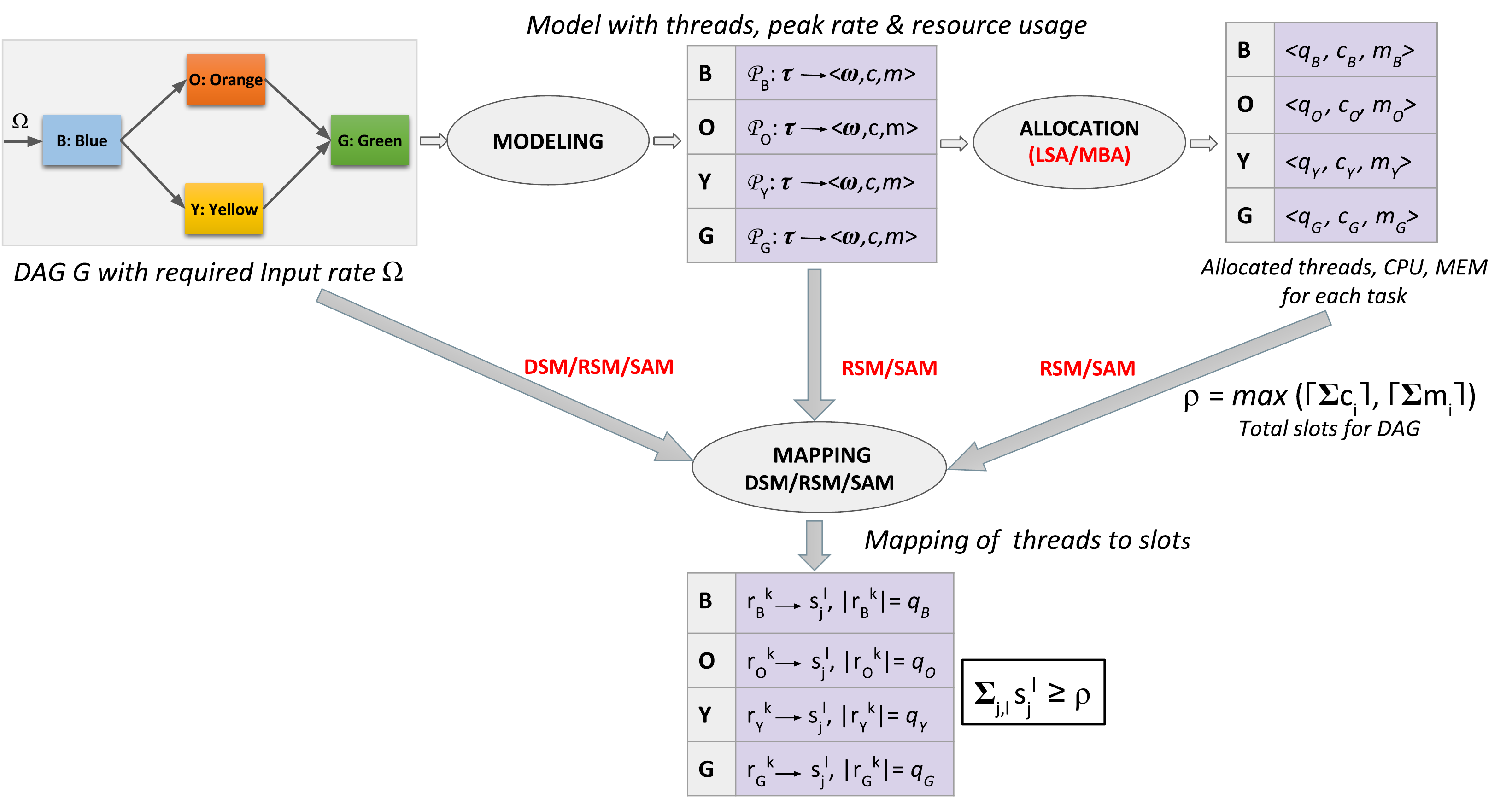}
	\caption{Illustration of \emph{Modeling}, \emph{Allocation} and \emph{Mapping} phases performed when scheduling a sample DAG }
	\label{fig:flow}
\end{figure}

We provide a high-level visual overview of the schedule planning in Fig.~\ref{fig:flow} for a given DAG $\mathcal{G}$ with four tasks, \texttt{Blue}, \texttt{Orange}, \texttt{Yellow} and \texttt{Green}, with a required input rate of $\Omega$. The procedure has three phases. In the \textit{Modeling} phase, we build a performance model $\mathcal{P}_i$ for tasks in the DAG that do not already have a model available. This gives a profile of the peak input tuple rates supported by a task with different numbers of threads, and their corresponding CPU and memory utilization, using a single resource slot. For e.g., the performance models for the four tasks in the DAG are given by $\mathcal{P}_B, \mathcal{P}_O, \mathcal{P}_Y$ and $\mathcal{P}_G$ in Fig.~\ref{fig:flow}.

In the \textit{Allocation} phase, we use the above performance model to decide the number of threads $q_i$ for each task required to sustain the tuple rate that is incident on it. The input rate to the task is itself calculated based on the DAG's input rate and the selectivity of upstream tasks. We use the Linear Scaling Allocation (LSA) and our Model Based Allocation (MBA) approaches for this. While LSA only uses the performance model for a single task thread, our MBA uses the full profile that is available. These algorithms also return the CPU\% and memory\% for the threads of each task that are summed up to get the total number of resource slots for the DAG. Fig.~\ref{fig:flow} shows the table for each of the four tasks after allocation, with the number of threads $q_b, q_O, q_Y$ and $q_G$, and their estimated resource usages $c$ and $m$ that are summed up to calculate the total resource slot needs for the DAG $\rho$.

Lastly, the \textit{Mapping} phase decides the number and types of VMs required to meet the resource needs for the DAG. It then maps the set of $r_i^k$ threads allocated for the DAG to the slots $s_j^l$ in the VMs, and the total number of these slots can be greater than the estimated $\rho$, depending on the algorithm. Here, we use the Default Storm Mapping (DSM), R-Storm Mapping (RSM) and our proposed Slot Aware Mapping (SAM) algorithms as alternatives. As shown in the figure, DSM is not resource aware and only uses the information on the number of threads $q_i$ and the number of slots $\rho$ for its round-robin mapping. RSM and SAM use the task dependencies between the DAG and the allocation table. RSM uses the performance of a single thread while SAM uses all values in the performance model for its mapping.

\subsection{Discussion}
As with any approach that relies on prior profiling of tasks, our approach has the short-coming of requiring effort to empirically build the performance model for each task before it can be used. However, this is mitigated in two respects.

First, as has been seen for scientific workflows, enterprise workloads and even HPC applications~\cite{de:fgcs:2009,scheidegger:sigmod:2008,medeiros:sigmod:2005}, many domains have common tasks that are reused in compositional frameworks by users in that domain. Similarly, for DSPS applications in domains like social network analytics, IoT or even Enterprise ETL (Extract, Transform and Load), there are common task categories and tasks such as parsing, aggregation, analytics, file I/O and Cloud service operations~\cite{filgueira:escience:2015,shukla:tpctc:2016,lu:ucc:2014}. Identifying and developing performance models for such common tasks for a given user-base -- even if all tasks are not exhaustively profiled -- can help leverage the benefits of more efficient and predictable schedules for streaming applications.

Second, the effort in profiling a task is small and can be fully automated, as we describe in the next section. It also does not require access to the eventual DAG that will be executed. This ensures that as long as we can get access to the individual task, some minimal characteristics of the input tuple streams, and VMs with single resource slots comparable to slots in the eventual deployment, the time, costs and management overheads for building the performance model are mitigated.

\section{Performance Modeling of Tasks}
\label{sec:bm}
\subsection{Approach}
Performance modeling for a task builds a profile of the peak input tuple rate supported by the task, and its corresponding CPU and memory usage, using a single resource slot. It does this by performing a constrained parameter sweep of the number of threads and different inputs rates as a series of micro-benchmark trials. Algorithm~\ref{alg:bm} gives the pseudo-code for build such a performance model. For a given task $t$, we initialize the number of threads $\tau$ and the input rate $\omega$ to $1$, and iteratively increase the number of threads (lines~\ref{alg:bm:thread-start}--\ref{alg:bm:thread-end}), and for each thread count, the input rate (lines~\ref{alg:bm:rate-start}--\ref{alg:bm:rate-end}). The steps at which the thread count and rates are increased, $\Delta_\tau$ and $\Delta_\omega$, can either be fixed, or be a function of the iteration or the prior performance. This determines the granularity of the model -- while a finer granularity of thread and rate increments offers better modeling accuracy, it requires more experiments, and costs time and money for VMs.

\begin{algorithm}
	\footnotesize
	\caption{Performance Modeling of a Task}\label{alg:bm}
	\begin{algorithmic}[1]
		\Procedure{PerfModel(Task $t$)}{}
		
		\State $\mathcal{P} \gets new~Map(~)$ \Comment{\emph{Holds the mapping from threads to input rate, resource usage}}
		\State $\lambda_\omega = 0$ \Comment{\emph{Slope of the last window of peak stable rates}}
		\State $\tau = 1$ \Comment{\emph{Number of threads being tested}}\\
		
		\Comment{\emph{Increase the number of threads until $\tau_{max}$, or slope of peak supported rate remains stable or drops}}
		\While{$\tau < \tau_{max}$ and $\lambda_\omega \le \lambda_\omega^{max}$} \label{alg:bm:thread-start}\\
		
		\Comment{\emph{For each value of $\tau$, increase input rate in steps of $\Delta_\omega$ until trial is unstable, or max rate $\omega_{max}$ reached}}
		\For{$\omega \gets 1$ \textbf{ to } $\omega_{max}$ \textbf{ step } $\Delta_\omega$} \label{alg:bm:rate-start}\\
		\Comment{\emph{Run DSPS with task. Check if rate is supported. Get CPU and memory\%.}}
		\State $\langle c, m, isStable \rangle \gets$ \textsc{RunTaskTrial($t,\tau,\omega$)} \label{alg:bm:trial}
		\If{$isStable = false$} \Comment{\emph{If rate not supported, break}}
		\State \textbf{break}\label{alg:bm:unstable}
		\EndIf
		\State $\mathcal{P}.put(\tau \rightarrow \langle \omega, c, m \rangle)$ \Comment{\emph{Add or update mapping from $\tau$ to peak rate, resource usage}}
		\EndFor \label{alg:bm:rate-end}
		\State $\lambda_\omega \gets $\textsc{Slope($\mathcal{P}, \omega$)} \Comment{\emph{Get slope of the last window of peak stable rates before $\omega$}}
		\State $\tau \gets \tau + \Delta_\tau$ \Comment{\emph{Increment thread count}}\label{alg:bm:thread-end}
		\EndWhile\\
		\Return $\mathcal{P}$
		\EndProcedure
	\end{algorithmic}
\end{algorithm}

For each combination of $\tau$ and $\omega$, we run a trial of the task (line~\ref{alg:bm:trial}) that creates a sequential 3-task DAG, with one source task that generates input tuples at the rate $\omega$, the task $t$ in the middle with task threads set to $\tau$, and a sink task to collect statistics. The threads for task $t$ are assigned one independent resource slot on one VM while on a second VM, the source and sink tasks run on one or more separate resource slots so that they are not resource-constrained. This trial is run for a certain interval of time that goes past the ``warm-up'' period where initial scheduling and caching effects are overcome, and the DAG executes in a uniform manner. For e.g., in our experiments, the warm up period was seen to be $\le 5~mins$. During the trial, a profiler collects statistics on the CPU and memory usage for that resource slot, and the source and sink collect the latency times for the tuples. Running the source and sink tasks on the same VM avoids clock skews. At the end of the trial, these statistics are summarized and returned to the algorithm.

In running these experiments, two key termination conditions need to be determined for automated execution and generation of the model. For a given number of threads, we need to decide when we should stop increasing the input rate given to the task. This is done by checking the latency times for the tuples processed in the trial. Under stable conditions, the latency time for tuples are tightly bound and fall within a narrow margin beyond the warm-up period. However, if the task is resource constrained, then it will be unable to keep up its processing rate with the input rate causing queuing of input tuples. As the queue size keeps increasing, there will be an exponential growth in the end-to-end latency for successive tuples. 

To decide if the task execution is stable, we calculate the slope $\lambda_L$ of the tuple latency values for the trial period past the warm-up and check if it is constant or less than a small positive value, $\lambda_L^{max}$. If $\lambda_L \le \lambda_L^{max}$, this execution configuration is stable, and if not, it is unstable. Once we reach an input rate that is not stable, we stop the trials for these number of threads for the task, and move to a higher number of threads (line~\ref{alg:bm:unstable}). In our experiments, using a tight slope value of $\lambda_L^{max} \approx 0.001$ was possible, and none of the experiments ran for over $12~mins$. 

The second termination condition decides when to stop increasing the number of threads. Here, the expectation is that as the thread count increases there is an improvement, if any, in the peak rate supported until a point at which it either stabilizes or starts dropping. We maintain the peak rate supported for previous thread counts in the $\mathcal{P}$ hashmap object. As before, we take the slope $\lambda_\omega$ of the rates for the trailing window of thread counts to determine if the slope remains flat at $0$ or turns negative. Once the rate drops or remains flat for the window, we do not expect to see an improvement in performance by increasing the thread count, and terminate the experiments. In our experiments, we set $\lambda_\omega^{min} \approx -0.001$. 

\subsection{Performance Modeling Setup}
\label{sec:bm:setup}

We identify $5$ representative tasks, shown in Table~\ref{tbl:bm}, to profile and build performance models for. They also empirically motivate the need for fine-grained control over thread and resource allocation. These tasks have been chosen to be diverse, and representative of the categories of tasks often used in DSPS domains such as social media analytics, IoT and ETL pipelines~\cite{gedik:tpds:2014,wang:bigdatabench:2014}.


\begin{itemize}
	\item \emph{Parse XML.} It parses an array of in-memory XML strings for every incoming tuple. Parsing is often required for initial tasks in the DAG that receive text or binary encoded messages from external sources, and need to translate them to a form used by downstream tasks in the DAG. XML was used here due its popular usage though other formats like JSON or Google Protocol Buffers are possible as well. 
	Our XML parsing implementation uses the Java SAX parser that allows serial single-pass parsing even over large messages at a fast rate. 
	Parsing XML is CPU intensive and requires high memory due to numerous string operations (Table ~\ref{tbl:bm}).
	
	\item \emph{Pi Computation.} This task numerically evaluates the approximate value of $\pi$ using an infinite series proposed by \textit{Viete} \cite{dence:pi:1993}. Rather than running it non-deterministically till convergence, we evaluate the series for a fixed number of iterations, which we set to 15. This is a CPU intensive floating-point task, and is analogous to tasks that may perform statistical and predictive analytics, or computational modeling.
	
	\item \emph{Batch File Write.} It is an accumulator task that resembles both window operations like aggregation or join, and disk I/O intensive tasks like archiving data. The implementation buffers a $100~byte$ string in-memory for every incoming tuple for a window size of $10,000$ tuples, and then writes the batched strings to a local file on a HDD attached to VM. The number of disk operations per second is proportional to the input message rate. 
	
	\item \emph{Azure Blob Download.} Streaming applications may download metadata annotations, accumulated time-series data, or trained models from Cloud storage services to use in their execution. Microsoft Azure Blob service stores unstructured data as files in the Cloud. This task downloads a file with a given name from the Azure Blob storage service using the native Azure Java SDK. In our implementation, a $2~MB$ file is downloaded and stored in-memory for each input tuple, making it  
both memory and network intensive.
	
	\item \emph{Azure Table Query.}  Some streaming applications require to access historic data stored in databases, say, for aggregation and comparison. Microsoft Azure Table service is a NoSQL columnar database in the Cloud. 
          Our task implementation queries a table containing $2,000,000$ records, each with $20$ columns and $~\approx{200~byte}$ record size~\cite{data:taxi}, using the native Azure Java SDK. The query looks up a single record by passing a randomly generated record ID corresponding to a unique row key in the table.
	
\end{itemize}

As can be seen, these tasks cover a wide range of operations. These span from text parsing 
and floating-point operations 
to both local and Cloud-based file and table operations. 
There is also diversity in these tasks with respect to the resources they consume as shown in Table~\ref{tbl:bm}, be they memory, CPU, Disk or Network, and some rely on external services with their own Service Level Agreement (SLA).

\begin{table}[h]
	\caption{Characteristics of representative tasks for which performance modeling is performed}
	\label{tbl:bm}
	\small
	\centering
	\begin{tabular}{ p{3.0cm}||p{1.5cm} p{1.5cm} p{1.5cm} p{1.5cm} p{1.5cm}}
		\hline        
		\textbf{Task}& \textbf{CPU bound?}& \textbf{Memory bound?}&\textbf{N/W bound?}&\textbf{Disk I/O bound?}&\textbf{External Service?} \\        
		\hline        
		\hline        
		Parse XML   &\checkmark & \checkmark &  &&\\        
		\hline        
		Pi Computation       &\checkmark & &  &&\\        
		\hline        
		Batched File Write  & & & &\checkmark&\\        
		\hline        
		Azure Blob Download   &    & \checkmark&\checkmark   &&\checkmark\\
		\hline        
		Azure Table Query &      &    & \checkmark & &\checkmark\\        
		\hline        
	\end{tabular}
\end{table}

We wrap each of these tasks as a \emph{bolt} in the Apache Storm DSPS. We compose a \emph{topology} DAG consisting of 1 \emph{spout} that generates synthetic tuples with 1 field (\texttt{message-id}) at a given constant rate determined by that trial, 1 bolt with the task that is being modeled, and 1 \emph{sink} bolt that accepts the response from the predecessor. For each input tuple, The task bolt emits one output tuple after executing its application logic, keeping the selectivity $\sigma = 1:1$.

Apache Storm~\footnote{Apache Storm v1.0.1, released on 6 May 2016} is deployed on a set of standard \emph{D type VMs} running Ubuntu 14.04 in Microsoft Azure's Infrastructure as a Service (IaaS) Cloud, in the Southeast Asia data-center. Each VM has $2^{i-1}$ resource slots, where $i$ corresponds to the VM size, $D_i \in \{D_1,D_2,D_3,D_4\}$. Each slot has a one Intel Xeon E5-2673~v3 core $@$2.40~GHz processor with hyper-threading disabled
~\footnote{Azure A-SERIES, D-SERIES and G-SERIES: Consistent Performances and Size Change Considerations, \url{https://goo.gl/0X6yT2}}
, $3.5 GB$ memory and $50 GB$ SSD, for e.g., the D3 VM has $4$~cores, $14$~GiB RAM and $200$~GiB SSD. A separate $50 GB$ HDD is present for I/O tasks like Batch File Write.

For the performance modeling, we deploy the spout and the sink on separate slots of a single D2 size VM, and the task bolt being evaluated on one D1 size VM. The spout and sink have $2$ threads each to ensure they are not the bottleneck, while the number of threads for the task bolt and the input rate to this topology is determined by Algorithm~\ref{alg:bm}. Each trial is run for $12~mins$, to be conservative. We measure the CPU\% and memory\% using the Linux \texttt{top} command, and record the peak stable rate supported for each of these task bolts for specific numbers of threads. The experiments are run multiple times, and the representative measurements are reported.

\subsection{Performance Modeling Results}

\begin{figure*}[t]
	\centering
	\subfloat[Parse XML]{
		\includegraphics[width=0.30\textwidth]{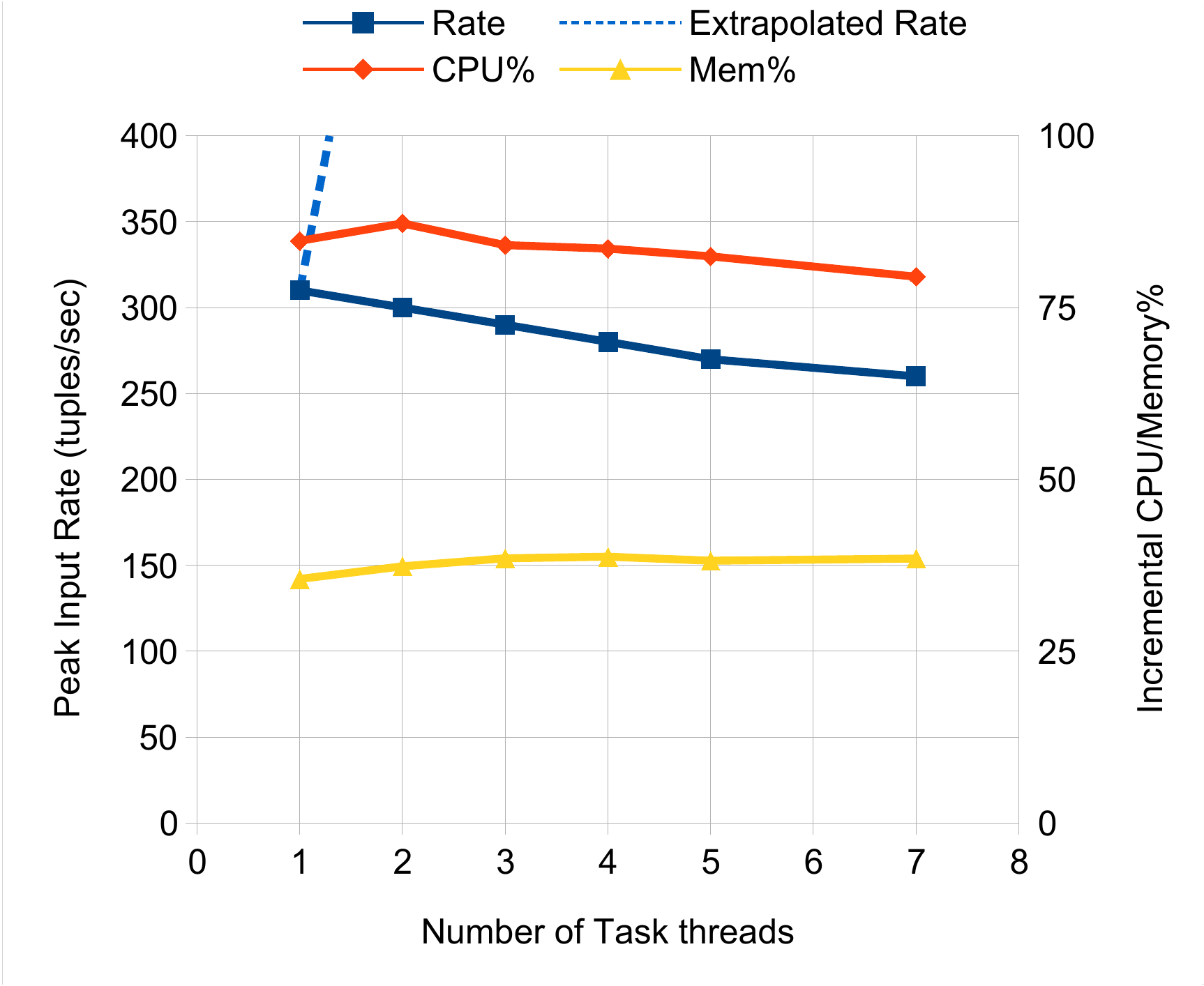}
		\label{fig:bm:xml}
	}
	\subfloat[Pi Computation]{
		\includegraphics[width=0.30\textwidth]{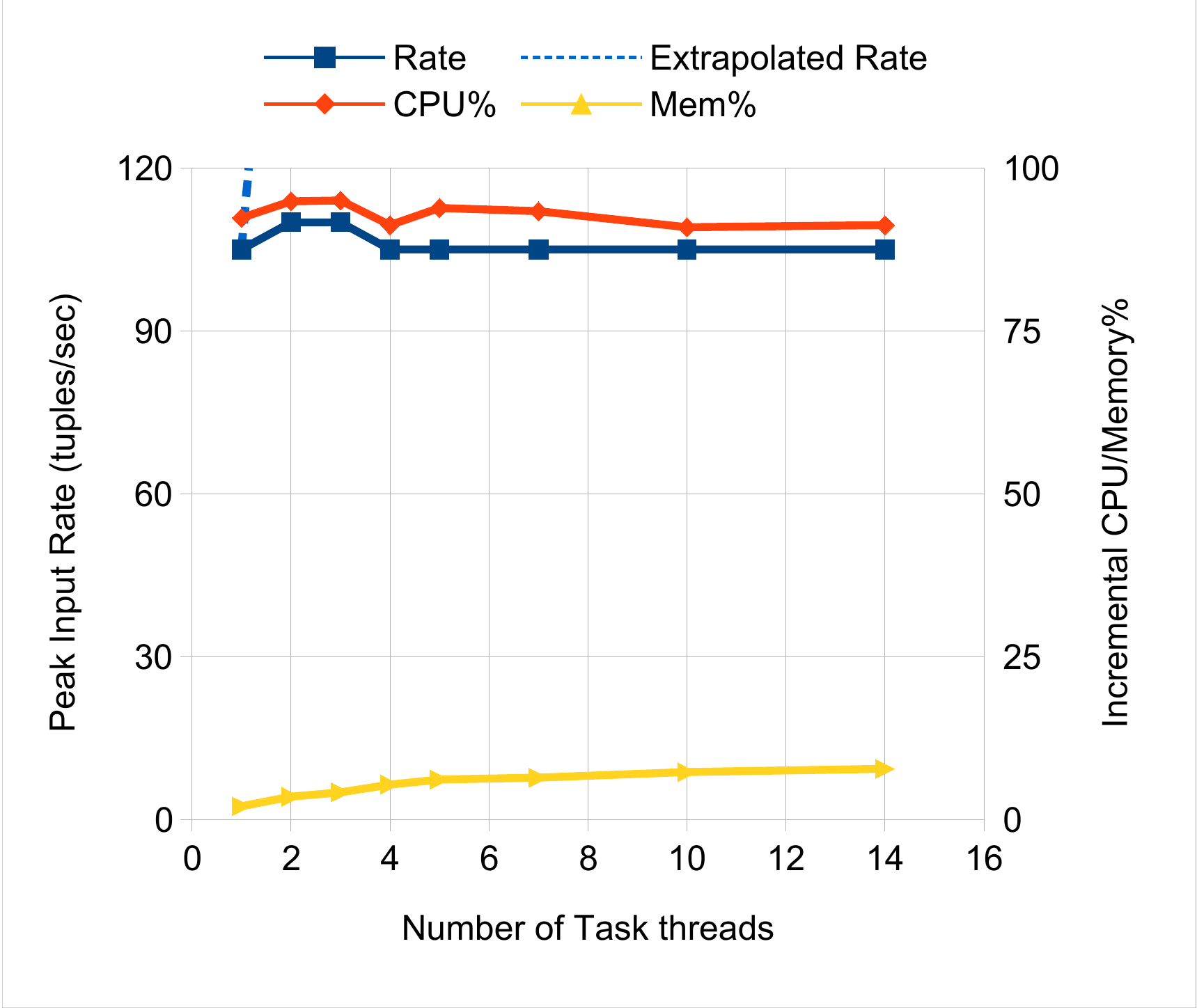}
		\label{fig:bm:pi}
	}\\
	\subfloat[Batched File Write]{
		\includegraphics[width=0.30\textwidth]{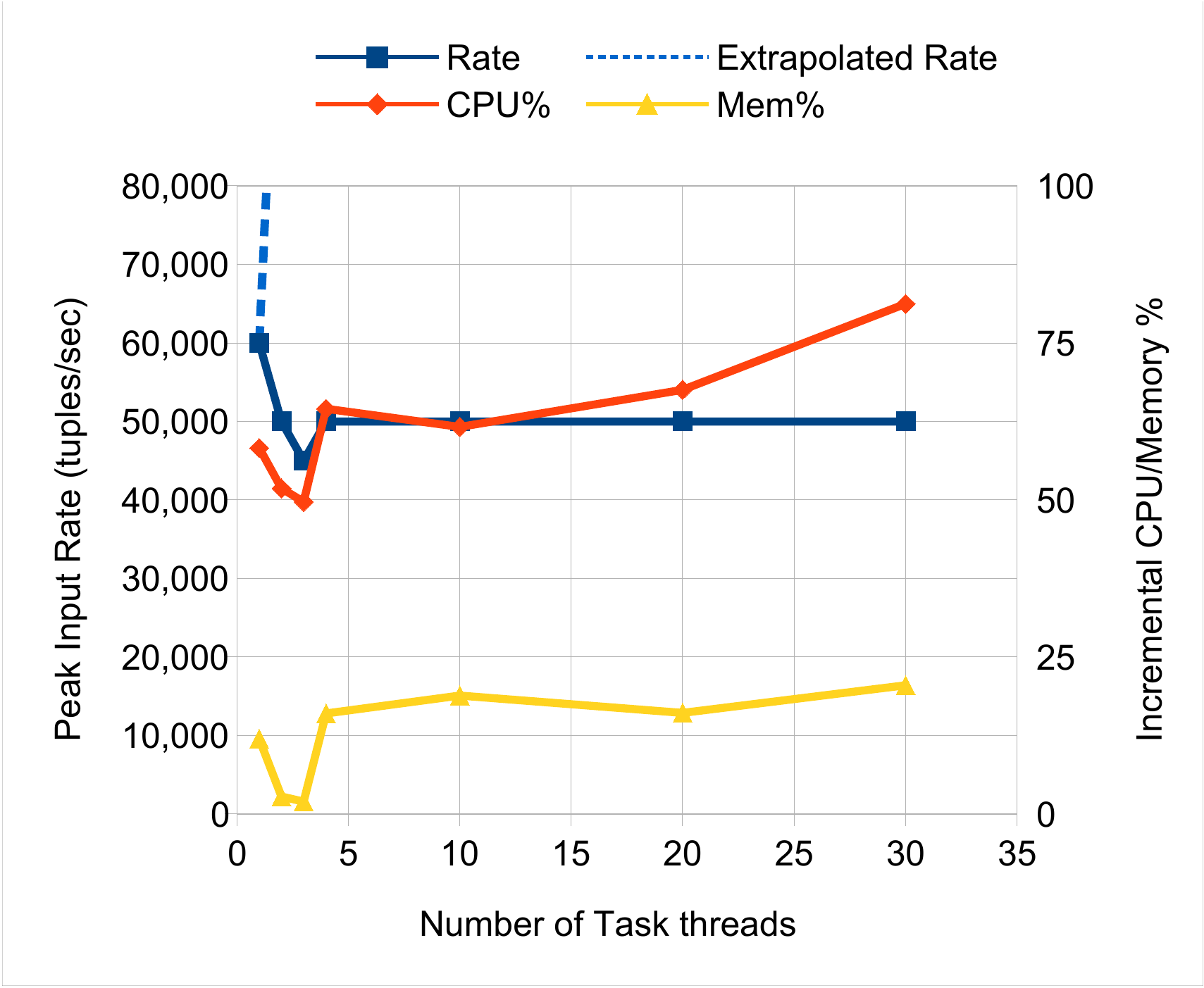}
		\label{fig:bm:file}
	}
	\subfloat[Azure Blob Download]{
		\includegraphics[width=0.30\textwidth]{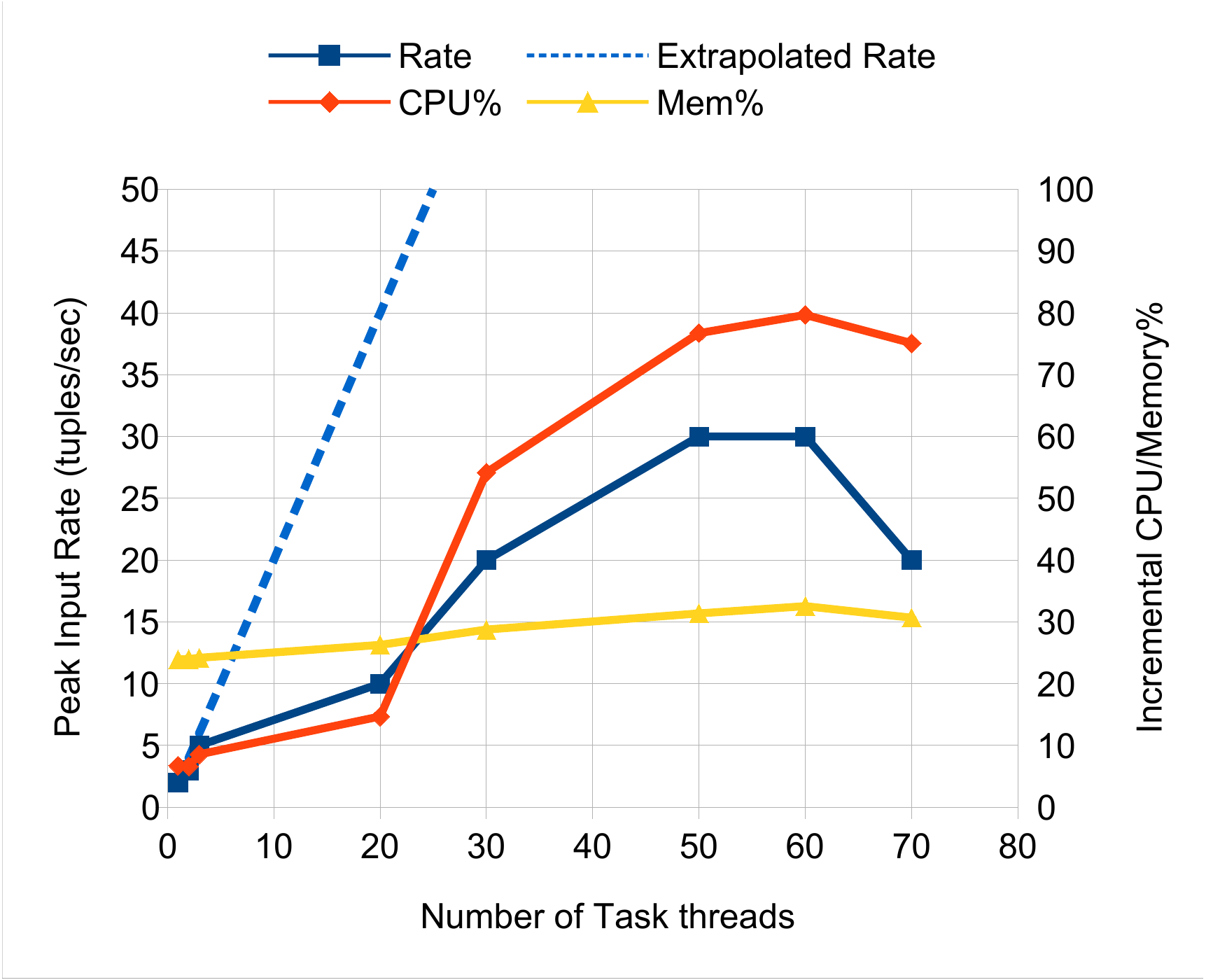}
		\label{fig:bm:blob}
	}
	\subfloat[Azure Table Query]{
		\includegraphics[width=0.33\textwidth]{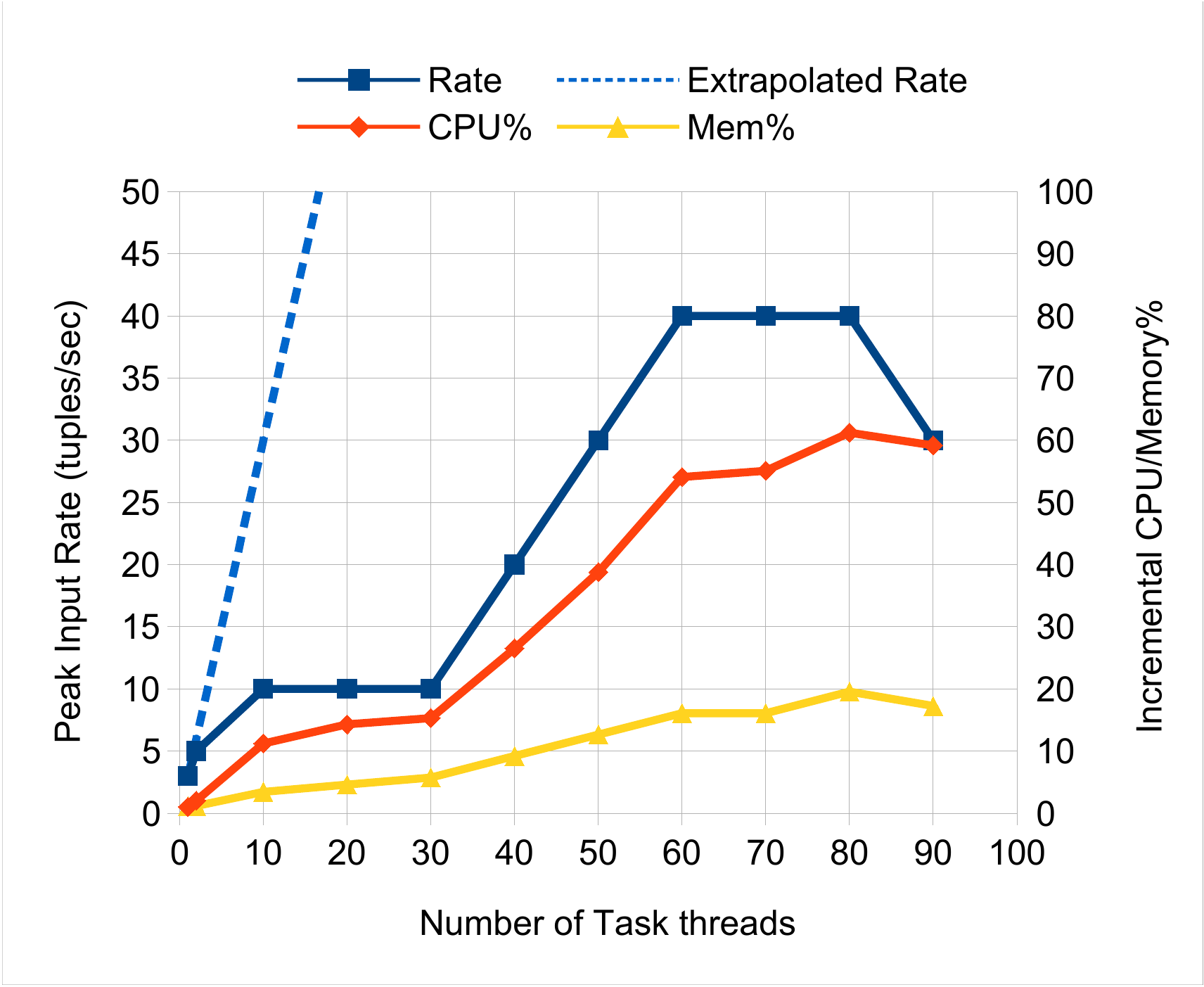}
		\label{fig:bm:table}
	}
	\caption{
		\emph{Peak input rate supported} (primary Y Axis) on one resource slot for the specified task bolt, with an increase in the \emph{number of threads} (X Axis). The \emph{\% incremental CPU and Memory usage} for these numbers of threads at the peak rate is shown in the secondary Y Axis.
	}
	
	\label{fig:bm}
\end{figure*}

The goal of these experiments is to collect the performance model data used by the scheduling algorithms. However, we also supplement it with some observations on the task behavior. Fig.~\ref{fig:bm} shows the performance model plots for the $5$ tasks we evaluate on a single resource slot. On the primary Y Axis (left), the plots show the peak stable rate supported (tuples/sec), and the corresponding CPU and memory utilization on the secondary Y Axis (right), as the number of threads for the task increases along the X Axis. The CPU\% and memory\% are given as a fraction of the utilization above the base load on the machine when no application topology is running, but when the OS and the stream processing platform are running. So a $0\%$ CPU or memory usage in the plots indicate that only the base load is present, and a $100\%$ indicates that the full available resource of the slot is used.

We see from Fig.~\ref{fig:bm:xml} that the \textit{Parse XML} task is able to support a peak input rate of about $310~tuples/sec$ with just 1 thread, and increasing the number of threads actually reduces the input throughput supported, down to about $255~tuples/sec$ with 7 threads. The CPU usage even for 1 thread is high at about $85\%$. Here, we surmise that since a single thread is able to utilize the CPU efficiently, increasing the threads causes additional overhead for context-switching and the performance deteriorates linearly. We also see that the Parse XML task uses about $35\%$ of memory due to Java's memory allocation for string manipulation, which is higher than for other tasks. 

\textit{Pi Computation} is a floating-point heavy task and uses nearly $90\%$ CPU at the peak rate of about $105~tuples/sec$ using a single thread. However, unlike XML Parse where the peak rate linearly reduces with an increase in threads, we see Pi manage to modestly increase the peak rate with 2 threads, to about $110~tuples/sec$, with a similar increase in CPU usage. 
However, beyond 2 threads, the performance drops and remains flat with the CPU usage being flat as well. This behavior was consistent across multiple trials, and is likely due to the Intel Haswell architecture's execution pipeline characteristics~\footnote{Intel's Haswell Architecture Analyzed: Building a New PC and a New Intel, Anand Lal Shimpi, Oct 2012, \url{http://www.anandtech.com/show/6355/intels-haswell-architecture/8}}. The memory usage is minimal at between $2-10\%$. 

\textit{Batch File Write} is an aggregation task that is disk I/O bound. It supports a high peak rate of about $60,000~tuples/sec$ with 1 thread, which translates to writing $6~files/sec$, each $1~MB$ in size. This peak rate decreases with an increase in the number of threads, but is non-linear. There is a sharp drop in the peak rate to $45,000~tuples/sec$ with 3~threads, but this increases and stabilizes at $50,000~tuples/sec$ with more threads. 
The initial drop can be attributed to the disk I/O contention, hence the drop in CPU usage as well, but beyond that the thread contention may dominate, causing an increase in CPU usage even as the supported rate is stable 

The \emph{Azure Blob} and \emph{Azure Table} tasks rely on an external Cloud service to support their execution. As such, the throughput of these tasks are dependent on the SLA offered for these services by the Cloud provider, in addition to the local resource constraints of the slot. We see the benefit of having multiple threads clearly in these cases. The peak rate supported by both increases gradually until a threshold, beyond which the peak rate flattens and drops. Their CPU and memory utilization follow a similar trend as well. Blob's rate grows from about $2~tuples/sec$ with 1 thread to $30~tuple/sec$ with 50 threads, while Table's increases from $3~tuples/sec$ to $60~tuples/sec$ when scaling from 1 to 60 threads.
This closely correlates with the SLA of the Blob service which is $60~MB/sec$, and matches with the $30~files/sec$ of $\approx 2~MB$ each that are cumulatively downloaded~\footnote{\url{https://docs.microsoft.com/en-us/azure/storage/storage-scalability-targets}}. 
Both these tasks are also network intensive as they download data from the Cloud services. 

\emph{Summary.} The first three tasks show a flat or decreasing peak rate performance with some deviations, but with differing CPU and memory resource behavior. The last two exhibit a bell-curve in their peak rates as the threads increase. These highlight the distinct characteristics of specific tasks (and even specific CPU architectures and Cloud services they wrap) that necessitate such performance models to support scheduling. Simple rules of thumbs assuming static of linear scaling are inadequate, and we see later, can cause performance degradation and resource wastage.


\section{Resource Allocation}
\label{sec:allocation}
Resource allocation determines the number of resource slots $\rho$ to be allocated for a DAG $\mathcal{G}:\langle \mathbb{V},\mathbb{E} \rangle$ for a given input rate $\Omega$, along with the number of threads $q_j$ required for each task $t_j \in \mathbb{V}$. In doing so, the allocation algorithm needs to be aware of the input rate to each task that will inform it of the resource needs and data parallelism for that task. We can define this \emph{input rate} $\omega_j$ for a task $t_j$ based on the input rate to the DAG, the connectivity of the DAG, and the selectivity of each input stream to a task, using a recurrence relation as follows:
\[
\omega_{j}=\begin{cases}
\Omega & \text{if } \not\exists e_{ij} \in \mathbb{E} \\
\sum\limits_{e_{ij} \in \mathbb{E}} \big( \omega_{i} \times \sigma_{ij} \big) & \text{otherwise}
\end{cases}
\]

%
In other words, if task $t_j$ is a source task without any incoming edges, its input rate is implicitly the rate to the DAG, $\Omega$. Otherwise, the input rate to a downstream task is the sum of the tuple rates on the out edges of the tasks $t_i$ immediately preceding it. This output rate is itself given by the product of those predecessor tasks' input rates $\omega_i$ and their selectivities $\sigma_{ij}$ on the out edge connecting them to task $t_j$. This recurrence relationship can be calculated in the topological order of the DAG starting from the source task(s). Let the procedure $\textsc{GetRate}(\mathcal{G}, t_j, \Omega)$ evaluate this for a given task $t_j$.

Next, the allocation algorithm will need to determine the threads and resources needed by each task $t_j$ to meet its input rate $\omega_j$. Algorithms can use prior knowledge on resource usage estimates for the task, which may be limited to the CPU\% and memory\% for a single thread of the task, irrespective of the input rate, or approaches like ours that use a more detailed performance model.

%
Say the following functions are available as a result of the performance modeling algorithm, Alg.~\ref{alg:bm}, or some other means. $\mathcal{C}_i(q)$ and $\mathcal{M}_i(q)$ respectively return the incremental CPU\% and memory\% used by task $t_i$ when running on a single resource slot with $q$ threads. Further, let $\mathcal{I}_i(q)$ provide the peak input rate that is supported by the task $t_i$ on a single slot for $q$ number of threads. Lastly, let $\mathcal{T}_i(\omega)$ be the inverse function of $\mathcal{I}_i(q)$ such that it gives the smallest number of threads $q$ adequate to satisfy the given input rate $\omega$ on a single resource slot. Since the $\omega$ values returned by $\mathcal{I}_i(q)$ for integer values of $q$ would be at coarse increments, $\mathcal{T}_i$ may offer an over-estimate depending on the granularity of $\Delta_\omega$ and $\Delta_\tau$ used in Alg.~\ref{alg:bm}. 

Next, we describe two allocation algorithms -- a baseline which uses simple estimates of resource usage for tasks, and another we propose that leverages the more detailed performance model available for the tasks in the DAG. 

\subsection{Linear Scaling Allocation (LSA)} 

The Linear Scaling Allocation (LSA) approach uses a simplifying assumption that the behavior of a single thread of a task will linearly scale to additional threads of the task. This scaling assumption is made both for the input rate supported by the thread, and the CPU\% and memory\% for the thread. For e.g., the R-Storm~\cite{peng:middleware:2015} scheduler assumes this additive behavior of resource needs for a single task thread as more threads are added, though it leaves the selection of the number of threads to the user. Other DSPS schedulers make this assumption as well ~\cite{schneider:ipdps:2009}~\cite{cardellini:hpcs:2016}.

Algorithm~\ref{alg:lsa} shows the behavior of such a linear allocation strategy. It first estimates the input rate $\omega_i$ incident at each task $t_i$ using the $\textsc{GetRate}$ procedure discussed before. It then uses information on the peak rate $\bar{\omega_i}$ sustained by a single thread of a task $t_i$ running in one resource slot, and its CPU\% and memory\% at that rate, $\mathcal{C}_i(1)$ and $\mathcal{M}_i(1)$, as a baseline. Using this, it tries to incrementally add more threads until the input rate required, in multiples of the peak rate, is satisfied (line~\ref{alg:lsa:greater}).  
When the remaining input rate to be satisfied is below the peak rate (line~\ref{alg:lsa:smaller}), we linearly scale down the CPU and memory utilization, proportional to the required rate relative to the peak rate.

The result of this LSA algorithm is the thread counts $\tau_i$ per task $t_i \in \mathbb{T}$. 
In addition, the sum of the CPU and memory allocation for all tasks, rounded up to the nearest integer, gives the nominal lower bound on the resource slots $\rho$ required to support this DAG at the given rate. 

\begin{algorithm}[t]
\footnotesize
	\caption{Linear Scaling Allocation (LSA)}\label{alg:lsa}
	\begin{algorithmic}[1]
		\Procedure{AllocateLSA} {$\mathcal{G} : \langle \mathbb {T},\mathbb {E} \rangle,~\Omega$}
		\For{$t_{i} \in \mathbb {T}$} \Comment{\emph{For each task in DAG...}}
		\State $\omega_{i} = \textsc{GetRate}(\mathcal{G}, t_i, \Omega)$ \Comment{\emph{Returns input rate on task $t_i$ if DAG input rate is $\Omega$}}
		\State $\bar{\omega_{i}} = \mathcal{I}_i(1)$ \Comment{\emph{Peak rate supported by task $t_{i}$ with 1 thread }}
		\State $\tau_{i} \gets 0$  \Comment{\emph{Allocated thread count for task $t_i$}}
		\State$c_{i} \gets 0$ \Comment{\emph{Estimated CPU\% for $\tau_i$ threads of task $t_i$}}
		\State$m_{i} \gets 0$ \Comment{\emph{Estimated Memory\% for $\tau_i$ threads of task $t_i$}}
		
		\While{$\omega_{i} \ge \bar{\omega_{i}}$} \label{alg:lsa:greater}\\
		\Comment{\emph{One additional thread added for $t_i$, with increase in cumulative rate supported and resources used}}
		\State $\tau_i \gets \tau_i + 1$
		\State $\omega_{i} \gets  \omega_{i} - \bar{\omega_{i}}$
		\State $c_{i} \gets c_{i} + \mathcal{C}_i(1)$
		\State $m_{i} \gets m_{i} + \mathcal{M}_i(1)$
		\EndWhile
		
		\If{$\omega_{i} > 0$} \Comment{\emph{Trailing input rate below $\bar{\omega_{i}}$. Add thread but scale down the resources needed.}} \label{alg:lsa:smaller}
		\State $\tau_i \gets \tau_i + 1$
		\State $c_{i} \gets c_{i}+\mathcal{C}_i(1) \times \cfrac{\omega_i}{\bar{\omega_{i}}}$
		\State $m_{i} \gets m_{i}+\mathcal{M}_i(1) \times \cfrac{\omega_i}{\bar{\omega_{i}}}$
		\State $\omega_{i} \gets  0 $
		\EndIf
		
		\EndFor\\
		\Return{$\langle \tau_{i}, c_{i}, m_{i} \rangle~~\forall t_{i}\in \mathbb{T}$} \Comment{\emph{Return number of threads, CPU\% and Memory\% allocated to each task}}
		\EndProcedure
	\end{algorithmic}
\end{algorithm}

\subsection{Model-based Allocation (MBA)}
\label{key} 
While the LSA approach is simple and appears intuitive, it suffers from two key problems that make it unsuitable for may tasks. \emph{First, the assumption that the input rate supported will linearly increase with the number of threads is not valid.} Based on our observations of the performance models from Fig.~\ref{fig:bm}, we can see that for some tasks like Azure Blob and Table, there is a loose correlation between the number of threads and the input rate supported. But even this evens out at a certain number of threads. Others like Parse XML, Pi Computation and Batch File Write see their peak input rate supported remain flat or decrease as the threads increase \emph{on a single resource slot} due to contention. One could expect a linear behavior if the threads run on different slots or VMs, but that would increase the slots required (and the corresponding cost).

\emph{Second, the assumption that the resource usage linearly scales with the number of threads, relative to the resources for 1 thread, is incorrect.} This again follows from the performance models, and in fact, the behavior of CPU and memory usage themselves vary for a task. For e.g., in Fig.~\ref{fig:bm:pi} for Pi, the CPU usage remains flat while the memory usage increases even as the rate supported decreases with the number of threads. For Azure Table query in Fig.~\ref{fig:bm:table}, the CPU and memory increase with the threads but with very different slopes.

Our Model-based Allocation algorithm, shown in Algorithm~\ref{alg:mba}, avoids such inaccurate assumptions and instead uses the performance models measured for the tasks to drive its thread and resource allocation. Here, the intuition is to select the sweet spot of the number of threads such that the peak rate $\widehat{\omega_i}$ among all number of threads (for which the model is available) is the highest for task $t_i$ (lines~\ref{alg:mba:greater:start}--\ref{alg:mba:greater:end}). This ensures that we maximize the input rate that we can support from a single resource slot for that task. At the same time, when we allocate these many threads to saturate a slot, we also disregard the actual CPU\% and memory\% and instead treat the whole slot as being allocated ($100\%$ usage). This is because that particular task cannot make use of additional CPU or memory resources available in that slot due to a resource contention, and we do not wish additional threads on this slot to exacerbate this problem.

When the residual input rate to be supported for a task falls below this maximum peak rate (line~\ref{alg:mba:smaller}), we instead select smallest number of threads that are adequate to support this rate, and use the corresponding CPU and memory\%. If a single thread is adequate (line~\ref{alg:mba:1}), just as for LSA, we scale down the resources needed proportion to the residual input rate relative to the peak rate using 1 thread.



\begin{algorithm}[t]
	\footnotesize
	\caption{Model Based Allocation (MBA)}\label{alg:mba}
	\begin{algorithmic}[1]
		\Procedure{AllocateMBA} {$\mathcal{G} : \langle \mathbb {T},\mathbb {E} \rangle ),~\Omega$}
		\For{$t_{i} \in \mathbb {T}$} \Comment{\emph{For each task in DAG...}}
		\State $\omega_{i} = \textsc{GetRate}(\mathcal{G}, t_i, \Omega)$ \Comment{\emph{Returns input rate on task $t_i$ if DAG input rate is $\Omega$}}
		\State $\widehat{\omega_{i}} = \max_{j}\big\{\mathcal{I}_i(j)\big\}$ \Comment{\emph{Maximum of peak rates supported by task $t_{i}$ with any number of threads }}
		\State $\widehat{\tau_i} = \mathcal{T}(\widehat{\omega_{i}})$ \Comment{\emph{Number of threads needed to support rate $\widehat{\omega_{i}}$ for task $t_{i}$}}
		\State $\tau_{i} \gets 0$  \Comment{\emph{Allocated thread count for task $t_i$}}
		\State$c_{i} \gets 0$ \Comment{\emph{Estimated CPU\% for $\tau_i$ threads of task $t_i$}}
		\State$m_{i} \gets 0$ \Comment{\emph{Estimated Memory\% for $\tau_i$ threads of task $t_i$}}
		
		\While{$\omega_{i} \ge \widehat{\omega_{i}}$} \label{alg:mba:greater:start}\\
		\Comment{\emph{Add threads for $t_i$ corresponding to maximum peak rate. Increase cumulative rate and resources.}}
		\State $\tau_i \gets \tau_i + \widehat{\tau_i}$
		\State $\omega_{i} \gets  \omega_{i} - \widehat{\omega_{i}}$
		\State $c_{i} \gets c_{i} + 1.00$ \Comment{\emph{Assign $100\%$ of resource slot to these threads}}
		\State $m_{i} \gets m_{i} + 1.00$
		\EndWhile \label{alg:mba:greater:end}
		\If{$\omega_{i} > 0$} \Comment{\emph{Trailing input rate below $\widehat{\omega_{i}}$ to be processed for $t_i$}} \label{alg:mba:smaller}
		\State $\tau_i' \gets \mathcal{T}(\omega_{i})$
		\State $\tau_i \gets \tau_{i} + \tau_i'$
		\If{$\tau_i' > 1$} 
		\State $c_{i} \gets c_{i}+\mathcal{C}_i(\tau_i')$
		\State $m_{i} \gets m_{i}+\mathcal{M}_i(\tau_i')$
		\Else \Comment{\emph{One thread adequate. Scale down resources needed}}\label{alg:mba:1}
		\State $c_{i} \gets c_{i}+\mathcal{C}_i(1) \times \cfrac{\omega_i}{\mathcal{I}_i(1)}$
		\State $m_{i} \gets m_{i}+\mathcal{M}_i(1) \times \cfrac{\omega_i}{\mathcal{I}_i(1)}$
		\EndIf
		\State $\omega_{i} \gets  0 $
		\EndIf
		\EndFor\\
		
		\Return{ $\langle \tau_{i},c_{i},m_{i} \rangle~~\forall t_{i}\in \mathbb {T}$} \Comment{\emph{Return number of threads, CPU\% and Memory\% allocated to each task}}
		\EndProcedure
	\end{algorithmic}
\end{algorithm}

The result of running MBA is also a list of thread counts $\tau_i$ per task in the DAG, which will be used by the mapping algorithm. It also gives the CPU\% and memory\% per task which, as before, helps estimate the slots for the DAG: 
\[ \rho = \max\Big(\Big\lceil\sum_{t_i \in \mathbb{T}}(c_i) \Big\rceil, \Big\lceil \sum_{t_i \in \mathbb{T}}(m_i) \Big\rceil \Big) \]

\section{Resource Mapping} 
\label{sec:mapping}

As we saw, the allocation algorithm returns two pieces of information $\tau_i$, the number of threads per task, and $\rho$, the number of resource slots allocated. The goal of resource mapping is to assign these task threads, $r_i^k \in R$ where $|r_i^k| = \tau_i$, to specific resource slots to meet the input rate requirements for the DAG.

\subsection{Resource Acquisition}
\label{sec:acquire}
The first step is to identify and acquire adequate number and sizes of VMs that have the estimated number of slots. This is straight-forward. Most IaaS Cloud providers offer on-demand VM sizes with slots that are in powers of $2$, and with pricing that is a multiple of the number of slots. For e.g., the Microsoft Azure D-series VMs in the Southeast Asia data center have $1,2,3$ and $4$ cores for sizes D1--D4, with $3.5~GB$ RAM per core, and costing $\$0.098, \$0.196, \$0.392$ and $\$0.784$/hour, respectively~\footnote{Microsoft Azure Linux Virtual Machines Pricing, \url{https://azure.microsoft.com/en-in/pricing/details/virtual-machines/linux/}}. Amazon AWS and Google Compute Engine IaaS Cloud services have similar VM sizes and pricing as well. 
So the total price for a given slot-count $\rho$ does not change based on the mix of VM sizes used, and one can use a simple packing algorithm to select VMs $v_j \in V$ with sizes such that $\sum_{v_j \in V}{p_j} = \rho$, where $p_j$ is the number of slots per VM $v_j$. At the same time, having more VMs means higher network transfer latency, and end-to-end latency for the DAG will increase. Hence, minimizing the number of distinct VMs is useful as well rather than having many small VMs.

One approach that we take to balance pricing with communication latency is to acquire as many VMs `$n$' with the largest number of slots as possible, say, $\widehat{p}$, which cumulatively have $(n \times \widehat{p}) \le \rho$. Then, for the remaining slots, we assign to the smallest possible VM whose number of slots is $\ge (\rho - n \times \widehat{p})$. This may include some additional slots than required, but is bound by $(2^{\widehat{p} - 1}-1)$ if slots per VM are in powers of $2$, as is common. Other approaches are possible as well, based on the trade-off between network latency costs and slot pricing. For the $v_j \in V$ set of VMs thus acquired, let $s_j^l \in S$ be the set of slots in these VMs, where $|s_j^l| = p_j$ and $p_j \le \widehat{p}$, such that $\Big(\sum_{v_j \in V} p_j\Big) \ge \rho$.

The second step, that we emphasize in this article, is to map the threads for each task to one of the slots we have acquired, and determine the mapping function $\mathcal{M} : R \rightarrow S$. Next, we discuss two mapping approaches available in literature, that we term DSM and RSM, as comparison and propose our novel mapping SAM as the third. While DSM is not ``resource aware'', i.e., it does not consider the resources required by each thread in performing the mapping, RSM and our SAM are resource aware, and use the output from the performance models developed earlier.

\begin{figure}[t]
	\centering
	\includegraphics[width=0.67\textwidth]{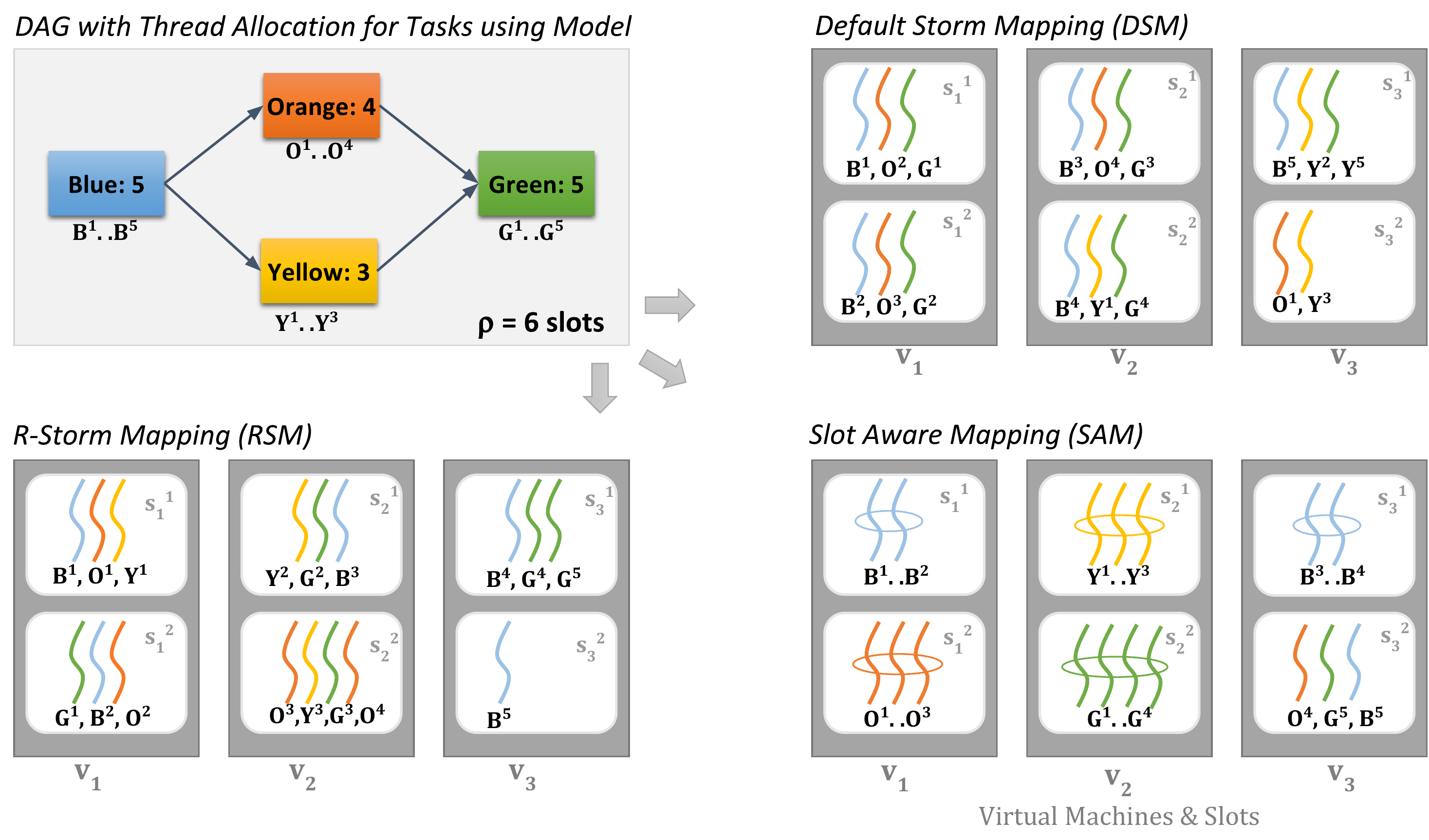}
	\caption{Mapping of a sample DAG to VMs and resource slots using DSM, RSM and SAM }
	\label{fig:mapping}
\end{figure}

\subsection{Default Storm Mapping (DSM)}
DSM is the default mapping algorithm used in Apache Storm, and uses a simple round-robin algorithm to assign the threads to slots. All threads are assumed to have a similar resource requirement, and all slots are assumed to have homogeneous capacity. Under such an assumption, this na\"{i}ve algorithm will balance the load of the number of threads across all slots. 
Algorithm~\ref{alg:dsm} describes its behavior. 
The task threads and resource slots available are provided as a set to the algorithm. The slots are considered as a list in some random order. The algorithm iterates through the set of threads in arbitrary order. For each thread, it picks the next slot in the list and repeats this in a   
round-robin fashion for each pending thread (line \ref{alg:dsm:rr}), wrapping around the slot list if its end is reached.

Fig.~\ref{fig:mapping} illustrates the different mapping algorithms for a sample DAG with four tasks, \texttt{\underline{B}lue, \underline{O}range, \underline{Y}ellow} and \texttt{\underline{G}reen}, and say $5, 4, 3$ and $5$ threads allocated to them, respectively, by some allocation algorithm. Let the resources estimated for them be $6$ slots that are acquired across three VMs with $2$ slots each.

Given this, the DSM algorithm first gets the list of threads and slots in some random order. For this example, let them be ordered as, $B^1,..., B^5, O^1,..., Y^1,..., \\G^1,..., G^5$, and $s_1^1, s_1^2, s_2^1, ..., s_3^2$. Firstly, the five blue threads are distributed across the first $5$ slots, $s_1^1 - s_3^1$ sequentially. Then, the distribution of the orange threads resumes with the sixth slot $s_3^2$, and wraps around to the first slot to end at $s_2^1$. The three yellow threads map to slots $s_2^2 - s_3^2$, and lastly, the five green threads wrap around and go to the first $5$ slots. 
As we see, DSM distributes the threads evenly across all the acquired slots irrespective of the resources available on them or required by the threads, with only the trailing slots having fewer threads. This can also help distribute threads of the same task to different slots to avoid them contending for the same type of resources. However, this is optimistic and, as we have seen from the performance models, the resource usages sharply vary not just across tasks but also based on the number of threads of a task present in a slot.


\begin{algorithm}[t]
		\footnotesize
	\caption{Default Storm Mapping (DSM)}\label{alg:dsm}
	\begin{algorithmic}[1]
		\Procedure{MapDSM} {$R,~S$}\Comment{\emph{Map each task thread in set $R$ to a slot in set $S$}}
		\State $\mathcal{M} \gets new~Map(~)$
		\State $S'[~] \gets \textsc{SetToList}(S)$ \Comment{\emph{Returns the slots in the set as a list, in some arbitrary order}}
		\State $n \gets 1$
		\For{\textbf{each} $r \in R$} \Comment{\emph{Round Robin mapping of threads to slots}}   \label{alg:dsm:rr}
		\State $m \gets n~\textbf{mod}~|S|$
		\State $s = S'[m]$ 
		\State $\mathcal{M}.put(r \rightarrow ~s)$ \Comment{\emph{Assign $n^{th}$ task thread to $m^{th}$ resource slot}} 
		\State $n \gets n+1$
		\EndFor\\
		\Return{$\mathcal{M}$}
		\EndProcedure
		
	\end{algorithmic}
\end{algorithm}

\subsection{R-Storm Mapping (RSM)}
The R-Storm Mapping (RSM) algorithm~\cite{peng:middleware:2015} was proposed earlier as a resource-aware scheduling approach to address the deficiencies of the default Storm scheduler. It has subsequently been included as an alternative scheduler for Apache Storm, as of v1.0.1. 
RSM offers three features that improve upon DSM. First, in contrast to DSM that balances the thread count across all available slots, RSM instead maximizes the resource usage in a slot and thus minimizes the number of slots required for the threads of the DAG. For this, it makes use of the resource usage for \emph{single threads} that we collect from the performance model ($\bar{c}_i, \bar{m}_i$), and the resource capacities of slots and VMs. A second side-effect of this is that it prevents slots from being over-allocated, which can cause an unstable DAG mapping at runtime. For e.g., DSM could place threads in a slot such that their total CPU\% or memory\% is greater than $100\%$. This is avoided by RSM. Lastly, RSM is aware of the network topology of the VMs, and it places the threads on slots such that the communication latency between adjacent tasks in the dataflow graph is reduced. 

At the heart of the RSM algorithm is a \textit{distance function} based on the available and required resources, and a network latency measure. This Euclidean distance between a given task thread $r_i^k \in R$ and a VM $v_j \in V$ is defined as: 
\[ d = w_M \times (M_j -  \bar{m_i})^2 + w_C \times (C_j -  \bar{c_i})^2 + w_N \times \textsc{NWDist}(\widehat{v},v_j) \] 
where $\bar{c_i} = \mathcal{C}_i(1)$ and $\bar{m_i} = \mathcal{M}_i(1)$ are the incremental CPU\% and memory\% required by a single thread of the task $t_i$ on one slot, and $C_j$ and $M_j$ are the CPU\% and memory\% not yet mapped across all slots of VM $v_j$. A network latency multiplier, from a \emph{reference VM}, $\widehat{v}$, to the candidate VM $v_j$, is also defined using the \textsc{NWDist} function. This reference VM is the last VM on which a task thread was mapped, and the network latency multiplier is set to $0.0$ if the candidate VM is the reference VM, $0.5$ if the VM is in the same rack as the reference, and $1.0$ if on a different rack. Lastly, the weights $w_C, w_M$ and $w_N$ are coefficients to tune this distance function as appropriated. 

Given this, the RSM Algorithm, given in Alg.~\ref{alg:rsm}, works as follows. 
It initializes the CPU\% and memory\% resources available for the candidate VMs to $100\%$ of their number of slots, the memory available per slot to $100\%$, and the number of threads to be mapped per task (lines~\ref{alg:rsm:init:start}--\ref{alg:rsm:init:end}). The initial reference VM is set to some VM, say $v_1 \in V$. Then, it performs one sweep where one thread of each task, in topological order rooted at the source task(s), is mapped to a slot (lines~\ref{alg:rsm:map:start}--\ref{alg:rsm:map:end}). This mapping in the order of BFS traversal increases the chance that threads of adjacent tasks in the DAG are placed on the same VM to reduce network latency.

During the sweep, we first check if the task has any pending threads to map, and if so, we test the VMs in the ascending order of their distance function, returned by function \textsc{GetSortedVMs}, to see if they have adequate resources for that task thread (lines~\ref{alg:rsm:dist}--\ref{alg:rsm:map:loop}). There are two checks that are performed: one to see if the VM has adequate CPU\% available for the thread, and second if any slot in that VM has enough memory\% to accommodate that thread. This differentiated check is because in Storm, the memory allocation per slot is tightly bound, while the CPU\% available across slots is assumed to be usable by any thread placed in that VM.

If no available slot meets the resource needs of a thread, then RSM fails. As we show later, this is not uncommon depending on the allocation algorithm. If a valid slot, $s_j'$, is found, the task thread is mapped to this slot, and the thread count and resource availability updated (lines~\ref{alg:rsm:post:start}--~\ref{alg:rsm:post:end}). The reference VM is also set to the current VM having that slot. Then the next threads in the sweep are mapped, and this process repeated till all task threads are mapped.

Fig.~\ref{fig:mapping} shows the RSM algorithm applied to the sample DAG. Mapping of the threads to slots is done in BFS ordering of tasks, $B, O, Y$ and $G$. 
For each thread of the task in this order, a slot on the VM with the minimum distance and available resources is chosen. Say in the first sweep, the threads  $B^1,O^1$ and $Y^1$ are mapped to the same slot $s_1^1$, and then next thread $G^1$ be mapped to new slot $s_1^2$ due to resource constraint on $s_1^1$ for this thread. The new slot $s_1^2$ is picked on same VM as it has the least distance among all VMs. In the second sweep, thread $B^2,O^2, Y^2$ and $G^2$ are mapped to slots $s_1^2$ and $s_2^1$, and likewise for the third sweep. In the fourth sweep, there are no threads for the Yellow task pending. Also, we see that thread $B^4$ is not mapped to slot $s_2^2$ due to lack of resources, instead going to $s_3^1$. However, a slot $s_2^2$ does have resources for a subsequent thread $O^4$, and the  
distance to $s_2^2$ is lesser than $s_3^1$. Thus RSM tries to perform a network-aware best fit packing to minimize the number of slots.

\begin{algorithm}[t]
	\footnotesize
	\caption{R-Storm Mapping (RSM)}\label{alg:rsm}
	\begin{algorithmic}[1]
		\Procedure{MapRSM} {$\mathcal{G}:\langle \mathbb {T},\mathbb {E} \rangle,~R,~V,~S$}
		\State $C_j = p_j \times 1.00,~~~ M_j = p_j \times 1.00,~~\forall v_j \in V$ \Comment{\emph{Initialize available CPU\%, Memory\% for all VMs}} \label{alg:rsm:init:start}
		\State $M_j^l = 1.00,~~\forall s_j^l \in S$ \Comment{\emph{Initialize available Memory\% for all slots of VMs}} 
		\State $\tau_i = |r_i^k|~~\forall r_i^k \in R$ \Comment{\emph{Initialize number of task threads to map for task $t_i$}} \label{alg:rsm:init:end}
		\State $\mathcal{M} \gets new~Map(~)$  \Comment{\emph{Initialize mapping function}}
		\State $\widehat{v} \gets v_1$ \Comment {\emph{Initialise the reference VM to the first VM in set}}
		\While{$\sum_{t_i \in \mathbb{T}}\tau_i > 0$}  \Comment{\emph{Repeat while tasks have unmapped threads}}
		\For{\textbf{each} $t_{i} \in \textsc{TasksTopoOrder}(\mathcal{G})$} \label{alg:rsm:map:start} 
		\If{$\tau_i \neq 0$} 
		\State \hskip-3em \Comment{\emph{Get list of VMs sorted by distance, based on their available resources for task $t_i$}}
		\State $V'[~] \gets \textsc{GetSortedVMs}(V, t_i, \widehat{v})$ \label{alg:rsm:dist}
		\State $s_j' \gets \varnothing$  
		\For{$v_j' \in V'[~] ~\&~ s_j' == \varnothing$} 
		\State $s_j' \gets s_j^l \in S ~|~ C_j \geq \bar{c}_i ~\&~ M_j^l \geq \bar{m}_i$   \Comment{\emph{Does VM have CPU\%, some slot in it have mem\% for 1 thread?}}
		\EndFor \label{alg:rsm:map:loop}
		\If{$s_j' == \varnothing$} \Return{``Error: Insufficient resources for task $t_i$''}
		\EndIf
		\State $r_i' \gets r_i^k \in R ~|~ \not\exists \mathcal{M}(r_i^k)$ \Comment{\emph{Pick one unmapped thread for task $t_i$}}
		\State $\mathcal{M}.put(r_i' \rightarrow s_j')$   \Comment{\emph{Assign the thread to the selected slot with available resources}}  \label{alg:rsm:map}
		\State $C_j \gets (C_j - \bar{c}_i),~~M_j \gets (M_j - \bar{m}_i),~~M_j' \gets (M_j' - \bar{m}_i)$  \Comment {\emph{Reduce available resources by 1 thread}}  \label{alg:rsm:post:start} 
		\State $\tau_i \gets \tau_i-1$  \label{alg:rsm:post:end}
		\State $\widehat{v} \gets v_j'$ \Comment{\emph{Update reference node to be the current mapped VM}}
		\EndIf
		\EndFor \label{alg:rsm:map:end}
		\EndWhile\\
		\Return{$\mathcal{M}$}
		\EndProcedure
		%
		
		%
	\end{algorithmic}
\end{algorithm}

\subsection{Slot Aware Mapping (SAM)} 
While the RSM algorithm is resource aware, it linearly extrapolates the resource usage for multiple threads of a task in a VM or slot based on the behavior of a single thread. As we saw earlier in the \S~\ref{sec:bm}, this assumption does not hold, and as we show later in \S~\ref{sec:results}, it causes inefficient mapping, non-deterministic performance and needs over-allocation of resources. Our Slot Aware Mapping (SAM) addresses these deficiencies by fully utilizing the performance model and the strategy used by our model based allocation. They key idea is to map a \emph{bundle of threads} of the same task exclusively to a single slot such that the stream throughput is maximized for that task on that slot based on its performance model, and the interference from threads of other tasks on that slot is reduced.

In Algorithm~\ref{alg:sam}, as for RSM, we initialize the resources for the VMs and slots. Further, in addition to the total slots $\rho$ required by the DAG, we also have the quantity of CPU\% and memory\% required by all the threads of each task available as $c_i$ and $m_i$. Recollect that the MBA algorithm returns this information based on the performance model. As for RSM, we iterate through the tasks in topological order (line~\ref{alg:sam:topo}). However, rather than map one thread of each task, we first check if the number of pending threads forms a \emph{full bundle}, which we define to be as $\widehat{\omega_i}$, the number of threads at the peak rate supported by the task on a single slot (line~\ref{alg:sam:full}). If so, we select an empty slot in the last mapped VM, or if none exist, in its neighboring one (line~\ref{alg:sam:fullslot}). We $\widehat{\omega_i}$ unmapped threads for this task and assign this whole bundle of threads to this exclusive slot, i.e., $100\%$ of its CPU and memory (line~\ref{alg:sam:full:map}). The resource needs of the task are reduced concomitantly, and this slot is fully mapped.

It is possible that the task has a \emph{partial bundle} of unmapped threads, having fewer than $\widehat{\omega_i}$ ones (line~\ref{alg:sam:partial}). In this case, we find the best-fit slot as the one whose sum of available CPU\% and memory\% is the smallest, while being adequate for the CPU\% and memory\% required for this partial bundle (line~\ref{alg:sam:bfslot}). We assign this partial bundle of threads to this slot and reduce the resources available for this slot by $c_i$ and $m_i$. At this point, all threads of this task will be assigned (line~\ref{alg:sam:partial:done}). 

Notice that slots co-locate threads from different tasks only for the last partial bundle of each task. So we have an upper bound on the number of slots with mixed thread types as $|\mathbb{V}|$. Since the performance models offers information on the behavior of the same thread type on a slot, this limits the interference between threads of different types, that is not captured by the model. In practice, as we show in the experiments, most slots have threads of a single task type. As a result, SAM has a more predictable resource usage and behavior for the mapped DAG.

It is possible that even in SAM, the resources allocated may not be adequate for the mapping (lines~\ref{alg:sam:full:na}, \ref{alg:sam:partial:na}), though the chances of this happening is smaller than for RSM since SAM uses a strategy similar to the allocation algorithm, MBA. This is a side-effect of the binning, when resource available in partly used slots are not adequate to fully fit a partial bundle.
Also, while we do not explicitly consider network distance unlike in RSM, the mapping of tasks in topological order combined with picking a bundle at a time achieves a degree of network proximity between threads of adjacent tasks in the DAG.
\begin{algorithm}[H]
	\footnotesize
	\caption{Slot Aware Mapping (SAM)}\label{alg:sam}
	\begin{algorithmic}[1]
		\Procedure{MapSAM} {$\mathcal{G}:\langle \mathbb {T},\mathbb {E} \rangle,~R,~V,~S$}
		\Comment{\emph{$c_{i}$ and $m_{i}$ are CPU\% and memory\% required by task $t_{i}$ from MBA}}
		\State $\tau_i = |r_i^k|~~\forall r_i^k \in R$ \Comment{\emph{Initialize number of task threads to map for task $t_i$}} 
		\State $\widehat{\tau_i} = \mathcal{T}(\widehat{\omega_{i}})$ \Comment{\emph{Number of threads needed to support peak rate $\widehat{\omega_{i}}$ for task $t_{i}$ on 1 slot}}
		\State $C_j^l = 1.00,~~~ M_j^l = 1.00,~~\forall s_j^l \in S$ \Comment{\emph{Initialize available CPU\%, Memory\% for all slots of VMs}} 
		
		\State $\mathcal{M} \gets new~Map(~)$  \Comment{\emph{Initialize mapping function}}
		\While{$\sum_{t_i \in \mathbb{T}}\tau_i > 0$}  \Comment{\emph{Repeat while tasks have unmapped threads}}
		\For{\textbf{each} $t_{i} \in \textsc{TasksTopoOrder}(\mathcal{G})$} \label{alg:sam:topo}
		\If{$\tau_i \geq \widehat{\tau_i}$} \Comment {\emph{At least 1 full bundle of threads remains for task $t_i$}} \label{alg:sam:full}
		\State $s_j' \gets \textsc{GetNextFullSlot(V)}$ \Comment{\emph{Returns next full slot in current or next VM}} \label{alg:sam:fullslot}
		\If{$s_j' == \varnothing$} \Return{``Error: Insufficient resources for task $t_i$''} \label{alg:sam:full:na}
		\EndIf
		\State $r_i'[~] \gets \{ r_i^k \} \in R ~|~ \not\exists \mathcal{M}(r_i^k),~~|r_i'| = \widehat{\tau_i}$ \Comment{\emph{Pick unmapped full bundle of $\widehat{\tau_i}$ threads for task $t_i$}} 
		\State $\mathcal{M}.putAll(r_i'[~] \rightarrow s_j')$   \Comment{\emph{Assign threads in bundle to the selected slot}}  \label{alg:sam:full:map}
		\State $\tau_i \gets \tau_i - \widehat{\tau_i},~~~c_{i} \gets c_{i} - 1.00,~~~ m_i \gets m_i - 1.00$  \Comment{\emph{Reduce resource needs for bundle}}
		\State $C_j' \gets 0.00,~~~M_j' \gets 0.00$  \Comment {\emph{Used all resources in slot}}  \label{alg:sam:full:end}
		\Else
		\If{$\tau_i > 0$} \Comment {\emph{Partial bundle of threads remains for task $t_i$}} \label{alg:sam:partial}
		\State $s_j' \gets \textsc{GetBestFitSlot}(V, c_i, m_i)$ \Comment{\emph{Find smallest slot with sufficient resources for partial bundle}} \label{alg:sam:bfslot}
		\If{$s_j' == \varnothing$} \Return{``Error: Insufficient resources for task $t_i$''} \label{alg:sam:partial:na}
		\EndIf
		\State $r_i'[~] \gets \{ r_i^k \} \in R ~|~ \not\exists \mathcal{M}(r_i^k)$ \Comment{\emph{Pick all remaining threads for task $t_i$}}
		\State $\mathcal{M}.putAll(r_i'[~] \rightarrow s_j')$
		\State $C_j' \gets C_j' - c_i,~~~M_j' \gets M_j' - m_i$  \Comment {\emph{Reduce resources in slot by partial bundle}}  
		\State $\tau_i \gets 0,~~~c_{i} \gets 0.00,~~~m_i \gets 0.00$  \Comment{\emph{All done for this task}} \label{alg:sam:partial:done}
		\EndIf
		\EndIf
		\EndFor
		\EndWhile\\         
		\Return{$\mathcal{M}$}
		\EndProcedure
	\end{algorithmic}
\end{algorithm}

Fig.~\ref{fig:mapping} shows the SAM algorithm applied to the sample DAG. Say a full bundle of the four tasks, $B, O, Y$ and $G$, have 2, 3, 3 and 4 threads, respectively. We iteratively consider each task in the BFS order of the DAG, similar to RSM, and attempt to assign a full bundle from their remaining threads to an exclusive slot. For e.g., in the first sweep, the full bundles $B^1..B^2, O^1..O^3, Y^1..Y^3, G^1..G^4$ are mapped to the four slots, $s_1^1, s_1^2, s_2^1, s_2^2$, respectively, occupying 2 VMs. In the next sweep, we still have a full bundle for the Blue task, $B^3..B^4$, that takes an independent slot $s_3^1$, but the Orange and Green tasks have only partial bundle consisting of one thread each. $O^4$ is mapped to a new slot $s_3^2$ as there are no partial slots, and $G^5$ is mapped to the same slot as it is the best-fit partial slot. All threads of the Yellow task are already mapped. In the last sweep, the only remaining partial bundle for the Blue task, $B^5$ is mapped to the partial slot $s_3^2$ as the best fit.

\section{Results and Analysis }
\label{sec:results}

\subsection{Implementation}
We validate our proposed allocation and mapping techniques, MBA and SAM, on the popular Apache Storm DSPS, open-sourced by Twitter. Streaming applications in Storm, also called \emph{topologies}, are composed in Java as a DAG, and the resource allocation -- number of threads per task (\emph{parallelism hint}) and the resource slots for the topology (\emph{workers}) -- is provided by the user as part of the application. Here, we implement our MBA algorithm within a script that takes the DAG and the performance model for the tasks as input, and returns the number of threads and slots required. We manually embed this information in the application, though this can be automated in future. We take a similar approach and implement the LSA algorithm as well, which is used as a baseline. 

A Storm cluster has multiple hosts or VMs, one of which is the master and the rest are compute VMs having one or more resource slots. When the application is submitted to the Storm cluster, the master VM runs the \emph{Nimbus scheduling service} responsible for mapping the threads of the application's tasks to worker slots. A \emph{supervisor service} on each compute VM receives the mapping from Nimbus and assigns threads of the DAG for execution. While it is possible to run multiple topologies concurrently in a cluster, our goal is to run each application on an exclusive on-demand Storm cluster with the exact number of required VMs and slots, determined based on the allocation algorithm. For e.g., one scenario is to acquire a Storm cluster on-demand from Azure's HDInsight Platform as a Service (PaaS)~\footnote{Apache Storm for HDInsight, \url{https://azure.microsoft.com/en-in/services/hdinsight/apache-storm/}}.
%
%
%

Nimbus natively implements the default round-robin scheduler (DSM) and recently, the scheduling algorithm of R-Storm (RSM) using the \texttt{DefaultScheduler} and \texttt{ResourceAwareScheduler} classes, respectively. We implement our SAM algorithm as a custom scheduler in Nimbus, \texttt{SlotAwareScheduler}. It implements the \texttt{schedule} method of the \texttt{IScheduler} interface which is periodically invoked by the Nimbus service with the pending list of threads to be scheduled. When a thread for the DAG first arrives for mapping, the SAM scheduler generates and caches a mapping for all the threads in the given DAG to slots available in the cluster. The host IDs and slot IDs available in the cluster is retrieved using methods in Storm's \texttt{Cluster} class. Then, the algorithm groups the threads by their slot ID as Storm requires all thread for a slot to be mapped at once.  
The actual mapping is enacted by calling the \texttt{assign} method of the \texttt{Cluster} class that takes the slot ID and the list of threads mapped to it.

\subsection{Experiment Setup}
\label{sec:setup}
The setup for validating and comparing the allocation and mapping algorithms are similar to the one used for performance modeling, \S~\ref{sec:bm:setup}. In summary, Apache Storm $v1.0.1$ is deployed on Microsoft Azure D-series VMs in the Southeast Asia data center. The type and number of VMs depend on the experiment, but each slot of this VM series has one core of Intel Xeon E5-2673~v3 CPU $@2.4~GHz$, $3.5~GB$ RAM and a $50~GB$ SSD. We use three VM sizes in our experiments to keep the costs manageable -- \texttt{D3} having $2^{(\underline{3}-1)} = 4$~slots, \texttt{D2} with $2$~slots, and \texttt{D1} with $1$~slot.

We perform two sets of experiments. In the first, we evaluate the resource benefits of our Model Based Allocation (MBA) combined with our Slot Aware Mapping (SAM), in comparison with the baseline Linear Storm Allocation (LSA) with the resource-aware R-Storm Mapping (RSM), for a given input rate (\S~\ref{sec:results:usage}). For the allocated number of resource slots, we acquire the largest size VMs (\texttt{D3}) first to meet the needs, and finally pick a \texttt{D2} or a \texttt{D1} VM for the remaining slot(s), as discussed in \S~\ref{sec:acquire}.  

In the second set of experiments (\S~\ref{sec:results:accy}), we verify the predictability of the performance of our MBA and SAM approaches, relative to the existing techniques. Here, we perform five experiments, using a combination of 2 allocation and 3 mapping algorithms. We measure the highest stable input rate supported by the DAGs using these algorithms on a fixed number of five \texttt{D3} VMs, and compare the observed rate and resource usage against the planned rate, and with the rate and usage estimated by our model. 

In all experiments, a separate \texttt{D3} VM is used as the master node on which the Nimbus scheduler and other shared services run. For the RSM implementation, we need to explicitly specify the memory available to a slot and to the VM, which we set to $3.5~GB$ and the number of slots times $3.5~GB$, respectively. For RSM and SAM, we set the available CPU\% per slot to 100\% and for the VM to be the number of VM slots times $100\%$. 

\subsection{Streaming Applications}
\label{sec:impl:topo}

\begin{figure*}[t]
	\includegraphics[width=0.99\textwidth]{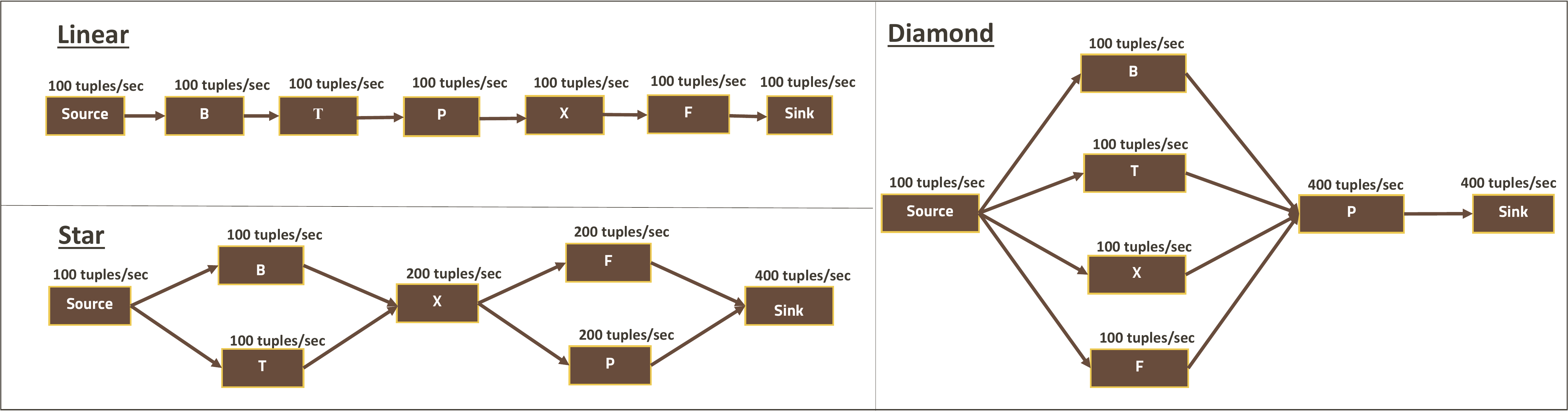}
	\caption{Micro DAG used in experiments. Tasks are referred to by their initials: B-Azure Blob Download, F-Batched File Write , P-Pi Computation, T-Azure Table Query, X-XMLparse. Since selectivity is 1:1, the input and output rates are the same, and indicated above the task}
	\label{fig:micro:dag}
\end{figure*}

\begin{figure*}[h]
	\centering
	\subfloat[Finance DAG]{
		\includegraphics[width=0.50\textwidth]{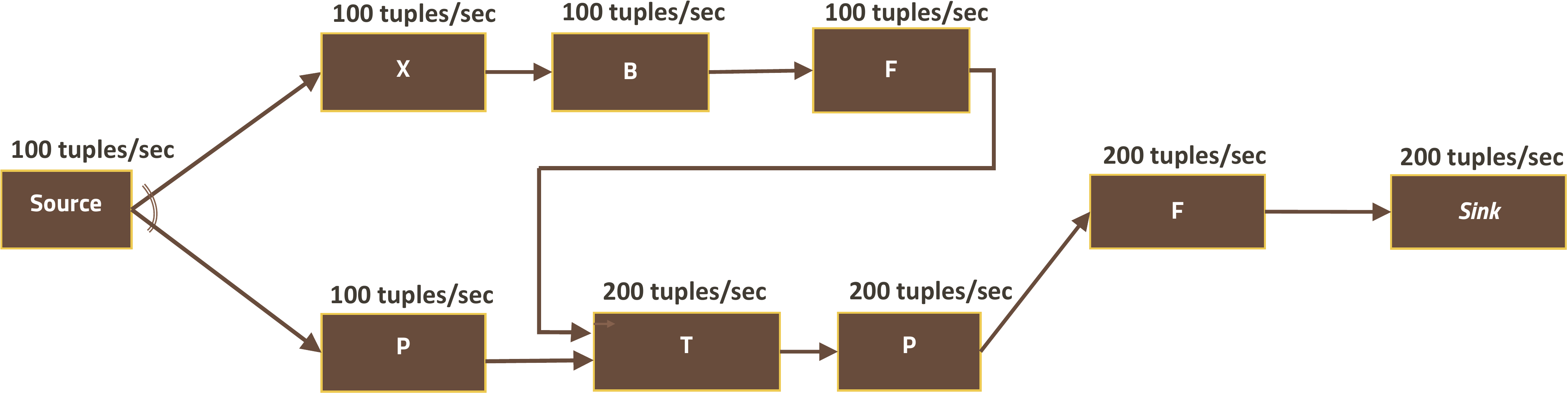}
		\label{fig:app:finance}
	}
	\subfloat[Traffic DAG]{
		\includegraphics[width=0.50\textwidth]{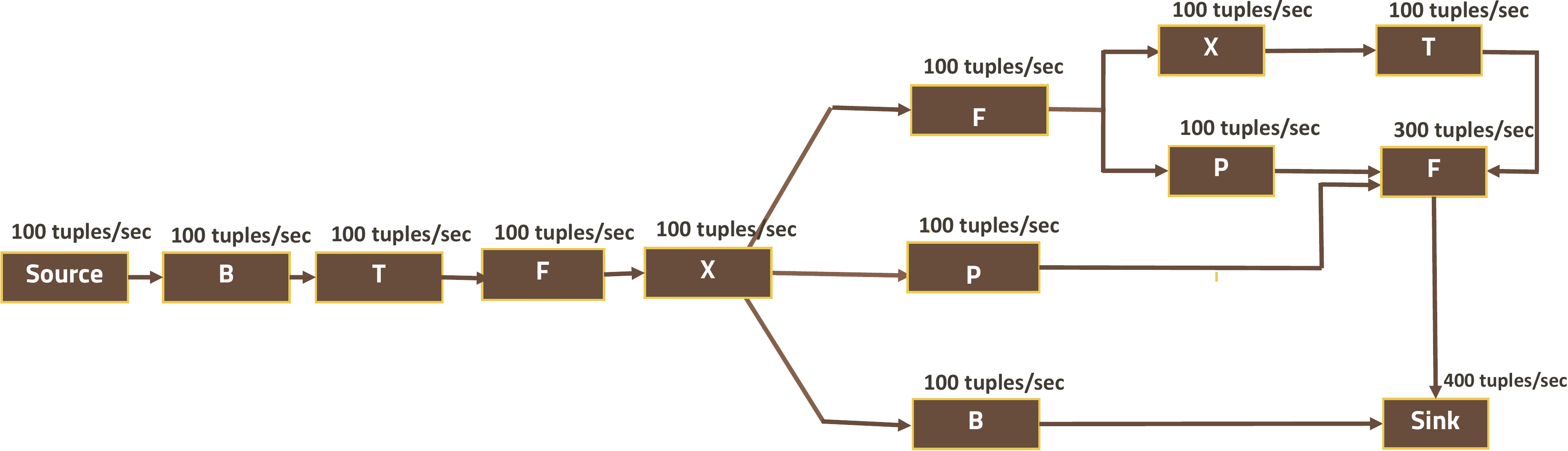}
		\label{fig:app:traffic}
	}\\
	\subfloat[Smart grid DAG]{
		\includegraphics[width=1.0\textwidth]{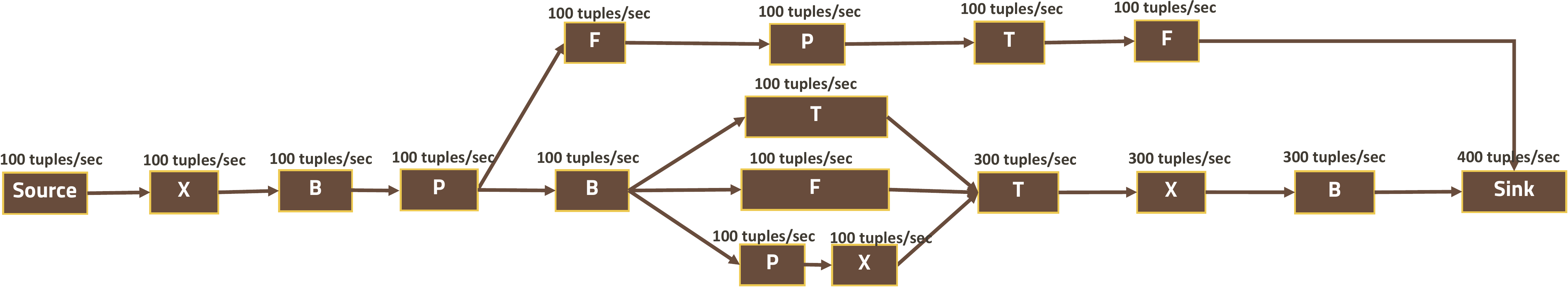}
		\label{fig:app:smartgrid}
		
	}
	\caption{Application DAGs [Notation: B-Azure Blob Download, F-Batched FileWrite , P-Pi Computation, T-Azure Table Query, X-XMLparse] }
	\label{fig:app:dag}
\end{figure*}

In our experiments, we use two types of streaming dataflows  -- \emph{micro-DAGs} and \emph{application DAGs}. The micro-DAGs capture common dataflow patterns that are found as sub-graphs in larger applications, and are commonly used in literature, including by R-Storm~\cite{peng:middleware:2015,millwheel:vldb:2013,xu:ic2e:2016}. These include \emph{Linear}, \emph{Diamond} and \emph{Star} micro-DAGs that respectively capture a sequential flow, a fan-out and fan-in, and a hub-and-spoke model (Fig.~\ref{fig:micro:dag}). While the linear DAG has a uniform input and output tuple rate for all tasks, the diamond exploits task parallelism, and the star doubles the input and output rates for the hub task. All three micro-DAGs have 5 tasks, in addition to the source and sink tasks, and we randomly assign the five different tasks that were modeled in \S~\ref{sec:bm} to these five DAG vertices, as labeled in Fig.~\ref{fig:micro:dag}. The figure also shows the input rates to each task based on a sample input rate of $100~tuples/sec$ to the DAG. All tasks have a selectivity of $\sigma = 1:1$. 

The application DAGs have a structure based on three real-world streaming applications that analyze traffic patterns from GPS sensor streams (\emph{Traffic})~\cite{biem:sigmod:2010}, compute the bargain index value from real-time stock trading prices (\emph{Finance})~\cite{gedik:sigmod:2008}, and perform data pre-processing and predictive analytics over electricity meter and weather data streams from Smart Power Grids (\emph{Grid})~\cite{simmhan:cise:2012}. In the absence of access to the actual application logic, we reuse and iteratively assign the five tasks we have modeled earlier to random vertices in these application DAGs and use a task selectivity of $\sigma = 1:1$.

These three applications DAGs have between $7-15$ logic tasks, and exhibit different composition patterns. Their overall DAG selectivity ranges from $1:2$ to $1:4$. We also see that the five diverse tasks we have chosen as proxies for these domain tasks are representative of the native tasks, as described in literature. For e.g., a task of the Traffic application does parsing of input streams, similar to our XML Parse task, and another archives data for historical analysis, similar to the Batch File Write task. The moving average price and bargain index value tasks in the Finance DAG are floating-point intensive like the Pi task. The Grid DAG performs parsing and database operations, similar to XML Parse and Azure Table, as well as time-series analytics that tend to be floating-point oriented. As a result, these are reasonable equivalents of real-world applications for evaluating resource scheduling algorithms.

Along with the application logic tasks, we have separate source and sink tasks for passing the input stream and logging the output stream for all DAGs. The source task generates synthetic tuples with a single opaque field of 10~bytes at a given constant rate. The sink task logs the output tuples and helps calculate the latency and other empirical statistics. Both these tasks are mapped by the scheduler just like the application tasks. Given the light-weight nature of these tasks, we empirically observe that  
a single thread for each of these tasks is adequate, with a static allocation of $10\%$ CPU and $15\%$ memory for the source and $10\%$ CPU and $20\%$ memory for the sink.

Each experiment is run for $15$~minutes and over multiple runs, and we report the representative values seen for the entire experiment. 

\subsection{Resource Benefits of Allocation and Mapping}
\label{sec:results:usage}
We compare our combination of MBA allocation and SAM mapping, henceforth called \textbf{MBA+SAM}, against LSA allocation with RSM mapping, referred to as \textbf{LSA+RSM}. The metric of success here is the ability to \emph{minimize the overall resources allocated} for a stable execution of the DAG at a given fixed input rate. We consider both micro-DAGs and applications DAGs. First the allocation algorithm determines the minimum number of resource slots required and then the mapping algorithm is used to schedule the threads on slots for the DAG. There may be cases where the resource-aware mapping algorithm is unable to find a valid schedule for the resource allocation, in which case, we incrementally increase the number of slots by $1$ until the mapping is successful. We report and discuss this as well. We then execute the DAG and check if it is able to support the expected input rate or not. 

\subsubsection{Micro DAG}
\label{exp:micro:slots}

\begin{figure*}[t]
	\centering
	\subfloat[Linear DAG]{
		\includegraphics[width=0.33\textwidth]{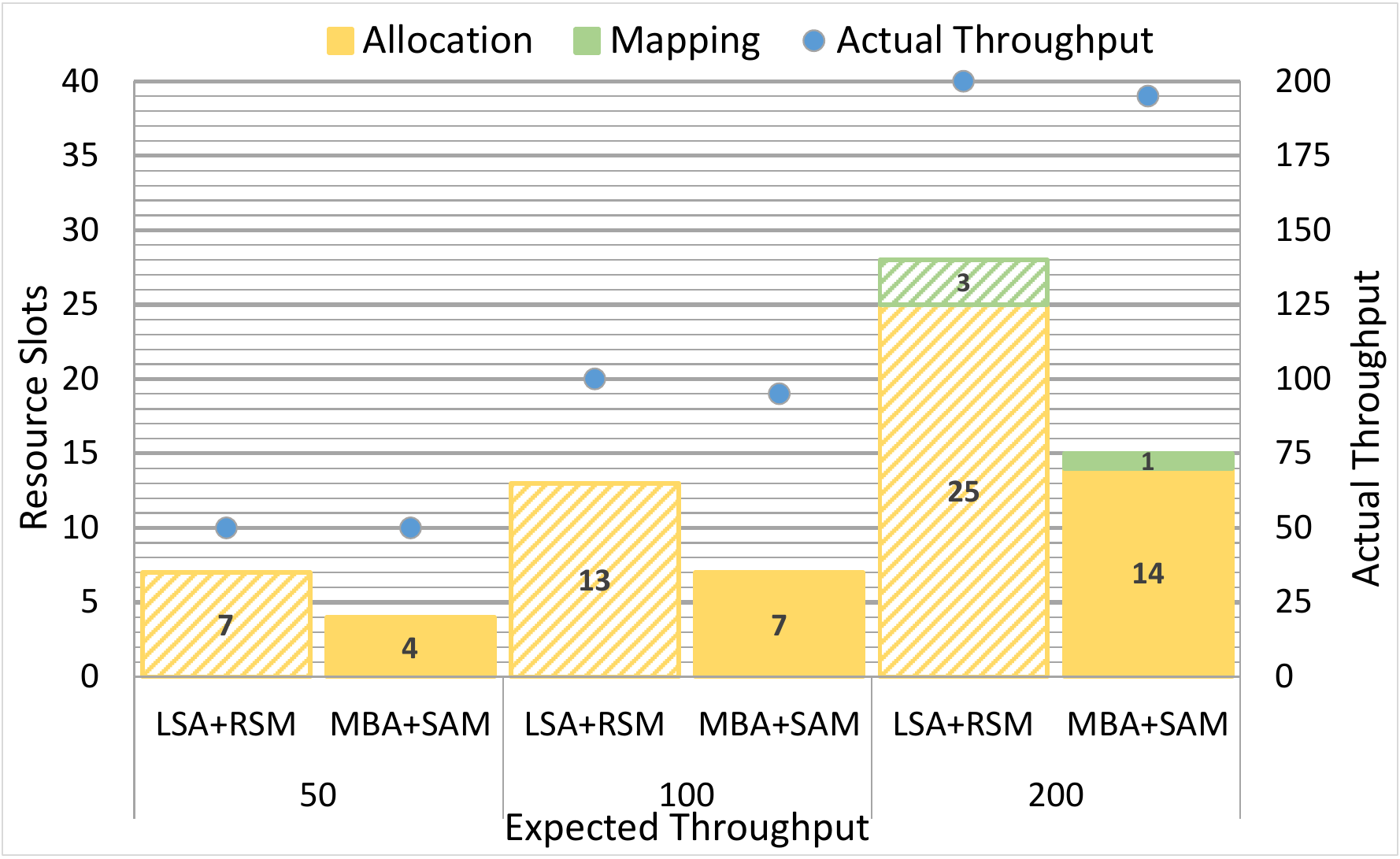} 
		\label{fig:slots:linear}
	}
	\subfloat[Diamond DAG]{
		\includegraphics[width=0.33\textwidth]{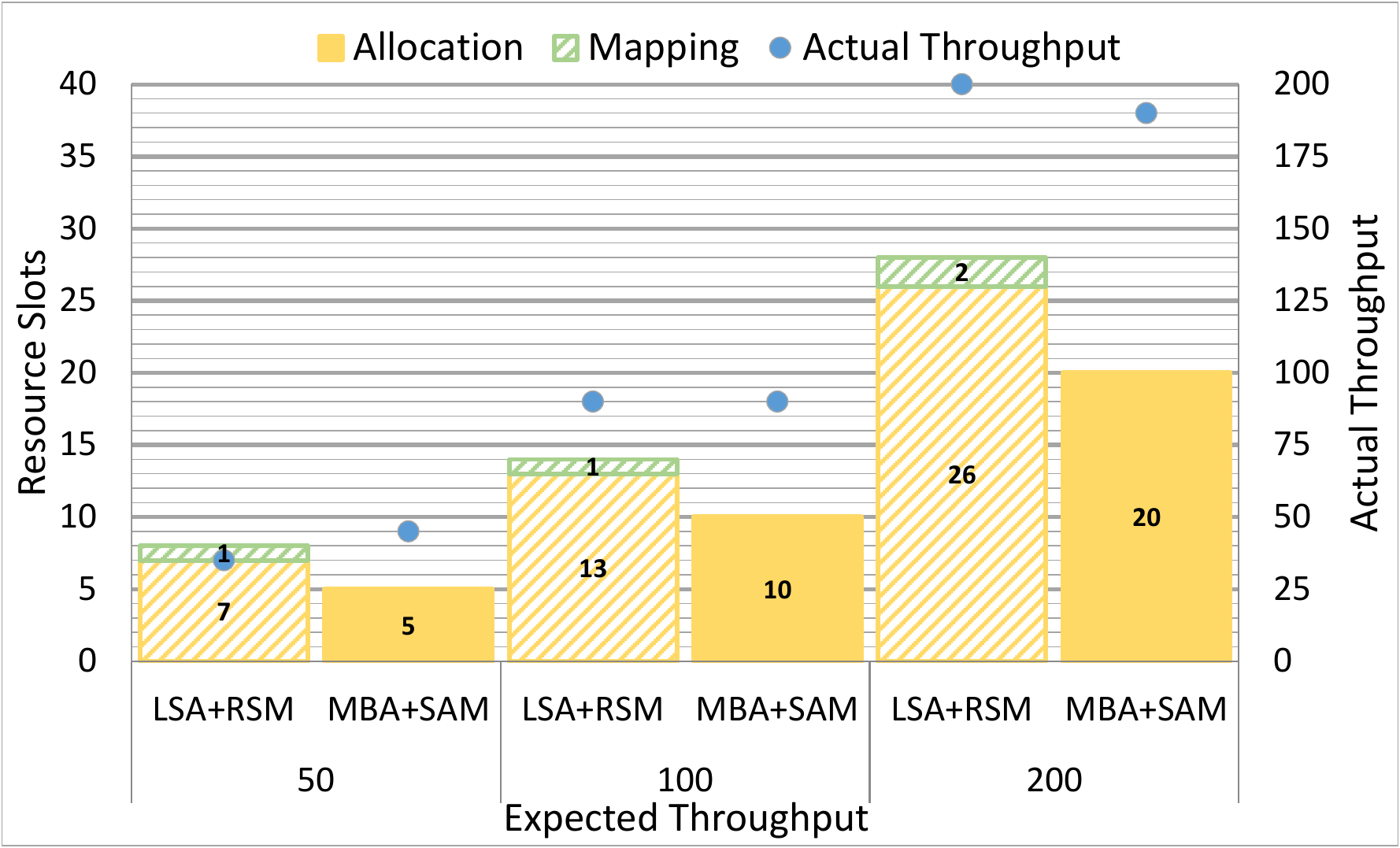}
		\label{fig:slots:diamond}
	}
	\subfloat[Star DAG]{
		\includegraphics[width=0.33\textwidth]{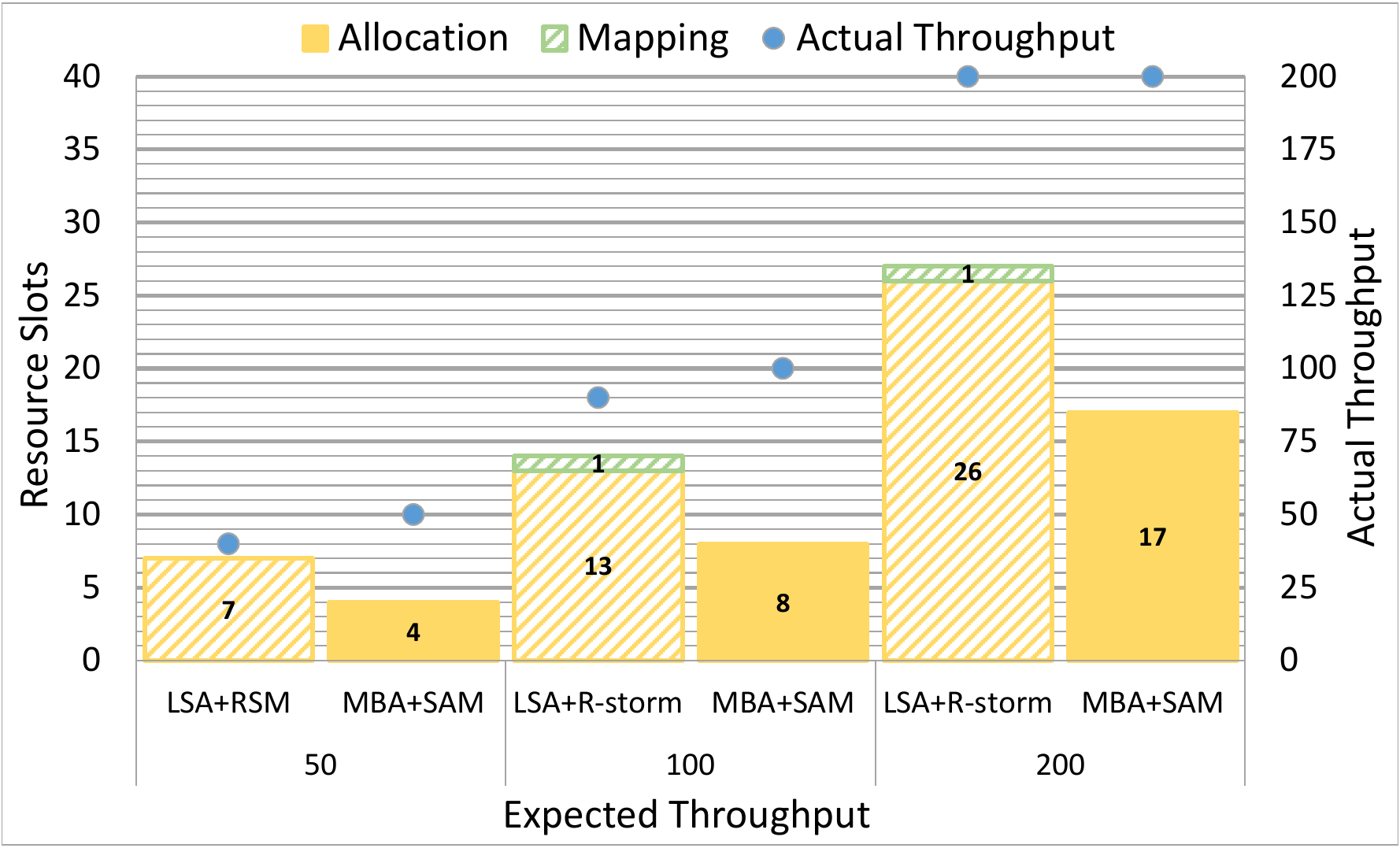}
		\label{fig:slots:star}
	}
	\caption{Micro DAGs: Required Slots on primary, Actual throughput on secondary Y-axis} 
	\label{fig:micro:slots}
\end{figure*}

The experiments are run for the micro-DAGs with input rates of $50, 100$ and $200~tuples/sec$. This allows us to study the changes in resource allocation and usage as the input rate changes. These specific rates are chosen to offer diversity while keeping within reasonable VM costs. For e.g., each run costs $\approx US\$1.00$, and many such runs are required during the course of these experiments. 


Fig.~\ref{fig:micro:slots} shows the number of resource slots allocated by the LSA and MBA algorithms (yellow bars, left Y axis) for the three different input rates to the three micro-DAGs. Further, it shows the additional slots beyond this allocation (green bars, left Y axis) that is required by the resource-aware RSM and SAM mapping algorithms to ensure that threads in a slot are not under-provisioned. The DAGs are run with the input rate they the schedule was planned for (i.e., $50, 100$ or $200~tuples/sec$), and if they were not stable, we incrementally reduced the rate by $5~tuples/sec$ until the execution is stable. The stable input rate, less than or equal to the planned schedule, is shown on the right Y axis (blue circle).



\emph{We can observe that LSA allocates more slots than MBA in all cases.} In fact, the resources allocated by LSA is nearly twice as that by MBA requiring, say, 7, 13 and 28 slots respectively for the Linear DAG for the rates of $50, 100$ and $200~tuples/sec$ compared to MBA that correspondingly allocates only 4, 7 and 15 slots. This is observed for the other two micro-DAGs as well.
The primary reason for this resource over-allocation in LSA is due to a linear extrapolation of resources with the number of threads. In fact, while MBA allocates $\approx 3\times$ more threads than LSA for the DAGs, 
the resource allocation for these threads by LSA is much higher. For e.g., LSA allocates $337\%$ CPU and $1196\%$ memory for its $50$ threads of the Blob Download task for the Linear DAG at $100~tuples/sec$ while MBA allocates only $315\%$ of CPU and $326\%$ of memory for its corresponding $170$ Blob Download threads. This alone translates to a difference between $\approx12$ slots allocated by LSA (based on memory\%) and $\approx3$ slots by MBA.


\emph{Despite the higher allocation by LSA, we see that RSM is often unable to complete the mapping without requiring additional slots.} This happens for 6 of the 9 combination of DAG and rates, for e.g., requiring $1$ more slot for the Diamond DAG with $50~tuples/sec$ (Fig.~\ref{fig:slots:diamond}, green bar) and $3$ more for the Linear DAG at $200~tuples/sec$ (Fig.~\ref{fig:slots:linear}). In contrast, our SAM mapping uses 1 additional slot, only in the case of Linear DAG at $50~tuples/sec$ (Fig.~\ref{fig:slots:linear}) and none other, despite being allocated fewer resources by MBA compared to LSA.

Both RSM and SAM are resource aware, which means they will fail if they are unable to pack threads on to allocated slots such that their expected resource usage by all threads on a slot is within the slot's capacity. RSM more often fails to find a valid bin-packing than SAM. This is because of its distribution of a task's threads across many slots, based on the distance function, which causes resource fragmentation. We see memory fragmentation to be more common, causing vacant holes in slots that are each inadequate to map a thread but are cumulatively are sufficient. 

For e.g., the Linear DAG at $200~tuples/sec$ is assigned $25$ slots by LSA. During mapping by RSM, the $100$ Blob Download threads dominate the memory and occupy $4 \times 23.9\%$ of memory in each of the $25$ slots, leaving only $8\%$ memory on each slot. This is inadequate to fit threads for other tasks like XML Parse which requires $22.98\%$ of memory for one of its thread, though over $25 \times 8\% = 200\%$ of fragmented memory is available across slots.

This happens much less frequently in SAM due to its preference for packing a slot at a time with a full bundle of threads in a single slot without any fragmentation. Fragmenting can only happen for the last partial thread bundle for each task. A full bundle also takes less resources according to the performance models than linear extrapolation from a single thread. For e.g., MBA packs $50$ threads of the same Blob Download task from above in a single slot. 





\emph{We see that in several cases, the DAGs are unable to support the rate that the schedule was planned for.} This reduction in rate is up to $30\%$ for LSA+RSM and up to $10\%$ for MBA+SAM. For e.g., in the Diamond DAG in Fig.~\ref{fig:slots:diamond}, the observed/expected rates in $tuples/sec$ for LSA+RSM is $35/50$ and $90/100$ while it is $90/100$ and $190/200$ for MBA+SAM. 
The reasons vary between the two approaches. 

In LSA+RSM, LSA allocates threads assuming a linear scaling of the rate with the threads but this holds only if each thread is running on an exclusive slot. As RSM packs multiple threads to a slot, the rate supported by these threads is often lower than estimated. For e.g., for the Diamond DAG in Fig.~\ref{fig:slots:diamond}, LSA allocates 18~threads for the Azure Table task for an input rate of $50~tuples/sec$ based on a single thread supporting $3~tuples/sec$. However, RSM distributes 2~threads each on 4~slots and the remaining 9~threads on 1~slot. As per our performance model for the Azure Table task, 2~threads on a slot support $5~tuples/sec$ and 9~threads support $10~tuples/sec$, to give a total of $4 \times 5 + 1 \times 10 = 30~tuples/sec$. This is close to the observed $35~tuples/sec$ supported by this DAG for LSA+RSM.

While SAM's model-based mapping of thread bundles mitigates this issue, it does suffer from an imbalance in message routing by Storm to slots. Storm's \emph{shuffle grouping} from an upstream task sends an equal number of tuples to each downstream thread. However, the individual threads may not have the same per-capita capacity to process that many tuples on its assigned slot, as seen from the performance models. 
This can cause a mismatch between tuples that arrive and those that can be processed on slots. 

For e.g., the Diamond DAG at $100~tuples/sec$ (Fig.~\ref{fig:slots:diamond}), MBA allocates 160~threads for the Azure Table task and SAM maps two full bundles of 60~threads each to 2 slots, and the remaining 40~threads on 1 partial slot. As per the model, SAM expects the threads in a full slot to support $40~tuples/sec$ and the partial slot to support $20~tuples/sec$. However, due to Storm's shuffle grouping, the full slots receive $37~tuples/sec$ while the partial slot receives $26~tuples/sec$. This problem does not affect RSM since it distributes threads across many slots achieving a degree of balance across slots. As future work, it is worth considering a slot-aware \emph{routing} in Storm as well~\footnote{\url{https://issues.apache.org/jira/browse/STORM-162}}.

Also Figs.~\ref{fig:micro:slots} show that as expected, \emph{the resource requirements increase proportionally with the input rate for both LSA+RSM and MBA+SAM}. Some minor variations exist due to rounding up of partial slots to whole, or marginal differences in fragmentation. For e.g., LSA+RSM assigns the Star DAG $\lceil 6.23 \rceil=7~slots$ and $\lceil 12.47 \rceil=13~slots$ for $50$ and $100~tuples/sec$ rates, respectively.




\emph{Lastly, we also observe that all three micro-DAGs acquire about the same number of slots, for a given rate, using LSA+RSM,} e.g., using about 7 slots for $50~tuples/sec$ for all three micro-DAGs. 
These three DAGs have the same 5 tasks though their composition pattern is different. However, for LSA, the memory\% of the Blob Download task threads dominates the resource requirements for each DAG, and the input rate to this task is the same as the DAG input rate in all cases. As a result, for a given rate, the threads and resource needs for this task is the same for all three DAGs at 25 threads taking $598\%$ memory for $50~tuples/sec$, while the memory\% for the entire Linear DAG is marginally higher at $623\%$ for this rate and its total CPU\% need is only  $242\%$. Hence, the resource needs for all other tasks, which are more CPU heavy, falls within the available 7~slots that is dominated by this Blob task, i.e., $\sum_{\text{All task threads}} CPU\% < \sum_{\text{Blob task threads}} Memory\%$. 

In case of MBA+SAM, there is diversity both in CPU and memory utilization, and the number of threads for each task for the different DAGs. So the resource requirements are not overwhelmed by a single task. For e.g., the same Blob task at $50~tuples/sec$ requires only $128\%$ memory according to MBA while the CPU\% required for the entire Linear DAG is $323\%$, which becomes the deciding factor for slot allocation.

\subsubsection{Application DAG }

\begin{figure*}[t]
	\centering
        \subfloat[Finance DAG]{
		\includegraphics[width=0.33\textwidth]{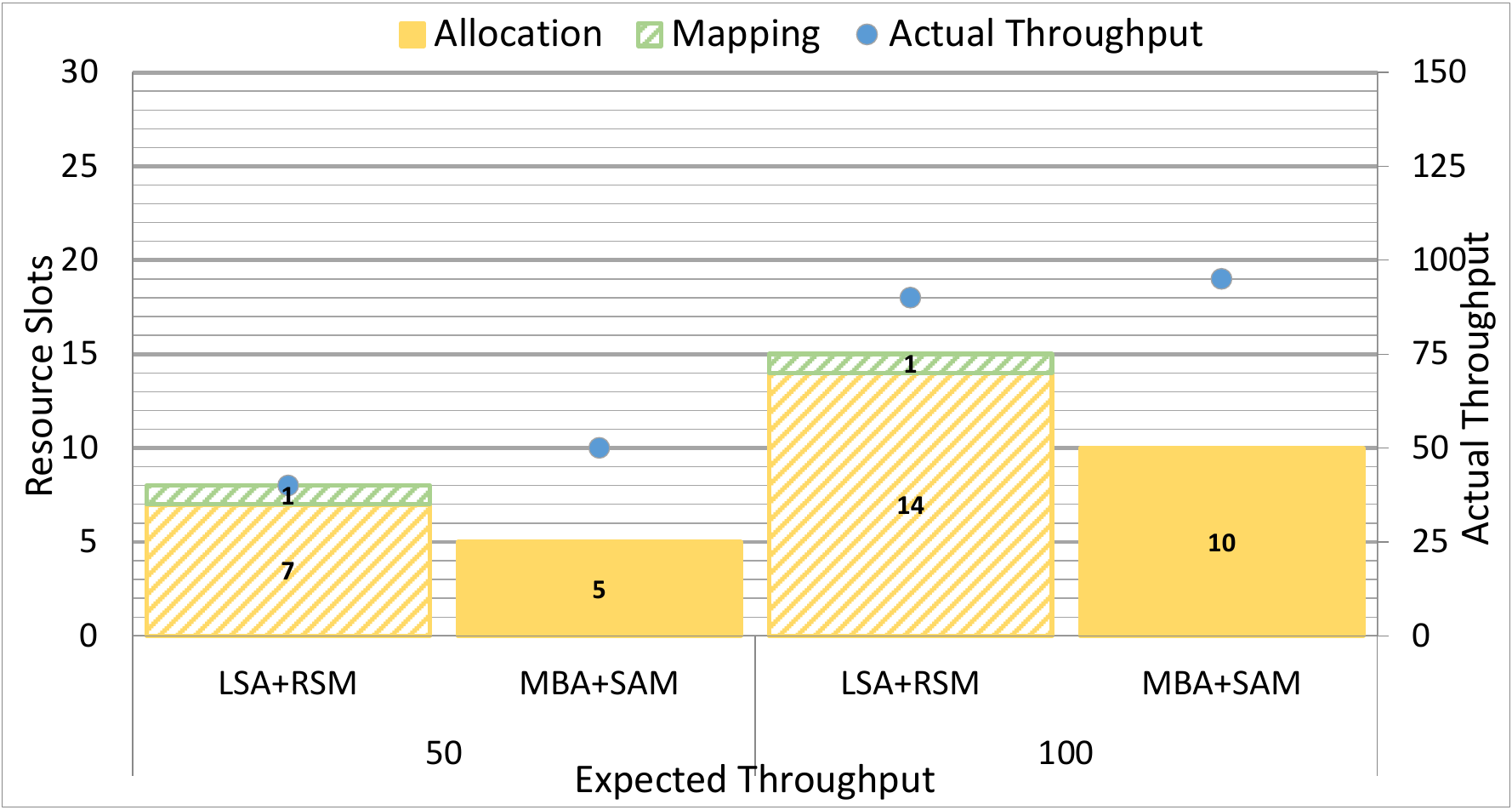}
		\label{fig:slots:finance}
	}	
	\subfloat[Traffic DAG]{
		\includegraphics[width=0.33\textwidth]{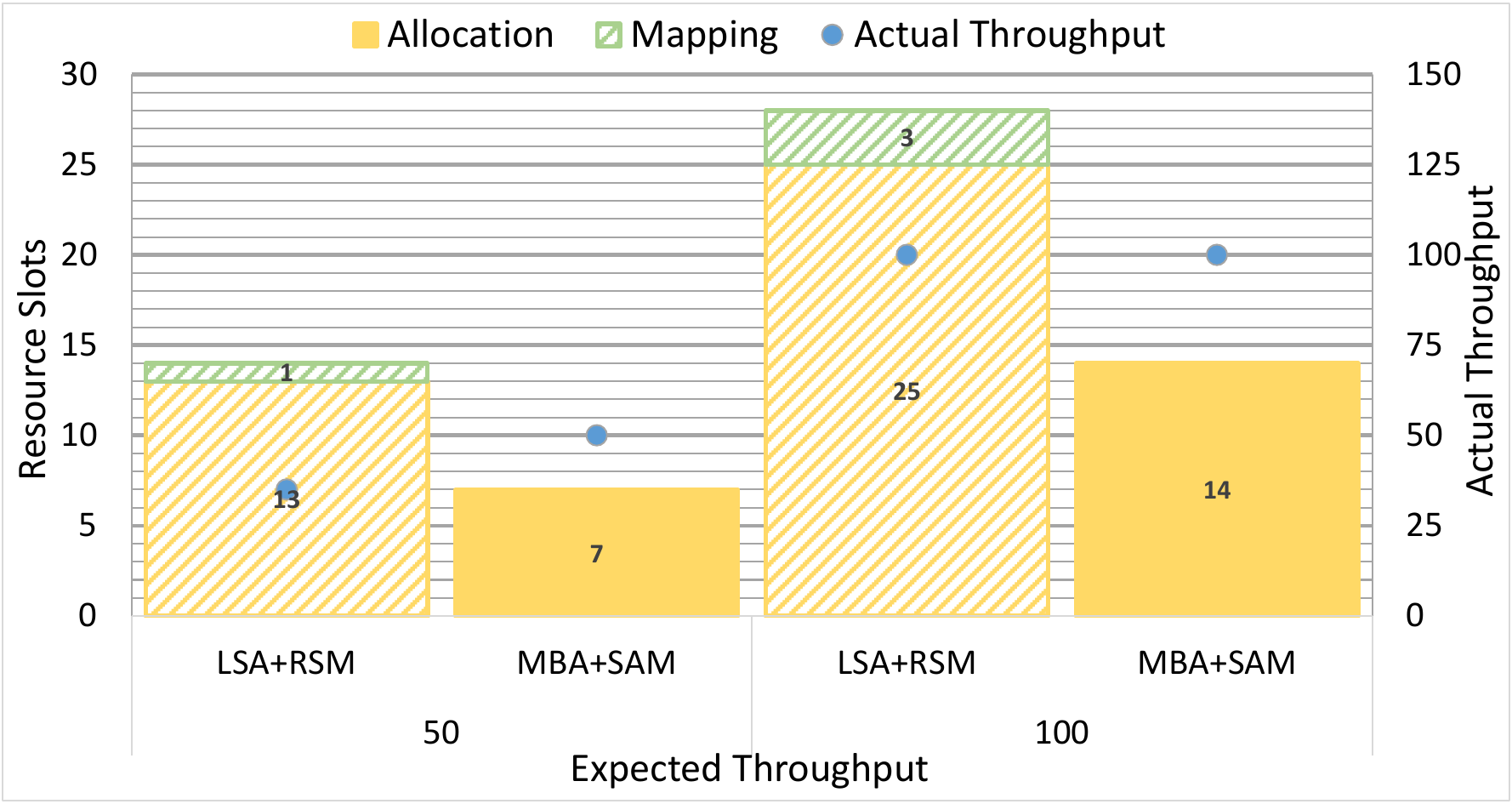}
		\label{fig:slots:traffic}
		
	}
	\subfloat[Smart Grid DAG]{
		\includegraphics[width=0.33\textwidth]{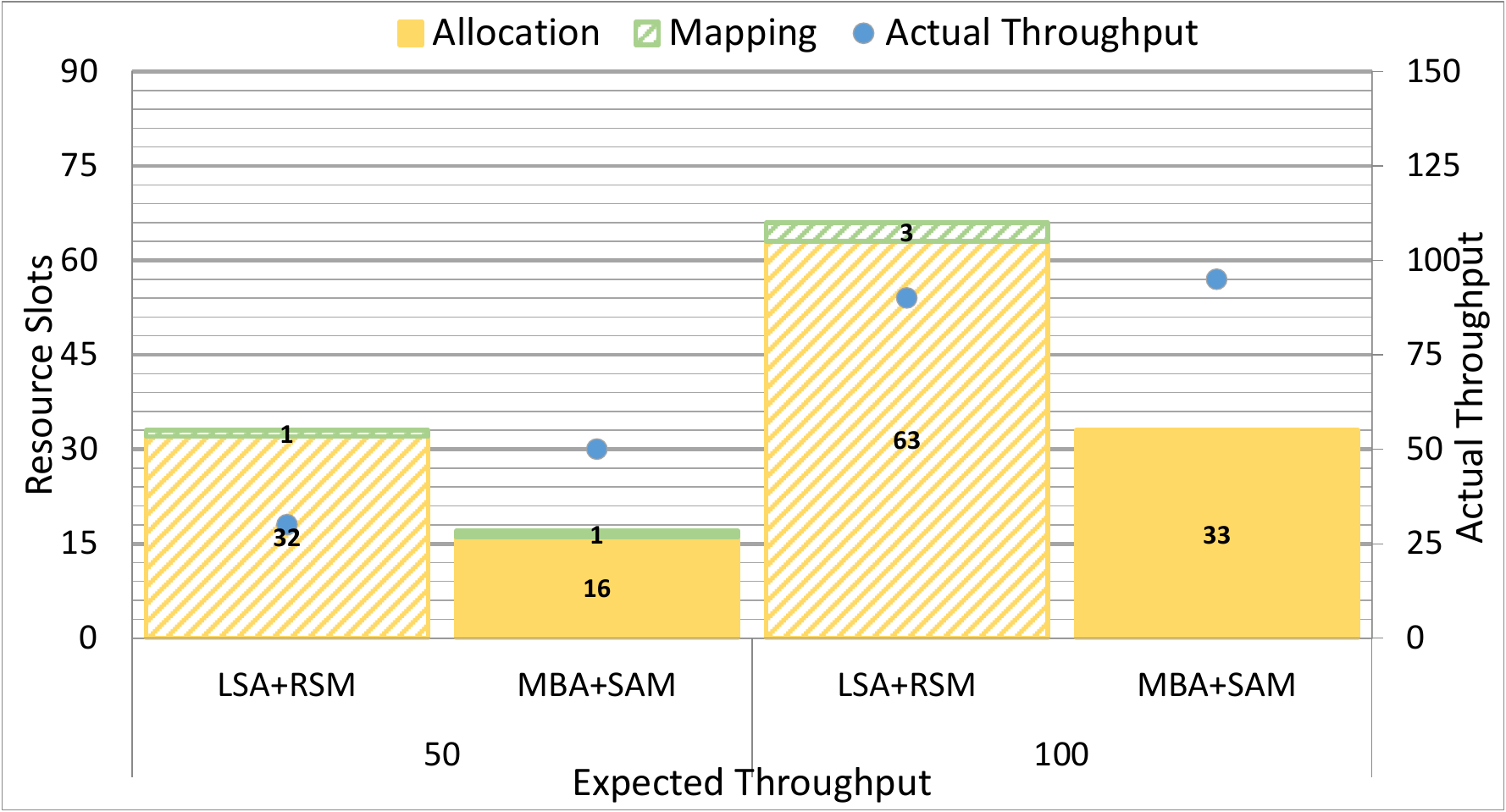}
		\label{fig:slots:smartgrid}
	}
	
	\caption{Application DAGs: Required Slots on primary, Actual throughput on secondary Y-axis }
	\label{fig:app:slots}
\end{figure*}

We perform similar experiments to analyze the resource benefits for more complex applications DAGs, limited to 
input rates of $50$ and $100~tuples/sec$ that require as much as 65~slots for LSA+RSM costing $\approx US\$2$ per run. Figs.~\ref{fig:app:slots} plot the results. 
Several of our observations here mimic the behavior of the scheduling approaches for the micro-DAGs, but at a larger scale. We also see more diversity across the DAGs, with Finance taking $3-5\times$ fewer resources than Grid for the same data rates.



As before for micro-DAGs, we see that MBA+SAM consistently uses $33-50\%$ fewer slots than LSA+RSM for all the application DAGs and rates. This is seen both for the allocation and in the incremental slots acquired by the mapping during packing. In fact, RSM acquires additional slots for all application DAGs allocated by LSA while MBA needs this only for Grid DAG at $50~tuples/sec$. 
The resource benefit for MBA+SAM relative to LSA+RSM is the least for the Finance DAG in Fig.~\ref{fig:slots:finance}, using $3$ fewer slots for $50~tuples/sec$ rate and $5$ fewer for $100~tuples/sec$ rate. This is because total CPU\% tends to dominate for MBA+SAM, and this is higher for Finance compared to the other two due to a higher input rate that arrives at the compute-intensive Pi task. 
On the other hand, LSA+RSM consumes over twice the slots for Traffic and Grid DAGs due having one less Pi task and one more Blob task, which is memory intensive. This is due to random mapping of our candidate tasks to application tasks. As we saw earlier for the micro-DAG, this memory intensive task tends to be sub-optimal when bin-packing by RSM, and that causes a the resource needs to grow for the Traffic and Grid DAGs.


That said, while the DAGs are complex for these application workloads, the fraction of additional slots required by mapping relative to allocation does not grow much higher. 
In fact, the additional slots required by RSM is small in absolute numbers, at $1-3$ slots, as the bin-packing efficiency improves with more slots and threads. 
This shows that the punitive impact of RSM's additional slot requirements is mitigated for larger application DAGs.


As for the micro-DAGs, several of the application DAGs are also unable to support the planned input rate. The impact worsens for LSA+RSM with its stable observed rate up to $40\%$ below expected, while this impact is much smaller for MBA+SAM with only up to $5\%$ lower rate observed despite the complex DAG structure. 

The reasoning for SAM is the same as for the micro-DAGs, where the shuffle grouping unformly distributes the output tuple rate across all threads. 
For RSM, an additional factor comes into play for these larger DAGs. In practice, this algorithm allows threads in a slot to access all cores in a VM while restraining their memory use to only that slot's limit.  
This means threads in a single slot of a \texttt{D3} VM can consume up to $400\%$ CPU, as long as their memory\% is $\le 100\%$. This causes more CPU bound threads like Pi and XMLParse to be mapped to a single slot, consuming $\approx300\%$ of a VM's CPU in the Grid DAG for $50~tuples/sec$. 
However, each slot has just a single worker thread responsible for tuple buffering and routing between threads and across workers. Having many CPU intensive task threads on the same slot stresses 
this worker thread and cause a slowdown, as seen for the Grid DAG which has an observe/expected tuple rates of $30/50$~\footnote{\url{http://stackoverflow.com/questions/20371073/how-to-tune-the-parallelism-hint-in-storm}}~\footnote{\url{http://mail-archives.apache.org/mod_mbox/storm-user/201606.mbox/browser}}. This consistently happens across all VMs where threads with high CPU\% are over-allocated to a single slot.
%
In MBA, the mapping of full bundles to a slot rather than over-allocating CPU\% means that we have a better estimate of the collective behaviour of threads on each worker slot and these side-effects are avoided.

As before, we see that the resource requirements increase proportionally as the rate doubles from $50~tuples/sec$ to $100~tuples/sec$ in most cases.
%
%
%
However, unlike the micro-DAGs where all the dataflows for a given input rate consumed about the same number of slots using LSA+RSM, this is not the case for the application DAGs. Here, the number of tasks of each type vary and their complex compositions cause much higher diversity in input rates to these tasks. For e.g., the same Table task in Traffic, Finance and Smart Grid DAGs in Figs~\ref{fig:app:dag} have input rates of $100, 200$ and $300~tuples/sec$. 
In fact, this complexity means that the resource usage does not just proportionally increase with the number of tasks either. 
%
%
This argues the need for non-trivial resource- and DAG-aware scheduling strategies for streaming applications, such as RSM, MBA and SAM.

\subsection{Accuracy of Models}
\label{sec:results:accy}

In the previous experiments, we showed that our MBA+SAM scheduling approach offered \emph{lower resource costs} than the LSA+RSM scheduler while meeting the planned input rate more often. In these experiments, we show that our model based scheduling approach offers \emph{predictable resource utilization}, and consequently \emph{reliable performance} that can be generalized to other DAGs and data rates. Further, we also show that it is possible to independently use our performance-model technique to accurately predict the resource usage and supported rates for other scheduling algorithms as well.

Rather than determine the allocation for a given application and rate, we instead design these experiments with a fixed number of VMs and slots -- five \texttt{D3} VMs with $20$ total slots, for the three micro-DAGs. 
We then examine the highest input rate that our performance model estimates will be supported by the given schedule, and what is actually supported on enacting the schedules.

The \emph{planned input rate} is the peak rate for which the DAG's resource requirements can be fulfilled with the fixed number of five D3 VMs, according to the allocation+mapping algorithm pair that we consider. 
%
%
For this, we independently run the allocation \emph{and} mapping algorithm plans outside Storm, adding incremental input rates of $10~tuples/sec$ until the resources required is just within or equal to $20$ slots according to the respective algorithm pairs.
Subsequently, for the threads and their slot mappings determined by the scheduling algorithm, we use our performance models  
to find the \emph{predicted rate} supported by that DAG. We also use our model to \emph{predict the CPU\% and memory\%} for the slots as well, and report the cumulative value for each of the $5$ VMs. 
The actual input rate for the DAG is obtained empirically by increasing the rate in steps of $10~tuples/sec$ as long as the execution remains stable. The actual CPU and memory utilization corresponding to the peak rate supported is reported for each VM as well. 
Besides comparing the predicted and actual input rates, we also compare the predicted and actual VM resource usage in the analysis since there is a causal effect of the latter on the former.

We further show that our model based allocation algorithm can be used independently with other mapping algorithms, besides SAM. To this end, we evaluate and compare the baseline combination of LSA allocation with DSM and RSM mappings available in Storm (LSA+DSM and LSA+RSM), against our MBA allocation with DSM, RSM and SAM mapping algorithms (MBA+DSM, MBA+RSM and MBA+SAM).


\subsubsection{Prediction and Comparison of Input Rates}
\label{sec:an:rate}
\begin{figure*}[t]
	\centering
	\subfloat[Linear DAG]{
		\includegraphics[width=0.33\textwidth]{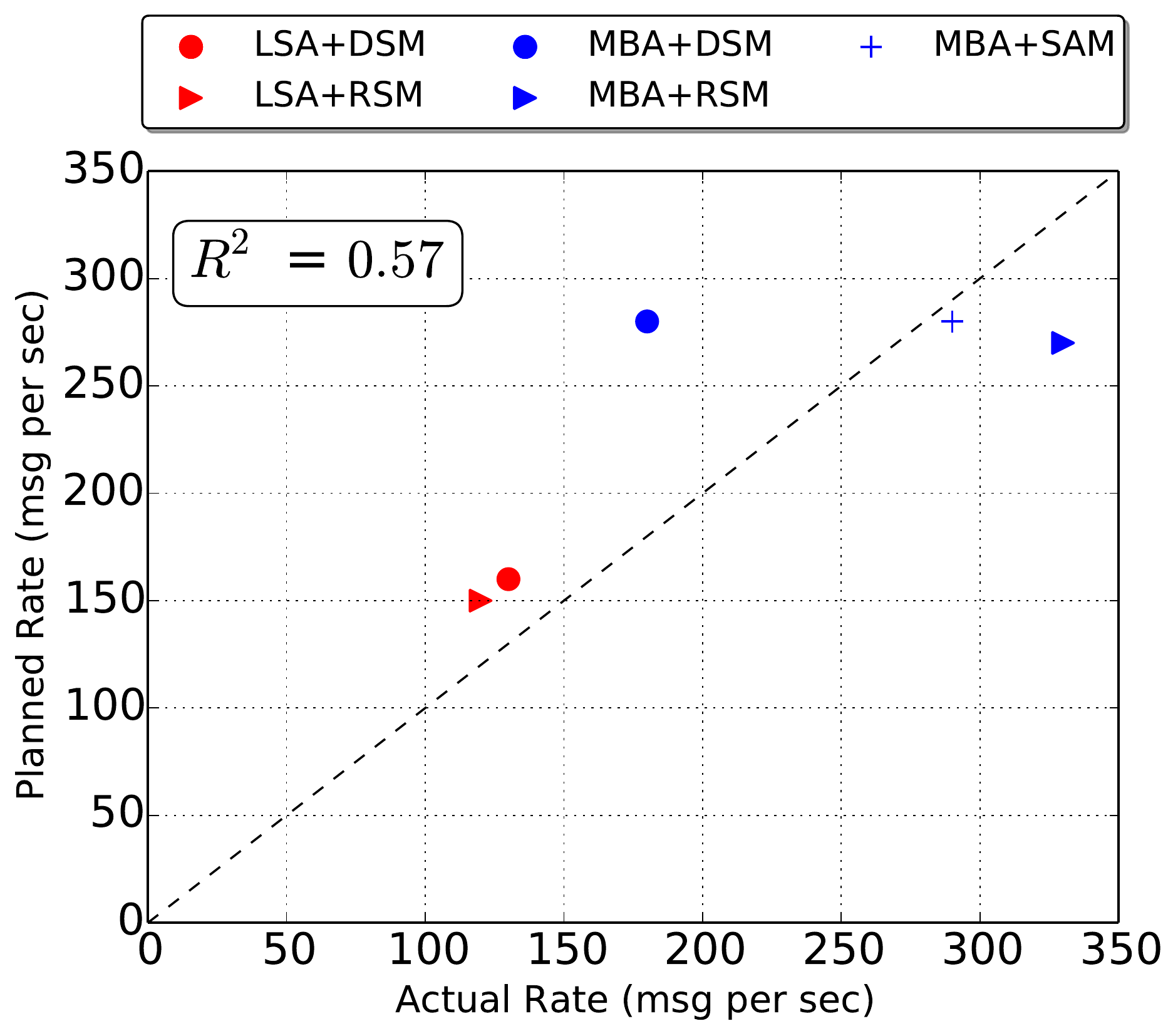}
	}
	\subfloat[Diamond DAG]{
		\includegraphics[width=0.33\textwidth]{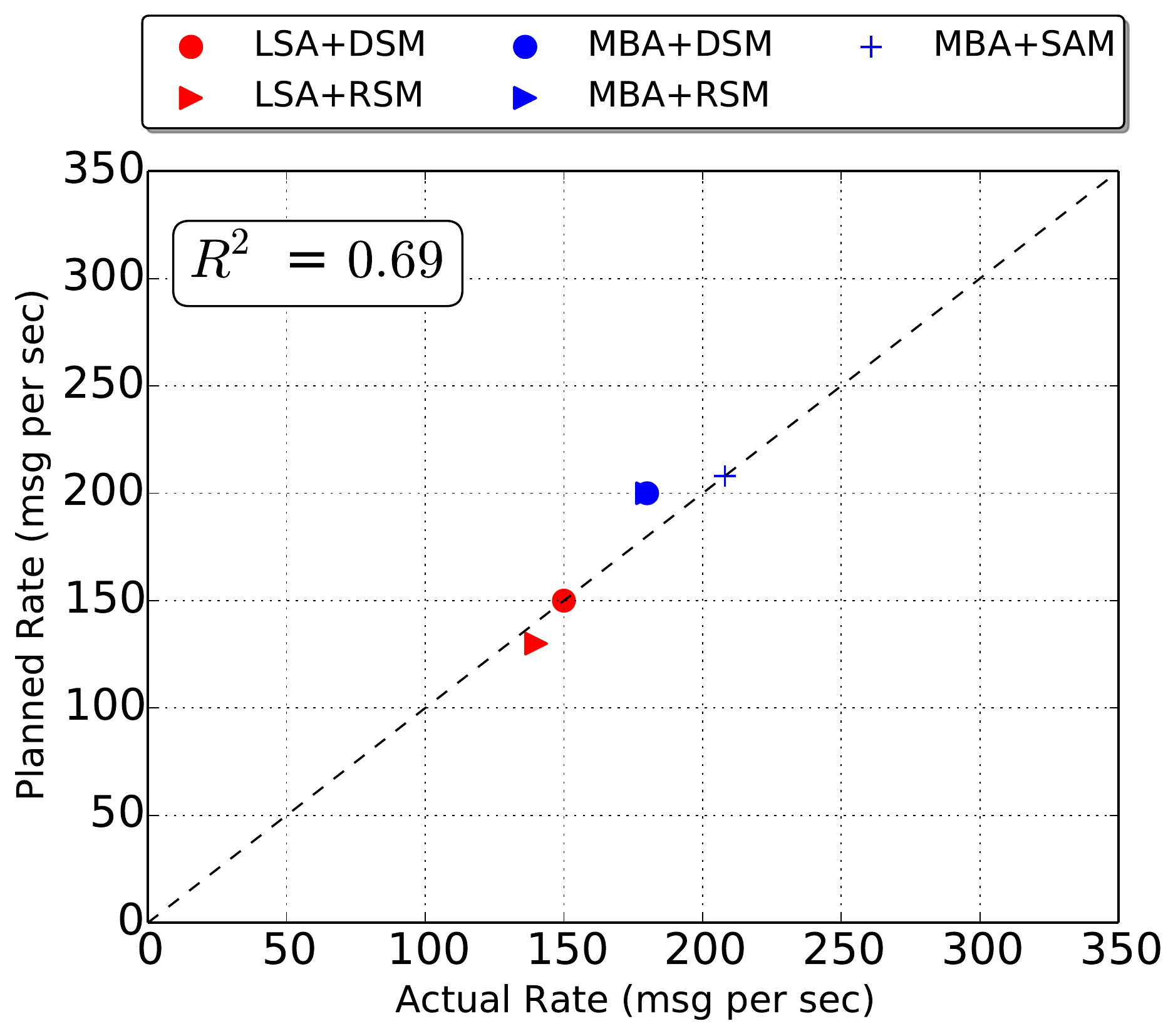}
	}
	\subfloat[Star DAG]{
		\includegraphics[width=0.33\textwidth]{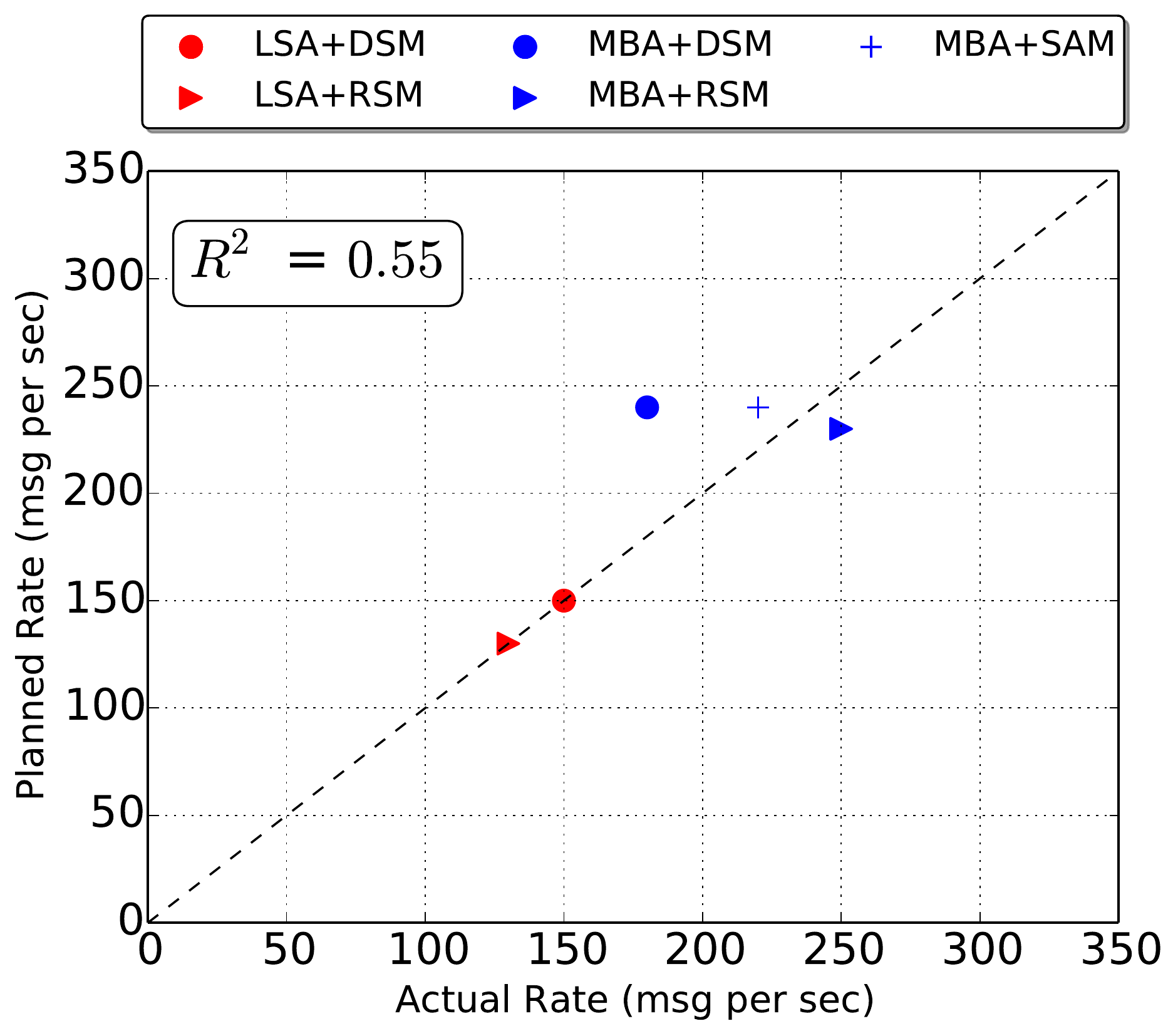}
	}
	\caption{Scatter plot of \emph{Planned} and \emph{Actual} input rates supported for the Micro-DAGs on 5 VMs using the scheduling strategy pairs 
      }
	\label{fig:plot:rate:planned}
\end{figure*}

\begin{figure*}[t]
	\centering
	\subfloat[Linear DAG]{
	\includegraphics[width=0.33\textwidth]{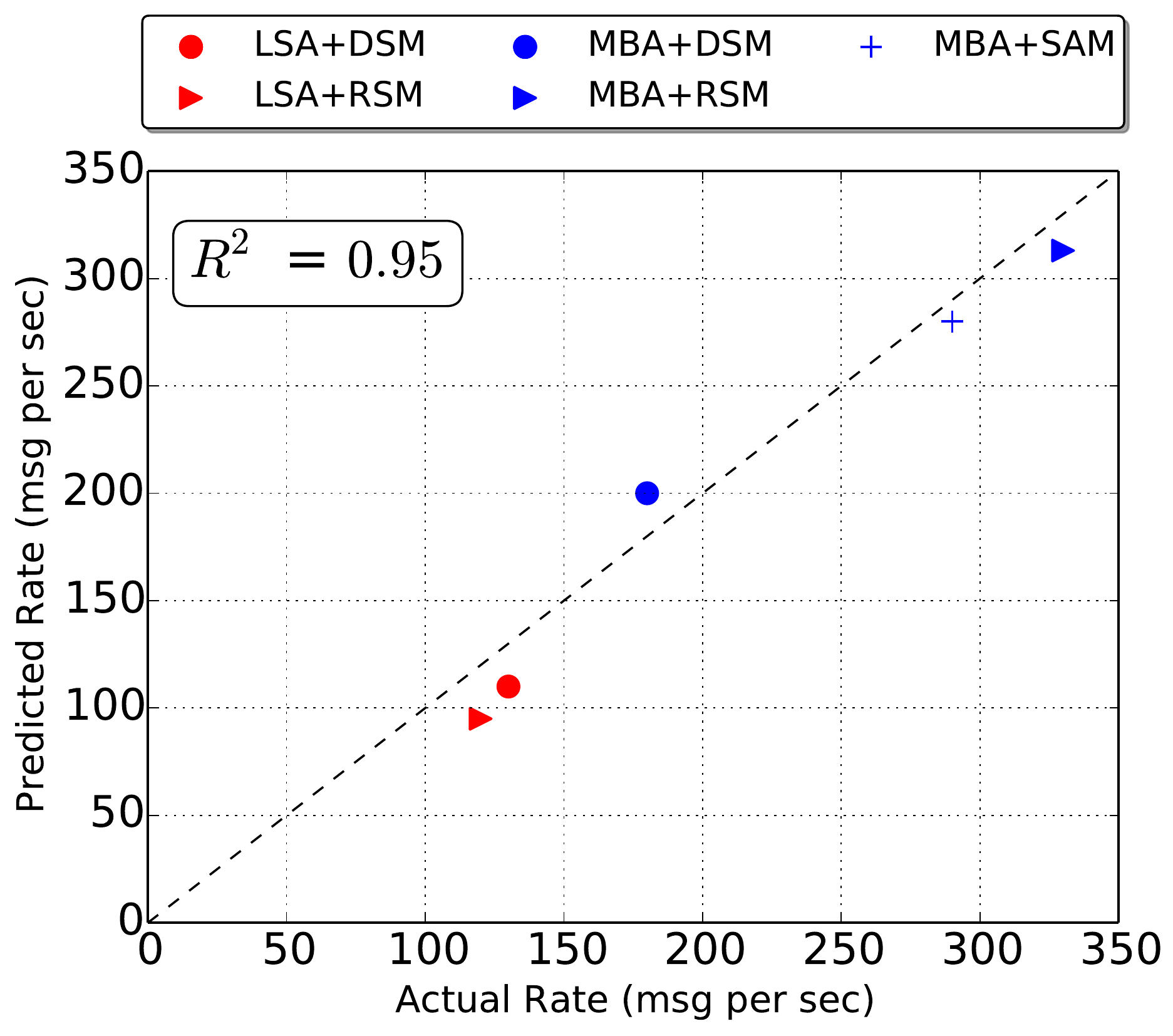}
		\label{fig:rate:linear}
	}
	\subfloat[Diamond DAG]{
		\includegraphics[width=0.33\textwidth]{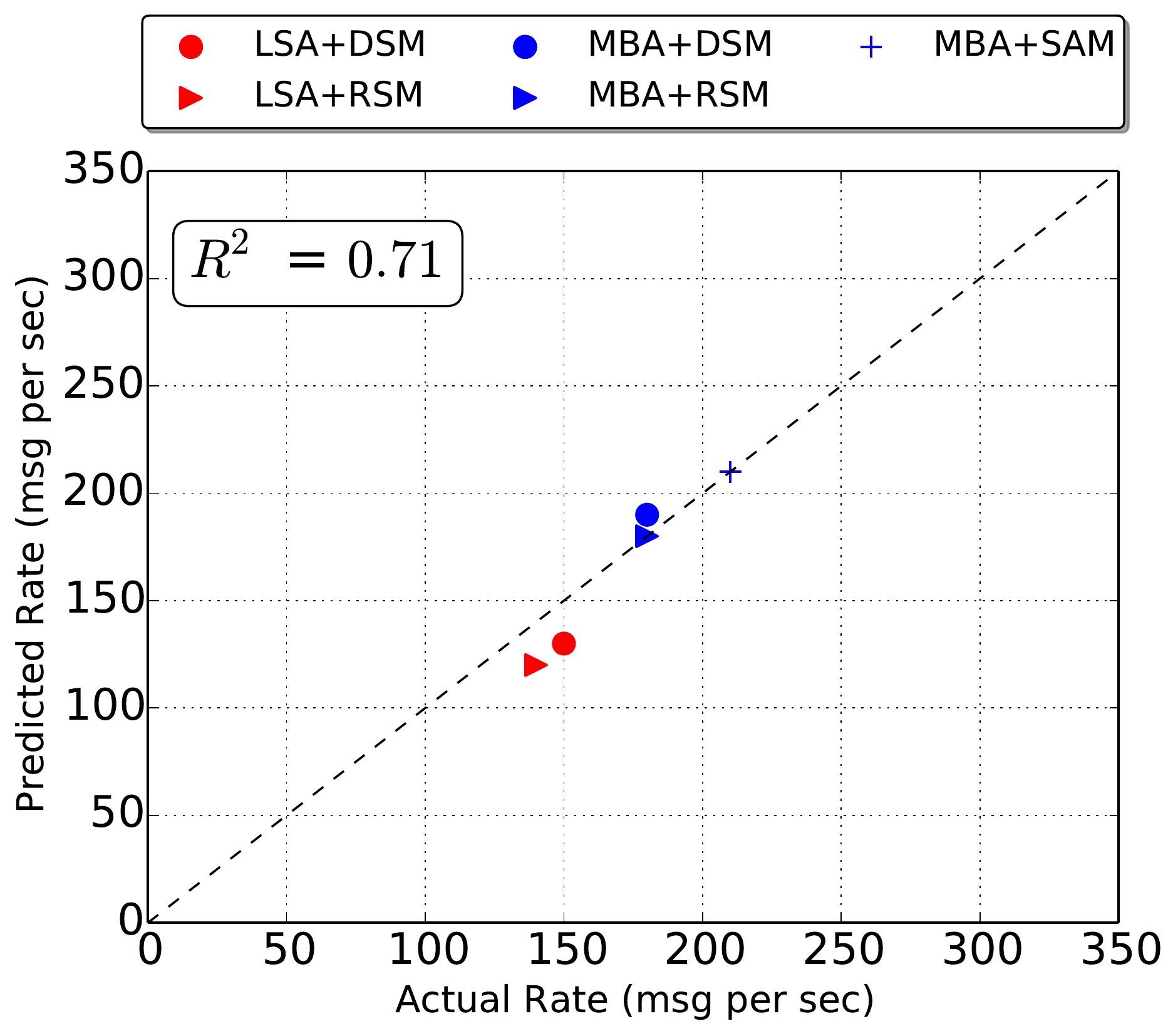}
		\label{fig:rate:diamond}		
	}
	\subfloat[Star DAG]{
		\includegraphics[width=0.33\textwidth]{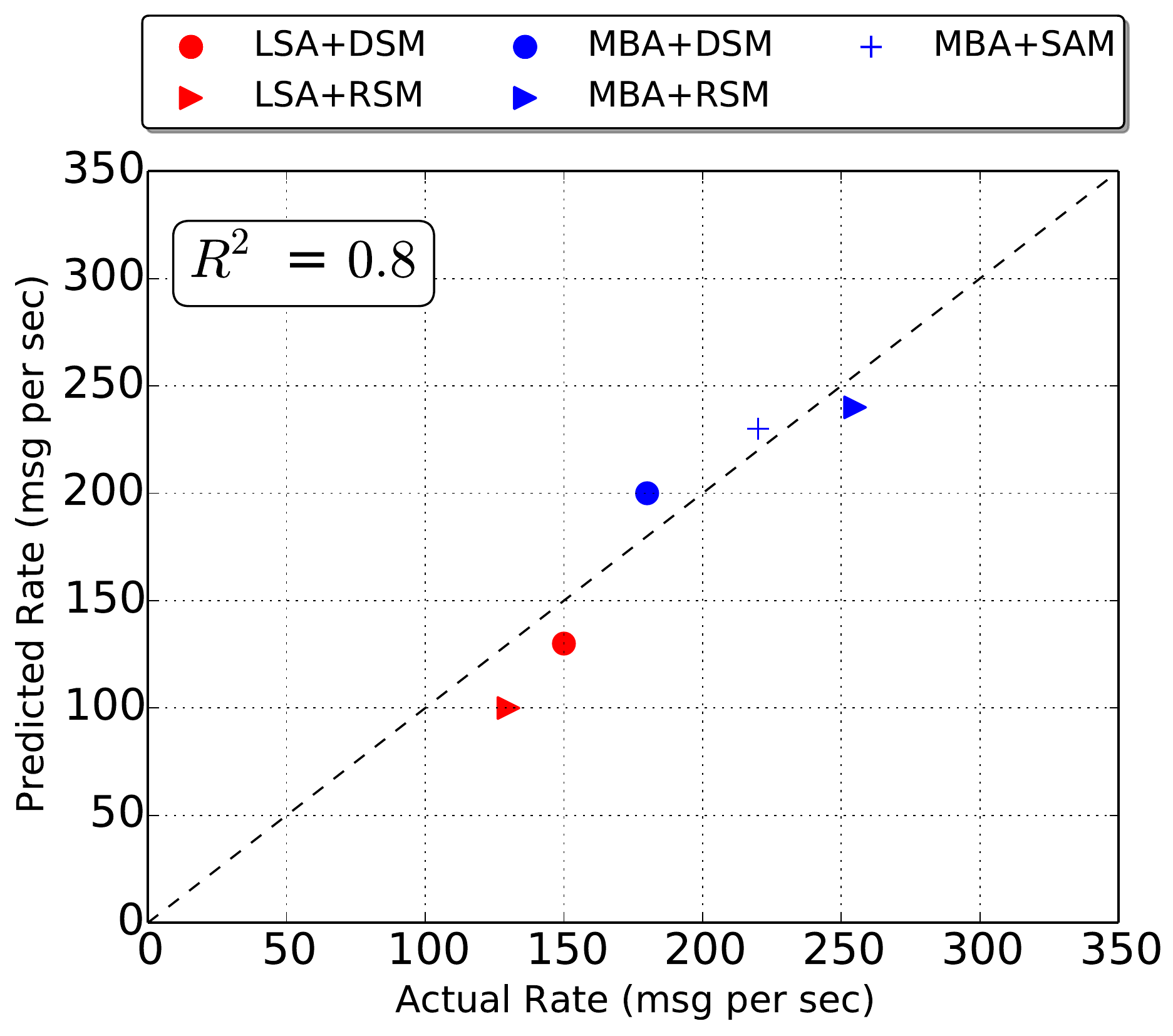}
		\label{fig:rate:star}		
	}
	\caption{Scatter plot of \emph{Predicted} and \emph{Actual} input rates supported for the Micro-DAGs on 5 VMs using the scheduling strategy pairs
	}
	\label{fig:plot:rate}
\end{figure*}

%
Fig.~\ref{fig:plot:rate:planned} shows a scatter plot comparing the \emph{Actual rate} (X axis) and the \emph{Planned rate} by the scheduling algorithm (Y axis) for the Linear, Diamond and Star micro-DAGs, while Fig.~\ref{fig:plot:rate} does the same for the \emph{Actual rate} (X axis) and our \emph{Model Predicted rate} (Y axis). We see that our performance model is able to \emph{accurately predict the input rate for these DAGs with a high correlation coefficient of $R^2$ ranging from $0.71 - 0.95$}. This is actually significantly better than the Planned rate by the schedulers for the three DAGs whose $R^2$ values fall between $0.55-0.69$. Thus, our performance model is able to accurately predict the input rates for the schedules from all 5 algorithm pairs, better than even the scheduling algorithms themselves.

%

%
There are also algorithm-specific behavior of the prediction models, which we analyze. The input rate predictions are more accurate for MBA+SAM (Fig.~\ref{fig:plot:rate}, blue `+'), falling close to the 1:1 line in all cases, since it uses the model both for allocation and for mapping. However, it is not $100\%$ accurate due to Storm's shuffle grouping that routes a different rate to downstream threads than expected.  

%
We also see our model underestimate the supported rate for LSA in Fig.~\ref{fig:plot:rate} by a small fraction. This happens due to the granularity of the model. With LSA, several slots have 3 table threads mapped to them. As we do not have exact performance models with 3 threads, we interpolate between the available thread values which estimates the rate supported at $6~tuples/sec$ while the observed rate is closer to $9~tuples/sec$. In such cases, the predictions can be made more accurate if the performance modeling is done at a finer granularity of thread counts ($\Delta_\tau$) in Algo.~\ref{alg:bm}.
%

%
The algorithms also show distinctions in the actual rates that they support the same quanta of resources for a DAG. When using MBA with DSM, the actual rate is often much smaller than the planned rate and, to a lesser extent, than the predicted rate as well. For e.g., the Linear DAG's planned rate is $280~tuples/sec$, predicted is $200~tuples/sec$ and actual is $180~tuple/sec$. Since DSM does a round-robin mapping of threads without regard to the model, it is unable to leverage the full potential of the allocation. 
In the case of the Linear DAG, the allocation estimates the planned performance for, say, the Blob Download task with $470~threads$ based on it being distributed in bundles of $50$ threads each on $9$ slots but DSM assigns them uniformly with $\approx{23~threads~per~slot}$. 
Hence, using MBA with DSM is not advisable, compared to RSM or SAM mapping approaches.

%
However, compared to LSA, MBA offers a higher predicted and actual input rate irrespective of the mapping, offering improvements of $20-175\%$. We observe from the plots that the cluster of points for MBA (in blue) is consistently at a higher rate than the LSA cluster (in red) despite both being allocated the same fixed number of resources. As discussed before, LSA allocates fewer data-parallel threads than MBA due to its linear-scaling assumption, and they are unable to fully exploit the available resources. This is consistent with the lower CPU\% and memory\% for LSA observed in Figs.~\ref{fig:plot:cpu} and \ref{fig:plot:mem}. For e.g., the Azure Table task in the Linear DAG is assigned only $54$ threads by LSA, with a planned rate of $160~tuples/sec$ whereas MBA assigns it $420$ threads with a planned rate of $280~tuples/sec$.

%
Interestingly, when using MBA, RSM is able to offer a higher actual input rate compared to SAM in two of the three DAGs, Linear and Star (Fig.~\ref{fig:plot:rate}), even as its planned rate is lower than SAM. For e.g., we see that RSM's distance measure is able find the sweet spot for distributing the $470~threads$ of Blob Download for the Linear DAG across 15 slots with $25-30$ threads each and 3 slots with under 10 threads each, to offer a predicted rate of $315~tuples/sec$ and actual rate of $330~tuples/sec$. SAM on the other hand favors predictable performance within exclusive slots and bundles $50~threads$ each on 9 slots and the rest in a partial slot to give a predicted and actual rate close to $280~tuples/sec$. This highlights the trade-off that SAM makes in favor of a predictable model-driven performance, while sacrificing some of performance benefits relative to RSM.

\subsubsection{Prediction and Comparison of CPU and Memory Utilization}

\begin{figure*}[t]
	\centering
	\subfloat[Linear DAG]{
		\includegraphics[width=0.33\textwidth]{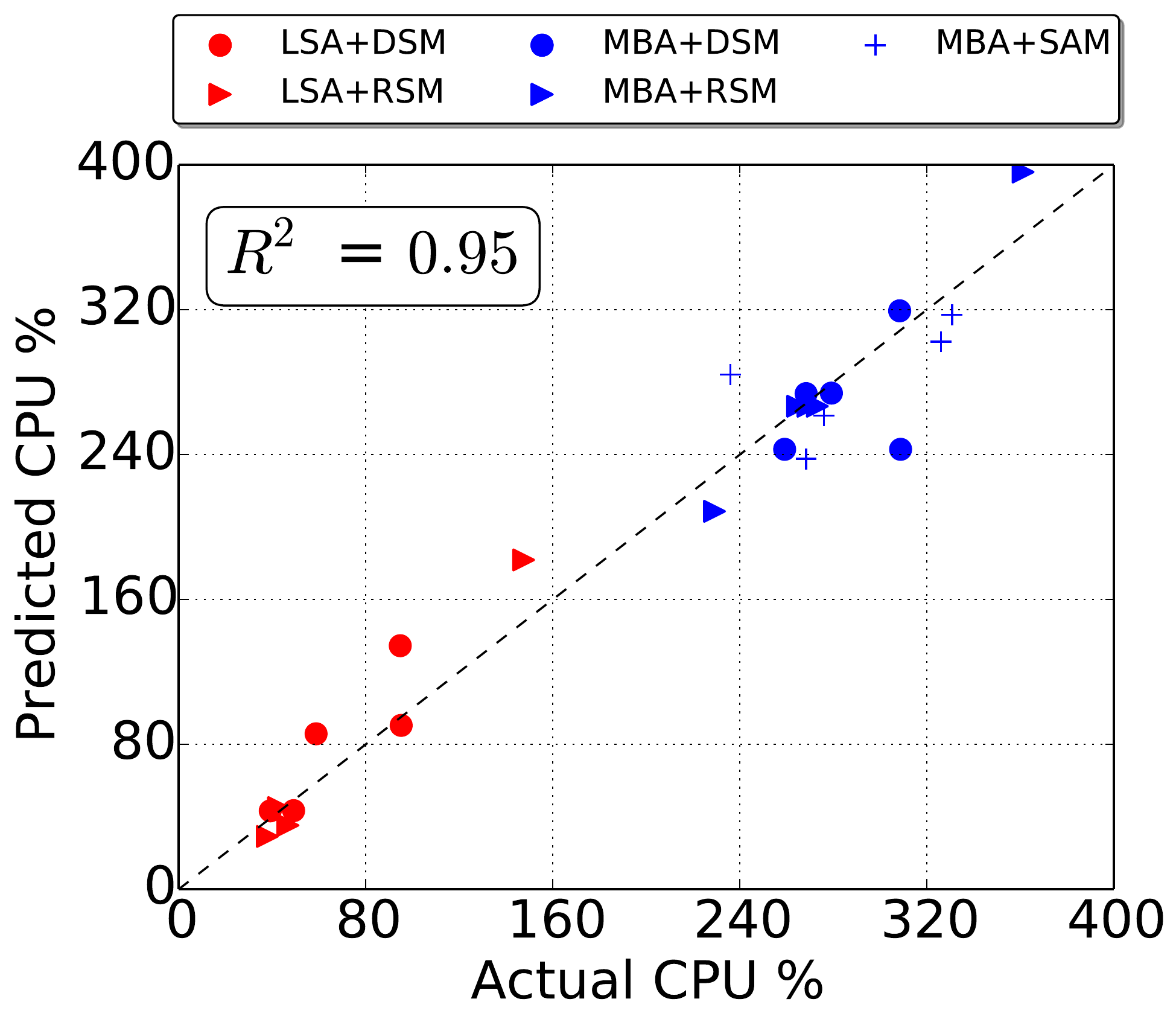}
		\label{fig:cpu:linear}
	}
	\centering
	\subfloat[Diamond DAG]{
		\includegraphics[width=0.33\textwidth]{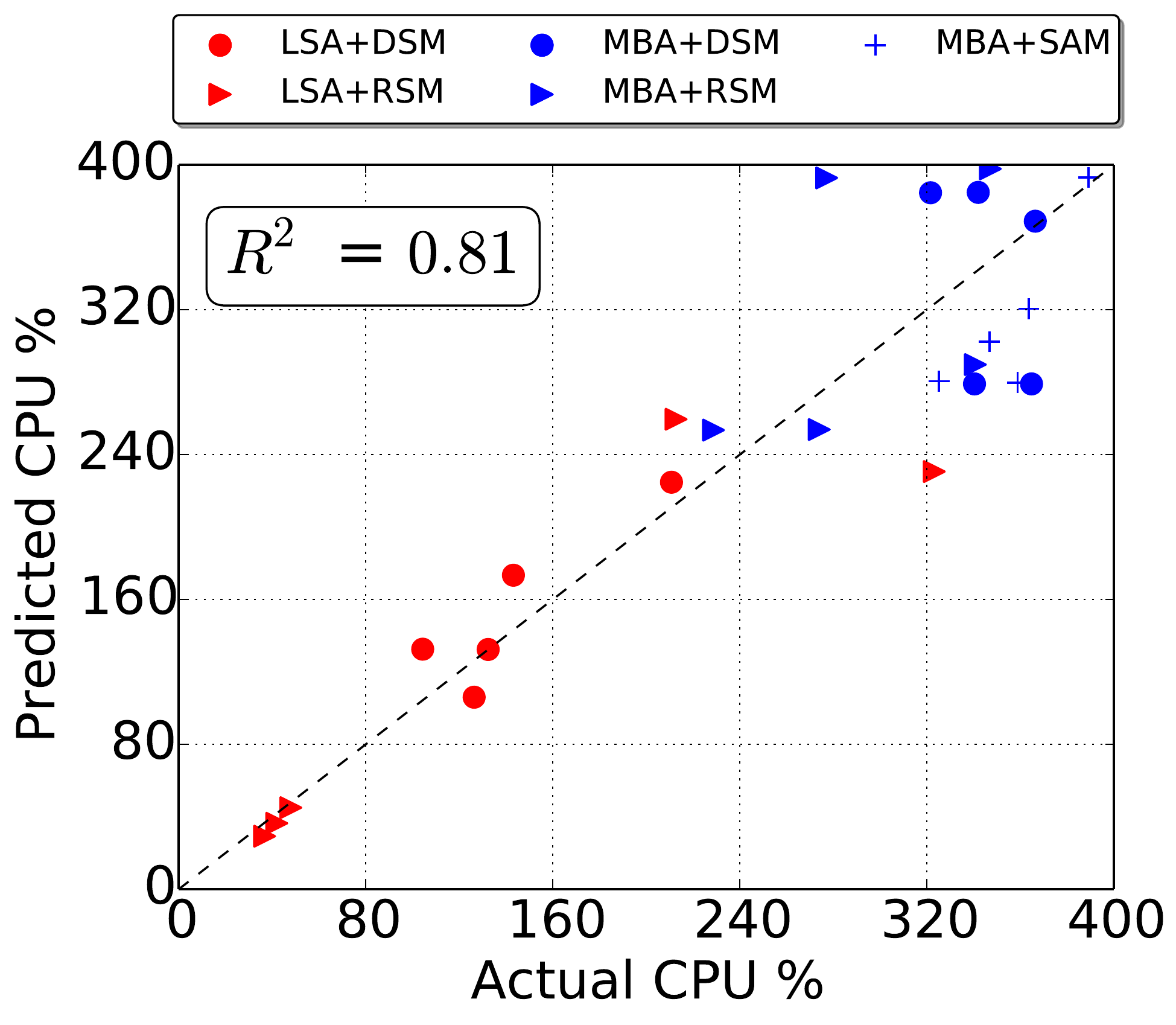}
		\label{fig:cpu:diamond}		
	}
	\centering
	\subfloat[Star DAG]{
		\includegraphics[width=0.33\textwidth]{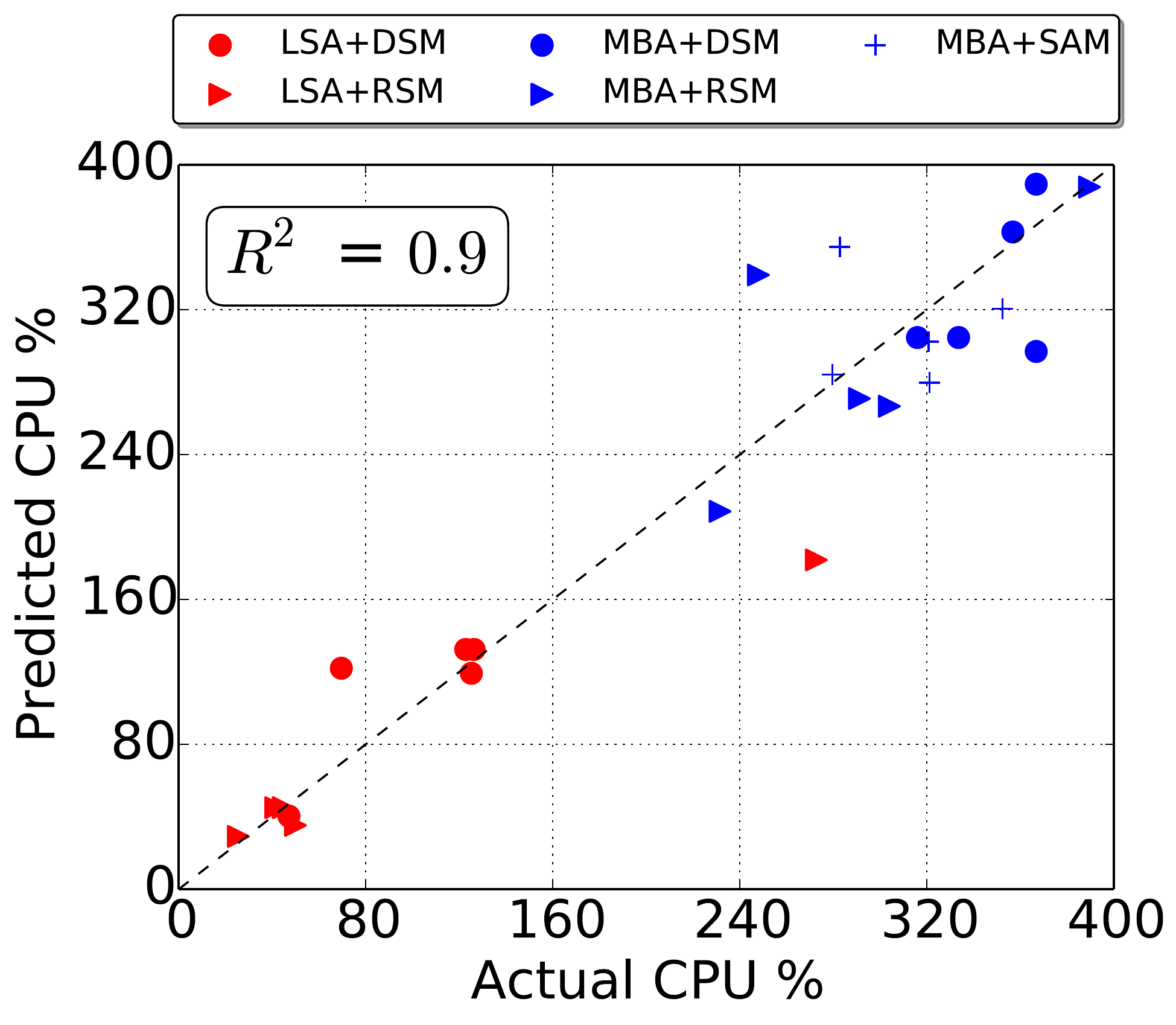}
		\label{fig:cpu:star}
	}
	\caption{Scatter plot of \emph{Predicted} and \emph{Actual} CPU\% per VM for the Micro-DAGs on 5 VMs using the 5 scheduling strategy pairs}
	\label{fig:plot:cpu} 
\end{figure*}

\begin{figure*}[t]
	\centering
	\subfloat[Linear DAG]{
		\includegraphics[width=0.33\textwidth]{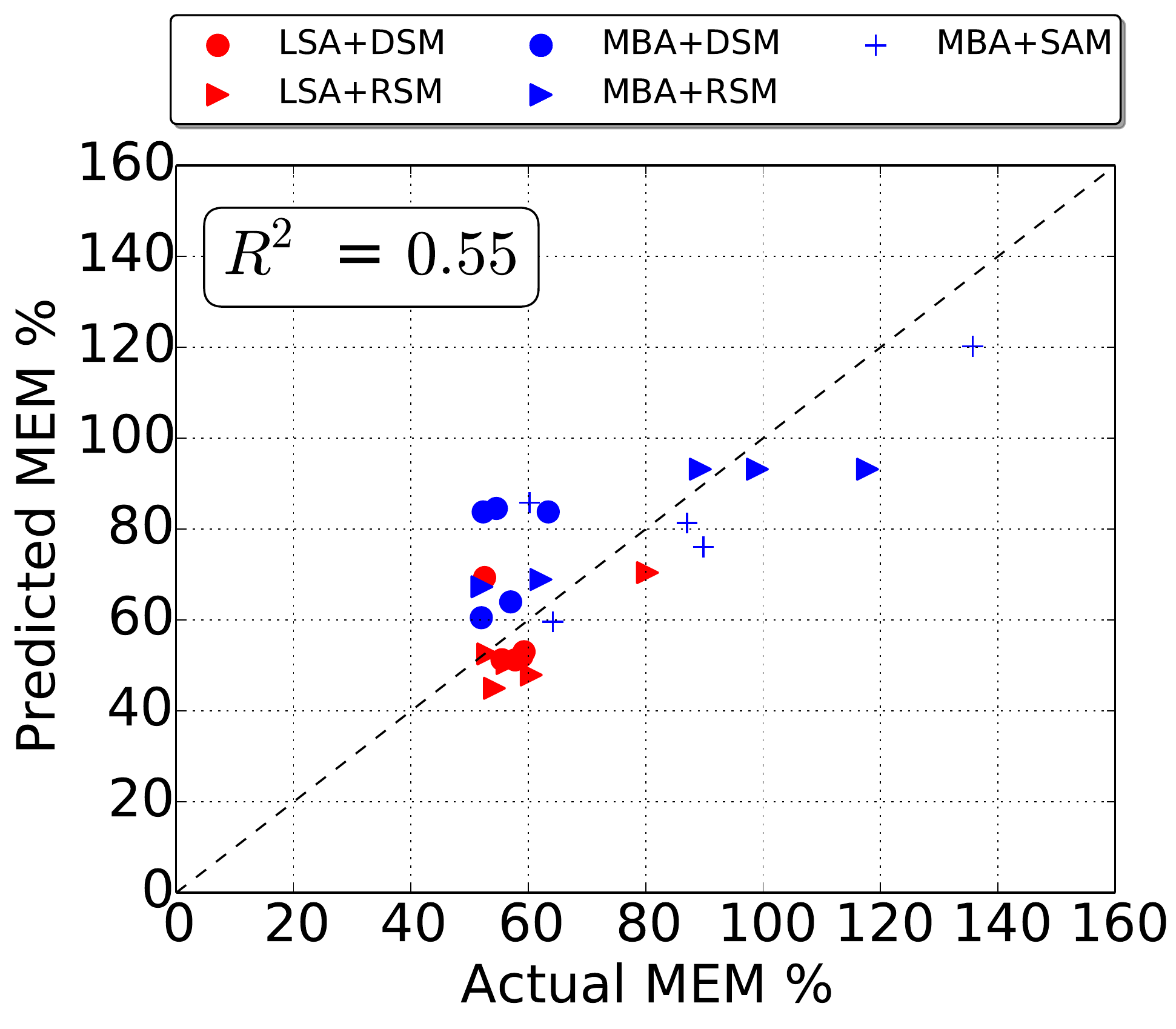}
		\label{fig:mem:linear}
	}
	\centering
	\subfloat[Diamond DAG]{
		\includegraphics[width=0.33\textwidth]{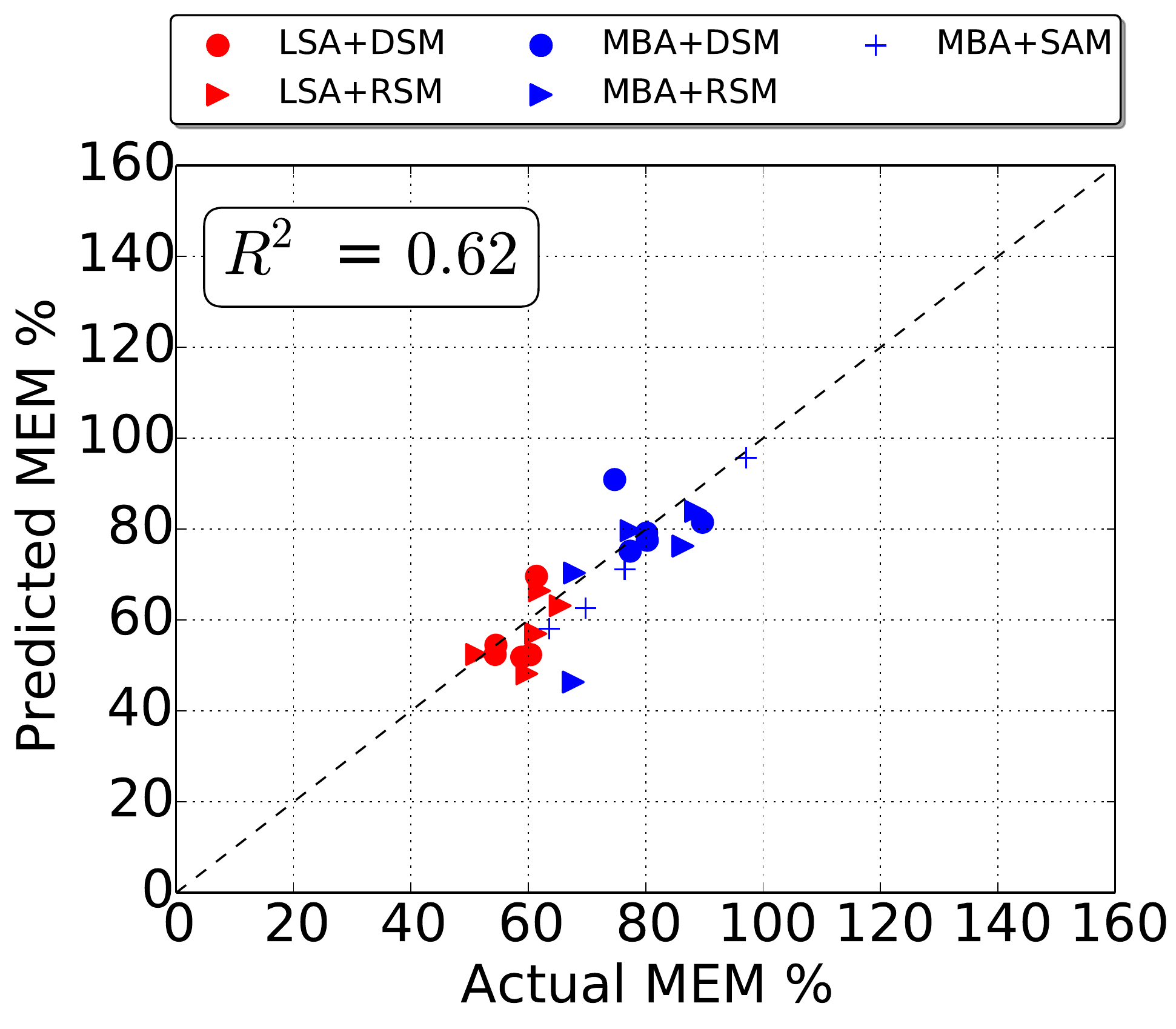}
		\label{fig:mem:diamond}		
	}
	\centering
	\subfloat[Star DAG]{
		\includegraphics[width=0.33\textwidth]{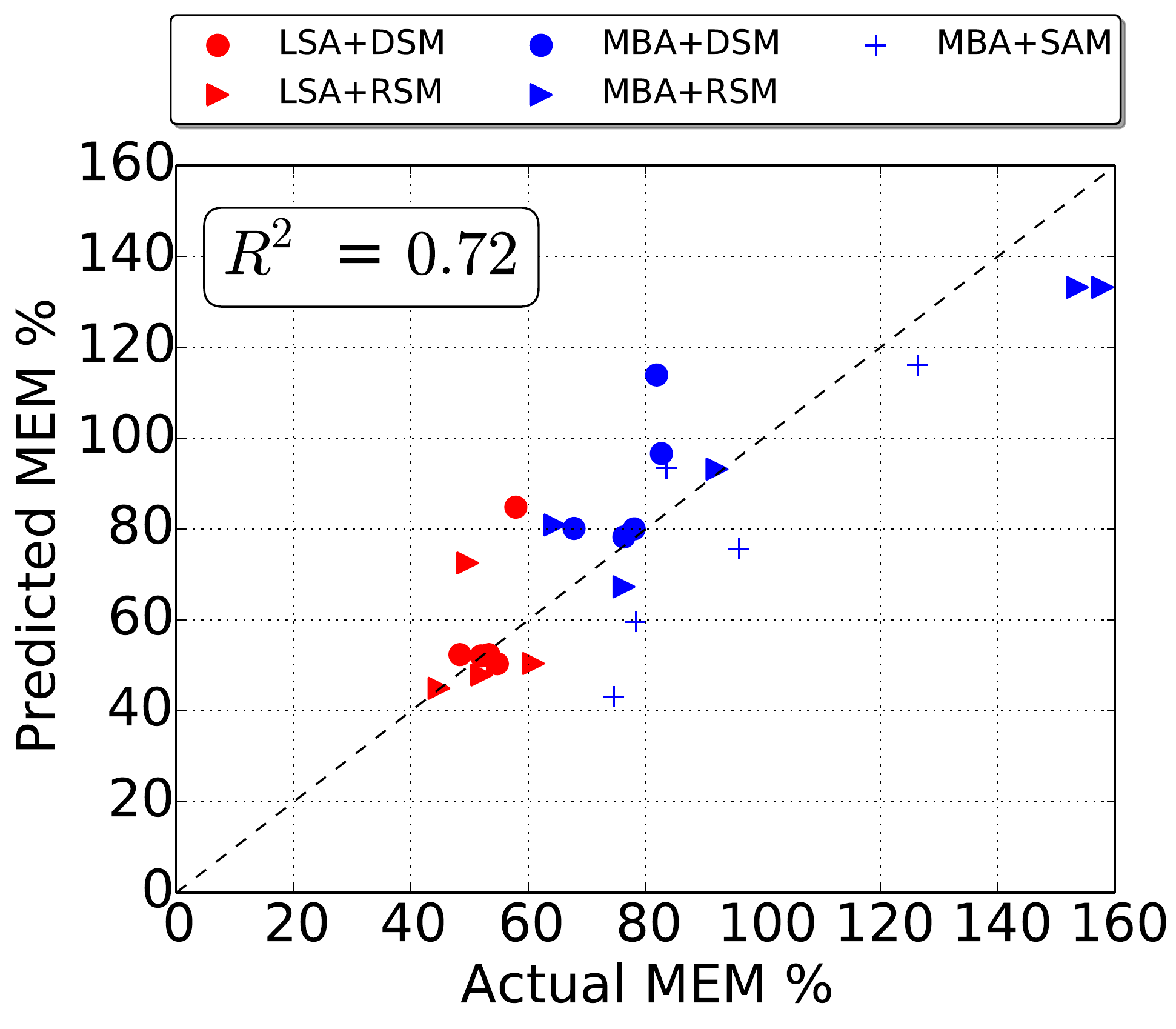}
		\label{fig:mem:star}
	}
	\caption{Scatter plot of \emph{Predicted} and \emph{Actual} Memory\% per VM for the Micro-DAGs on 5 VMs using the 5 scheduling strategy pairs}
	\label{fig:plot:mem} 
\end{figure*}

%
Figs.~\ref{fig:plot:cpu} and ~\ref{fig:plot:mem} show the Actual (X axis) and Predicted (Y axis) CPU\% and memory\% values for the three DAGs. Each scatter plot has a data point for each of the 5 VMs and for every scheduling algorithm pair. 
It is immediately clear from Figs.~\ref{fig:plot:cpu} that our performance model is able to \emph{accurately predict the CPU\% for each VM for these DAGs with a high correlation coefficient $R^2 \ge 0.81$.} This consistently holds for all three DAGs, scheduling algorithms, and for CPU utilization that ranges from $10-90\%$.

While for the Linear DAG, the CPU utilization accuracy is high, there are a few cases in the Diamond and Star DAGs where our predictions deviate from the observed for higher values of CPU\%. The under-prediction of CPU for Diamond DAG with MBA+SAM is because the VMs with Pi thread bundles receive a slightly higher input rate than expected due to Storm's shuffle grouping that impacts $4$ of the $5$ slots, and Pi's CPU model is sensitive to even small changes in the rate. 
For e.g., in Fig.~\ref{fig:cpu:diamond}, a VM with a predicted CPU use of $80\%$ for a predicted input rate of $110~tuples/sec$ ends up having an actual CPU usage of $88\%$ as it actually gets $116~tuples/sec$. 
This happens for the Star DAG with Pi and Blob threads we well. 
As mentioned before, enhancing Storm's shuffle grouping to be sensitive to resource allocation for downstream slots will address this skew while improving the performance as well. At the same time, just from a modeling perspective, it is also possible to capture the round-robin routing of Storm's shuffle grouping in the model to improve the predictability.


For Star DAG in Fig.~\ref{fig:cpu:star}, there is one VM whose predicted CPU\% is more than the actual for both MBA+RSM and MBA+SAM. We find that both these VMs have $2$ threads of the Parse task, each on a separate slot, that are meant to support a required input rate of $480~tuples/sec$.  
However, a single thread of Parse supports $310~tuples/sec$. Since these two threads receive less than the peak rate of input, our model proportionally scales down the expected resource usage and estimates it at $~47\%$ CPU usage. However, the actual turns out to be $~32\%$, causing an overestimate. As we mentioned, there is a balance between the costs for building fine-grained models and the accuracy of the models, and this prediction error is an outcome of this trade-off that causes us to interpolate.



For the Diamond DAG in Fig.~\ref{fig:cpu:diamond}, we see two VMs with expected CPU\% of $\approx100\%$ for MBA+RSM but the observed values that are much lesser. These correspond to two Pi threads that the MBA algorithm expects the pair to be placed on the same slot with $95\%$ combined usage while RSM actually maps them onto different VMs with $10\%$ fewer usage by each for 1 thread.

%

The prediction of memory utilization shown in Figs.~\ref{fig:plot:mem}, while not as good as the CPU\% is still valuable at $R^2 \ge 0.55$. Unlike the CPU usage that spans the entire spectrum from $0-400\%$ for each VM, the memory usage has a compact range with a median value of $\approx 60\%$. This indicates that the DAGs are more compute-bound than memory-bound. Due to this low memory\%, even small variations in predictions has a large penalty on the correlation coefficient. 

We do see a few outlying clusters in these memory scatter plots. In the Linear and Star DAGs, we see that MBA+DSM over-predicts the memory usage. This is because the round-robin mapping of DSM assigns single threads of XML Parse to different slots, each of which receive fewer than their peak supported input rate. As a result, our model proportionally scales down the resources but ends up with an over-estimate.  

On the other hand, we also see cases where we marginally under-predict the memory usage for these same DAGs for MBA+SAM. 
Here, the shuffle grouping that sends a higher rate than expected to some slots with full thread bundles, and consequently a lower to other downstream threads, causing the resource usage to be lower than expected. 

We also see broader resource usage trends for specific scheduling approaches that can impact their input rate performance. 
%
%
\emph{We see that plans that use LSA consistently under-utilize resources.} The CPU\% used is particularly bad for LSA, with the 5 VMs for LSA-based plans using an average of just $15-35\%$ CPU each while the MBA-based schedules use an average of $70-90\%$ per VM. 
This reaffirms our earlier insight that the allocation of the number of data-parallel threads by LSA is inadequate to utilize the resources available in the given VMs. 
Among DSM and RSM, we do see that RSM clearly has a better CPU\% when using LSA though the memory\% between DSM and RSM is relatively similar. The latter is because RSM ends up distributing memory intensive threads across multiple slots due to constraints on a slot, which has a pattern similar to DSM's round-robin approach. 
This shows the benefits of RSM over DSM, as is also seen in the input rates supported.

However, \emph{RSM has a more variable CPU\% and memory\% utilization across VMs irrespective of the allocation.} For e.g. in Fig.~\ref{fig:cpu:linear}, the Linear DAG has CPU\% that ranges from $10-40$ for LSA+RSM and from $55-90$ for MBA+RSM. This is because RSM tries to pack all slots in a VM as long as the cumulative CPU\% for the \emph{VM} and the memory\% \emph{per slot} is not exhausted. This causes the CPU\% of initially mapped VMs to grow quickly due to the best-fit distance measure, while the remaining VMs are packed with more memory-heavy tasks. 
This causes the skew. The DSM mapping uses a round-robin distribution of threads to slots and hence has is more tightly grouped. While SAM uses a best-fit packing similar to RSM, this is limited to the partial thread bundles, and hence its resource skew across VMs is mitigated.

\subsection{Comparison of Latency}
\begin{figure*}[t]
	\centering
		\includegraphics[width=0.80\textwidth]{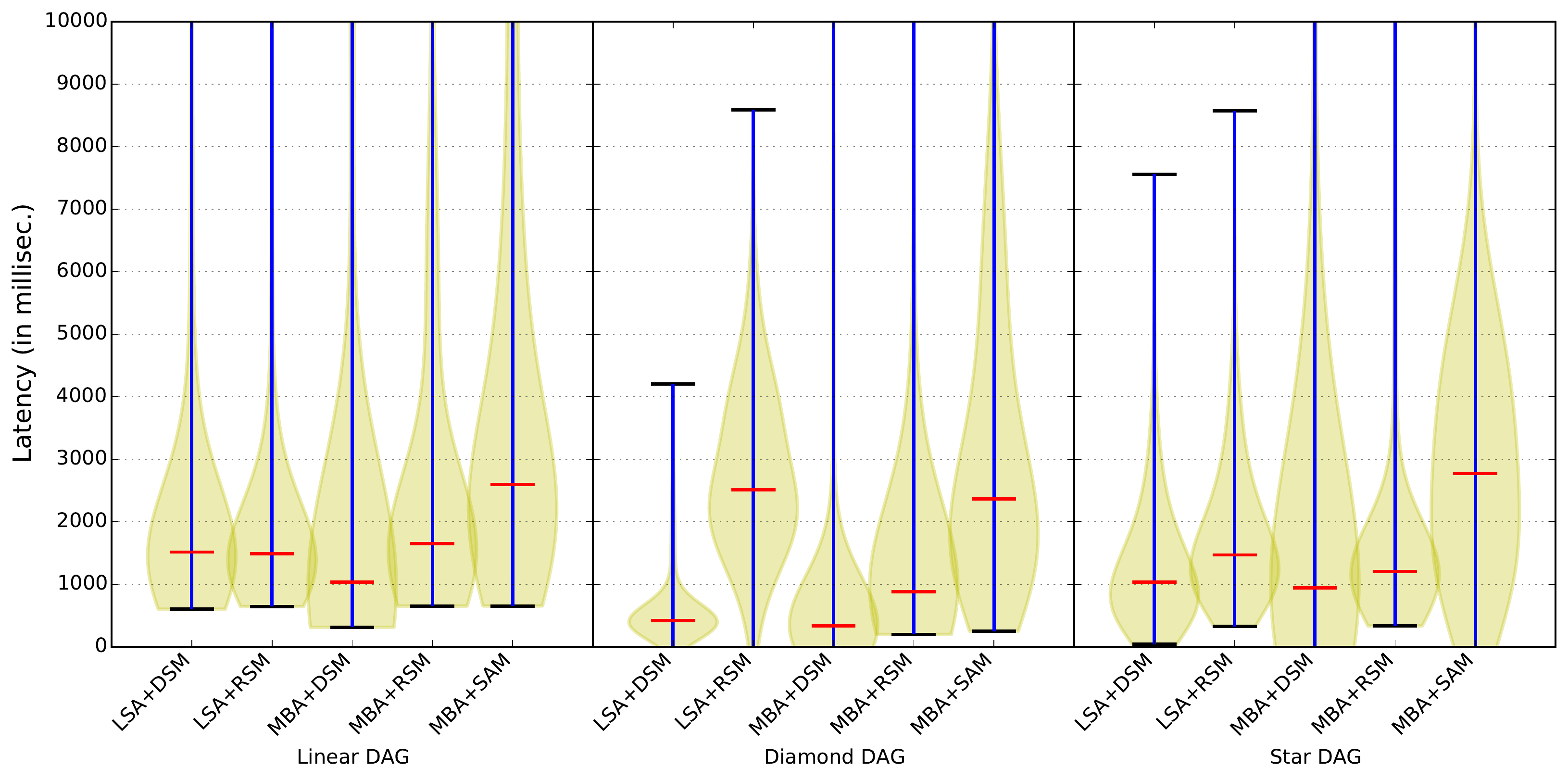}
	\caption{Violin Plot of observed \emph{latency per tuple}  for the Micro-DAGs on 5 VMs using the 5 scheduling strategy pairs}
			\label{fig:latency:dag}
\end{figure*}

Reducing the latency is not an explicit goal for our scheduling algorithms, though ensuring a stable latency is. However, some applications may require a low absolute latency value that is a factor in the schedule generator. So we also report the average latency distribution observed for the different scheduling algorithm pairs for the three micro-DAGs executed on a static set of 5 VMs. 

The \emph{average latency} of the DAG is the average of time difference between each message consumed at the source tasks and all its causally dependent messages that are generated at the sink tasks. The latency of a message depends on the input rate and resources allocated to the task. It includes the network, queuing and processing time of tuple. The average latency is relevant only for a DAG with a stable latency and resource configuration.

We have used separate spout and sink tasks for logging each input and output tuple with a timestamp, and use this to plot the distribution of the average latency for a DAG. Fig.~\ref{fig:latency:dag} shows a Violin Plot for the average latency for the three micro-DAGs executed on 5 VMs using both LSA and MBA based allocation with DSM, RSM, MBA mappings, at stable rate. These results are for the same experiment setup as \S~\ref{sec:results:accy}. 

We can make a few observations based on these plots. We see that the Diamond micro-DAG has a consistently lower latency, followed by the Star DAG and then the Linear DAG. As is evident from the dataflow structure, this is proportional to the number of tasks on the critical path of the DAG, from the source to the sink. This is $4$ for Diamond, $5$ for Star and $7$ for Linear.
%

The median observed latency values typically increase in the order: MBA+DSM $\le$ \{LSA+DSM, MBA+RSM\} $\le$ LSA+RSM $\le$ MBA+SAM.  
However, this has to be tempered by the input rates that these schedules support for the same DAG and resource slots. While MBA+DSM has a low latency, it supports the lowest rate among the three scheduling pairs that use MBA, though all support a higher rate than the LSA-based algorithm pairs. So this is suitable for low latency and average throughput. MBA+RSM has the next best latency given that RSM is network-aware and hence, able to lower the network transfer latency. This is positive given that it is also able to support a high input rate. The LSA+RSM schedule have the second worst latency and also the worst input rate, seen earlier. So this algorithm pair is not a good selection. Separately, we also report that the MBA schedules have a long tail distribution of latency values, indicating that the threads are running at peak utilization that is sensitive to small changes.

\section{Related Work}
\label{sec:related}


\subsection{Scheduling for Storm}

The popularity of Storm as a DSPS has led to several publications on streaming DAG scheduling that is customized for Storm. 
%
As discussed before, Storm natively supports the default round-robin scheduler and the R-Storm resource-aware scheduler. Both of these only participate in the mapping phase and not in thread or resource allocation.
The round-robin scheduler \cite{rychly:cisis:2014}does not take into account the actual VM resources required for a thread  
instead distributes them uniformly on all available VMs. 
%

In R-storm~\cite{peng:middleware:2015},  
the user is expected to provide the CPU\%, memory\% and network bandwidth for each task thread under a stable input message rate, along with the number of threads for each task. It uses its resource-aware distance function to pack threads to VMs with the goal of achieving a high resource utilization and minimum network latency costs. As we have shown earlier, this linear extrapolation is not effective in many cases. Further, R-Storm does not consider the input rates to the DAG in its model. This means the resource utilization provided by the user is not scaled based on the actual rate that is received at a task thread. Our techniques use both a performance model and interpolation over it to support non-linear behavior and diverse input rates that make it amenable to efficient scheduling even under dynamic conditions.

However, R-Storm is well suited for collections of dataflows that execute on large, shared Strom clusters with hundred's VMs that can be distributed across many racks in the data center. Here, the network latency between VMs vary greatly depending on their placement, and this can impact applications that have a high inter-task communication. 
Our algorithms do not actively consider the network latency other than scheduling the threads in BFS order, like R-Storm, to increase the possibility of collocating neighboring task threads in the DAG on the same slot. Consequently, our latency values suffer even as we offer predictable performance. Our target is streaming applications launched on a PaaS Storm cluster with tens of VMs that have network proximity, and for this scenario, the absence of network consideration is acceptable. That said, including network distance is a relatively simple extension to our model. 



Others have considered \emph{dynamic scheduling} for Apache Storm as well, where the scheduler adapts to changes in input rates at runtime. Latency is often the objective function they try to minimize while also limiting the resource usage. \cite{aniello:debs:2013} proposes an offline and an online scheduler which aim at minimizing the inter-node traffic by packing threads in decreasing order of communication cost 
into the same VM, while taking CPU capacity as constraint based on the resource usage at runtime. 
The goal here is on the mapping phase as well with the number of threads and slots for the tasks assumed to be known \emph{a priori}. 
The offline algorithm just examines the topological structure and does not take message input rate or resource usage into consideration for scheduling. It just place the threads of adjacent tasks on same slot and then slots are  assigned to worker nodes in round robin fashion. The online algorithm monitors the communication patterns and resource usage at run time to decide the mapping. It tries to reduce the inter-node and inter-slot traffic among the threads. The online algorithm have two phases, In the first phase threads are partitioned among the workers assigned to DAG, minimizing the traffic among threads of distinct workers and balancing the CPU on each worker. In the second phase these workers are assigned to available slots in the cluster, minimizing the inter-node traffic. Both these algorithms uses tuning parameters that controls the balancing of threads assigned per worker. These algorithms also have the effect of reducing intra-VM communication traffic besides inter-VM messaging.


\emph{T-Storm}~\cite{xu:icdcs:2014} also takes a similar mapping approach, but uses the summation of incoming and outgoing traffic rates for the tasks in descending order to decide the packing of threads to a VM. It further seeks to limit the messaging within a VM by just running one worker on each VM that is responsible for all threads on the VM. 
The algorithm monitors the load at run time and assigns the thread to available slot with minimum increamental traffic load. The number of threads
for each task is user defined and their distribution 
among worker nodes is controlled by some parameter (consolidation factor), which is obtained emperically. Also, algorithm does not gurantee that communicating threads will be mapped to the same node as ordering is done based on total traffic and not on traffic between threads.

Both these schedulers~\cite{xu:icdcs:2014,aniello:debs:2013} use only CPU\% as the packing criteria at runtime, and this can cause memory overflow for memory intensive tasks or when the message queue size grows.   
Their online algorithms also require active monitoring of the various tasks and slots at runtime, which can be invasive and entail management overheads. At the same time, despite their claim of adapting to runtime conditions, neither scheduler actually acquires new VM resources or relinquishes them, and instead just redistributes the existing threads on the captive set of VMs for load balancing and reducing the active slots. Thus, the input-rate capacity of the dataflow is bounded or the acquired captive resources can be under-utilized.  
Further, the use of a single worker per VM in T-Storm can be a bottleneck for queuing and routing when the total input and output rate of threads on that VM are high.
While we do not directly address dynamic scheduling, our algorithms can potentially be rerun when the input rate changes to rebalance the schedule, without the need for fine-grained monitoring. This can include acquiring and releasing VMs as well since we offer both allocation and mapping algorithms. We consider both memory and CPU usage in our model-based approach. A predictable schedule behavior is a stated goal for us rather than reducing absolute latency through reduced communication. 
The \emph{P-Scheduler}~\cite{eskandari:acsw:2016} uses the ratio of total CPU usage from all threads to the VM's CPU threshold to find the number of VMs required for the DAG at runtime. The goal here is to dynamically determine the number of VMs required at runtime based on CPU usage and then map the threads to these VMs such that the traffic among VMs is minimized. The mapping hierarchically partitions the DAG, with edge-weights representing tuple transfer rate. It first maps threads to VMs and then re-partitions threads within the VM to slots. This reduces the communication costs but the partitioning can cause unbalanced CPU usage, and the memory usage is not considered at all. 
While the algorithm does VM allocation, it does not consider thread allocation that can cause VMs to be under-utilized. 
It also requires a centralized global monitoring of the data rates between threads and CPU\% to perform the partitioning and allocation.
As mentioned before, our scheduling offers both VM and thread allocation in addition to mapping, consider input rate, CPU\% and memory\% for the decisions, and our model does not rely on active runtime monitoring.

There have been few works on resource and thread allocation for Storm. The \emph{DRS}~\cite{fu:icdcs:2015} is one such scheduler that models the DAG as open queuing network to find the expected latency of a task for a given number of threads. Its goal is to limit the expected task latency to within a user-defined threshold at runtime, while minimizing the total number of threads required and hence the resources. 
It monitors the tuple arrival and service rate at every task to find the expected latency using \emph{Erlang formula}~\cite{tijms:stochastic:1986}. The approach first finds the expected number of threads required for the task 
so that latency bound is met.
This is done by increasing a thread for the task which gives maximum latency improvement obtained by Erlang formula discussed in paper, but it requires that an upper bound on total number of threads to be set by user. Also paper assumes that only a fixed number of threads can run on a VM, independent of thread type.
The number of VMs are identified using total number of threads and number of threads that can run on a VM, already fixed by user. Mapping is done by default scheduler only.
DRS uses an analytical approach like us, but based on queuing theory rather than empirical models. They apply it for runtime application but do not consider mapping of the threads to VM slots. We consider both allocation and mapping, but do not apply them to a dynamic scenario yet. Neither approaches require intensive runtime monitoring of resources and rates. Like us, they consider CPU resources and threads for allocation and not network, but unlike us, they do not consider memory\%. Their experiments show a mismatch between expected and observed performance from failing to include network delays in their model while our experiments do not highlight network communication to be a key contributor, partly because of the modest rates supported on the DAGs and resources we consider. They also bound the total number of CPU slots and the number of threads per VM, which we relax. 
%

\subsection{Scheduling for DSPS}


While our scheduling algorithms were designed and evaluated in the context of Storm, it is generalizable to other DSPS as well. There has been a body of work on scheduling for DSPS, both static and adaptive scheduling on Cloud VMs, besides those related to Storm.  


Borealis~\cite{abadi:cidr:2005} an extnesion to Aurora~\cite{carney:vldb:2003} provides parallel processing of streams. It uses local and neighbor load information for balancing
load across the cluster by moving operators. They also differ from cloud based DSPS as they assume that only fixed amount resources are available beforehand. Some extensions to Borealis like ~\cite{xing:icde:2005,pietzuch:icde:2006}, does not use intra operator level parallelism and considers only dynamic mapping of tasks for load balancing. 
TelegraphCQ~\cite{chandrasekaran:sigmod:2003} uses adaptive routing using special operators like Eddy and Juggle to optimize
query plans. These special operators decides how to route data to different operators, reorders input
tuples based on their content. It also dynamically decides the optimal stream partitioning 
for parallel processing. These systems allocate queries to seperate nodes for scaling with the
number of queries and are not designed to run on cloud.

COLA~\cite{khandekar:middleware:2009} for System S, scalable distributed stream  processing system aims at finding best operator fusion (multiple operators within same process) possible for reducing inter process stream traffic.The approach first uses list scheduling (longest processing time) to get operators schedule then it checks for VM capacity(only CPU) if schedule is unfeasible, uses graph partitioning to split processing element to other VMs.Thus COLA also does not take memory intensive operators in to account.
Infosphere streams~\cite{biem:sigmod:2010} uses component-based programming model. It helps in composing and reconfiguring individual components to create different applications. The scheduler component~\cite{wolf:middleware:2008} finds the best partitioning of data-flow graph and distributes it across a set of physical nodes. It uses the computational profiles of the operators, the loads on the nodes for making its scheduling decisions.  
Apache S4~\cite{neumeyer:icdm:2010} follows the actor model and allocation is decided manually based on distinct keys in the input stream. The messages are distributed across the VMs based on hash function on all keyed attribute in input messages. S4 schedules parallel instances of operators but does not manage their parallelism and state. Since it does not support dynamic resource management thus unable to handle varying load.
IBM System S~\cite{amini:dmssp:2006} run jobs in the form of data-flow graphs. It supports intra-query parallelism but management is manual. It also supports dynamic application composition and stream discovery, where multiple applications can directly interact. This support for sharing of streams across applications is done by annotating the messages with already declared types in global type system. This enables sharing of applications written by different developers through streams.

Esc~\cite{satzger:cloudcomputing:2011} which process streaming  data as key-value pairs. Hash functions are used to balance load by dynamically mapping the keys  to the machines and function itself can be updated at run time. Hash function can also use the cpu,memory load based on the VM statistics for message distribution. Dynamic updation of the DAG based  on the custom rules from user is also supported for load balance. A processing element in a DAG can have multiple operators and can be created at run time as per need. There can be many workers for a processing element. Since it dynamically adjusts the required computational resources based on the current load of the system it is good fit for use cases with varying load, with deployment on cloud.

 \cite{kumbhare:sc:2013} have used variant called dynamic dataflows that adapts to changing performance of cloud resources by using alternate processing elements.The logic uses variable sized bin packing for allocating processing element over the VMs on cloud. Dynamic rates are managed  by allocating resources for alternate processing elements thus making tradeoff between cost and QoS on cloud.

In ~\cite{gedik:tpds:2014} have proposed elastic auto-parallelization for balancing the dynamic load in case of  SPL applications. The scaling is based on a control algorithm that monitors the congestion and throughput at runtime to adjust data parallelism. Each ope	rator maintains a local state for every stream partition. An incremental migration protocol is proposed for maintaining states while scaling, minimizing the amount of state transfer between hosts.

StreamCloud~\cite{gulisano:tpds:2012} 
modifies the parallelism level by splitting queries into sub queries minimizing the distribution overhead of parallel processing, each of which have utmost one stateful operator that stores its state in a \emph{tuple-bucket}, where the key for a state is a hash on a tuple. At the boundary between sub-queries, tuples are hashed and routed to specific stateful task instances that hold tuple-bucket with their hash key. This ensures consistent stateful operations with data-parallelism. It uses special operators called Load Balancers placed over outgoing edge of each instance of subcluster, LB does Bucket Instance Mapping to map buckets with instances of downstream clusters.

ChronosStream~\cite{wu:icde:2015}  hash-partitions computation states into collection of fine-grained slices and then distributes them to muliple nodes for scaling. Each slice is a computationally independent unit associated with a subset of input streams and and can be transparently relocated without affecting the consistency. The elasticity is achieved by migrating the workload to new nodes using a lightweight transactional migration protocol based on states.

ElasticStream~\cite{ishii:cloud:2011} considers a hybrid model for processing streams as it is impossible to process them on local stream computing environment due to finite resources. The goal is to 
minimize the charges for using the cloud while satisfying the SLA, as a trade-off between the application’s latency and charges uisng linear programming. The approach dynamically adjusts the resources required with dynamic rates in place of  over-provisioning with fixed amount of resources.
The implementation done on System S is able to assign or remove computational resources dynamically.

Twitter Heron~\cite{kulkarni:twitter:2015} does user defined thread allocation and mapping by Aurora scheduler. In the paper ~\cite{li:vldb:2015} proposed an analytical model for resource allocation and dynamic mapping to meet latency requirement while maximizing throughput, for processing real time streams on hadoop.  Stela ~\cite{xu:ic2e:2016} uses effective throughput percentage (ETP) as the metric to decide the task to be scaled when user requests scaling in/out with given number of machines. The number of threads required for the tasks and their mapping to slots is not being discussed in the paper.

\subsection{Parallel Scheduling}

Our model-based approach is similar to scheduling strategies employed in parallel job and thread scheduling for HPC applications. 

The Performance Modeling frameworks~\cite{snavely:sc:2002,bailey:ecpp:2005,snavely:apc:2004}  for large HPC systems predicts application performance from a function of system profiles (e.g., memory performance, communication
performance). These profiles can be analysed to improve the application performance by understanding the tuning parameters. Also~\cite{bailey:ecpp:2005} proposes methods to reduce the time required for performance modelling, like combining the results of several
compilation, execution, performance data analysis cycles into a application signature,
so that these steps need not to be repeated each time a new performance question is asked.

Warwick Performance Prediction (WARPP)~\cite{hammond:icst:2009} simulator is used to construct application performance models for complex parallel scientific codes executing
on thousands of processing cores. It utilises coarse-grained compute
and network models to
enable the accurate assessment of parallel application behaviour at large scale. The simulator exposes six types of discrete events ranging from compute to I/O read,write to generate events representing the behaviour of a parallel application.
~\cite{gahvari:ics:2011} models the aplication performance for future
architectures with several millions or billions of cores. It considers algebraic multigrid (AMG), a popular and highly efficient iterative solver to discuss the model-based predictions. It uses local
computation and communication models as baseline for predicting the performance and its scalability on future machines. The paper~\cite{hong:sigarch:2009}
proposes simple analytical model to predict the execution time of massively parallel programs on GPU architecture. This is done by estimating the number of parallel memory requests by considering the number of running threads and memory bandwidth. The aplication execution time in GPU is dominated by the latency of memory instructions. Thus by finding the number of memory requests that can be executed concurrently (memory warp parallelism) the effective costs of memory requests is estimated.

~\cite{hovestadt:wjsp:2003} proposes Planning systems and compares them to Queuing systems for resource managament in HPC. Features like advance resource reservation, request diffusing can not be achieved using queuing because it considers only present resource usgae. Planning systems like CCS, Maui Scheduler does resource planning for present and future by assigning start time to all requests and using run time estimates for each job.

Recent works like~\cite{escobar:hipc:2016} uses statistical approach to predict application execution time using emperical analysis of execution time for small input sizes. The paper uses a collection of well known kernel benchmarks for modelling the execution time of each phase of an application. The approach collects profiles obtained by few short application runs to match phases to kernels and uses it for predicting the execution times accurately.

Our model-based mapping of a bundle of threads also has some similarities with \emph{co-scheduling}~\cite{ousterhout:icdcs:1982} or \emph{gang scheduling}~\cite{feitelson:jpdcs:1992} of threads in concurrent and parallel systems. In the former, a set of co-dependent processes that are part of a working set are scheduled simultaneously by the Operating System (OS) on multi-processors to avoid process thrashing. In gang scheduling, a set of interacting threads are scheduled to concurrently execute on different processors and coordinate through busy-waits. The intution is to assign the threads to dedicated processors so that all dependent threads progress together without blocking. Our allocation and mapping based on performance models tries to identify and leverage the benefits of co-scheduling coarse-grained thread bundles from the same task onto one CPU, with deterministic performance given by the models, and by separating out thread bundles from different tasks onto different slots to execute concurrently without resource contention. 

We also encounter issues seen in gang scheduling that may cause processor over or under-allocation if sizes of the gangs do not match the number processors, which is similar to the partial thread bundles mapped to the same slot in our case, causing interference but reusing partial slots~\cite{feitelson:jc:1990}. At the same time, we perform the thread to slot mapping once in the streaming dataflow environment, and do not have to remap unless the input rate changes. Hence, we do not deal with recurrent or fine-grained scheduling issues such as constantly having to schedule threads since the number of threads are much more than the CPU cores, paired gang scheduling for threads with interleaved blocking I/O and compute~\cite{wiseman:tpds:2003}, or admission control due to inadequate resources~\cite{batat:ipdps:2000}.





\section{Conclusion}
\label{sec:conclusions}

Based on these results, we see that LSA+RSM consistently allocates more resources than MBA+SAM, often twice as many slots due to its linear extrapolation of rate and resources. However, it still misses the planned input rate supported by $30-40\%$ in several cases due to unbalanced mapping by RSM where the rate does not scale as it expects. We see a $5-10\%$ drop for MBA+SAM due to the shuffle grouping that uniformly routes tuples to threads. Also, RSM often requires additional resources than ones allocated by LSA due to fragmentation during bin-packing, though this tends to be marginal. SAM has less fragmentation due to packing full bundles to exclusive slots. These hold for all DAGs, both micro and application, small and large.

The model-based prediction of input rates is much more accurate than the planned prediction, correlating with the actual rate with and $R^2 \ge 0.71$. The few outliers we see are due to the model expecting a different routing compared to Storm's shuffle grouping, and due to the interpolation of rates based on the granularity of the performance models. 
MBA is consistently is better than LSA in the input rate supported for the same quanta of resources, through MBA+DSM shows the least improvement. MBA+RSM is often better than MBA+SAM in actual rate though MBA+SAM gives a predictable observed rate. 

Our performance model is able to predict the resource utilization for individual VMs with high accuracy, having $R^2$ value $\geq0.81$ for CPU\% and $\geq0.55$ for MEM\%, independent of the allocation and mapping technique used. The few prediction errors we see are due to threads receiving fewer than the peak rate for processing, where our model proportionally scales down the estimated resource usage relative to a single-thread usage at the peak rate. The low memory\% also causes the error to be sensitive to even small skews in the prediction, giving a lower correlation coefficient value.

MBA consistently has a higher resource utilization than LSA, that is also reflected in the better input rate performance. While the resource usage across VMs for schedules based on MBA are close together, RSM shows a greater variation of its CPU\% and memory\% across VMs.

\section{Future Work}
The current slot aware mapping does not consider load aware shuffle groping, we can leverage it to have more accuracy for predicting supported input rate and resource requirement.
Also Dynamic resource allocation and mapping for the given distribution of input rate or monitored input rate at run time is part of our future work.







\bibliographystyle{plain}
\footnotesize{
	\bibliography{paper}
}

\end{document}